%% file: main.tex
\documentclass[11pt, preprintnumbers, superscriptaddress, nofootinbib]{revtex4} 

\usepackage{subfiles}
\usepackage{ifthen}
\usepackage{amsmath,amssymb}
\usepackage{physics}
\usepackage{mathtools}
\usepackage{slashed}
\numberwithin{equation}{section}
\usepackage{graphicx} 
\usepackage{color}

\newcommand{\BR}{\text{BR}}
\newcommand{\cba}{c_{\beta-\alpha}}
\newcommand{\sba}{s_{\beta-\alpha}}
\newcommand{\tb}{\tan\beta}

\newcommand{\mHp}{m_{H^{\pm}}}
\allowdisplaybreaks[1]

\newcounter{mainflag}
\begin{document}

\title{Radiative corrections to decay branching ratios of the CP-odd Higgs boson \\
in two Higgs doublet models}


\preprint{KEK-TH 2438}
\preprint{OU-HET 1150}
\preprint{TU 1162}

\author{Masashi Aiko}
\email{maiko@post.kek.jp}
\affiliation{KEK Theory Center, High Energy Accelerator Research Organization (KEK), Tsukuba, Ibaraki, 305-0801, Japan}

\author{Shinya Kanemura}
\email{kanemu@het.phys.sci.osaka-u.ac.jp}
\affiliation{Department of Physics, Osaka University, Toyonaka, Osaka 560-0043, Japan}

\author{Kodai Sakurai}
\email{kodai.sakurai.e3@tohoku.ac.jp}
\affiliation{Department of Physics, Tohoku University, Sendai, Miyagi 980-8578, Japan}

\begin{abstract}
We calculate radiative corrections to decay rates of CP-odd Higgs boson $A$ for various decay modes in the four types of two Higgs doublet models with the softly broken discrete $Z_{2}$ symmetry.
The decay branching ratios are evaluated at the next-to-leading order for electroweak corrections and the next-to-next-to-leading order for QCD corrections.
We comprehensively study the impact of the electroweak corrections on the decay rates and the branching ratios.
We find that the radiative corrections can sizably modify the branching ratios, especially for the $A\to Zh$ decay mode in the nearly alignment scenario, where coupling constants of the SM-like Higgs boson $h$ are close to those in the standard model.
We also show correlations between the branching ratios of $A$ and the scaling factor of the SM-like Higgs boson coupling including higher-order corrections.
In addition, we show characteristic predictions on the decay pattern depending on the types of Yukawa interaction, by which we can discriminate the types of Yukawa interaction in future collider experiments.
\end{abstract}

\maketitle
\newpage
\tableofcontents
\newpage

\subfile{01introduction}

%
\subfile{02model}

%
\subfile{03decay_rate}

%
\subfile{04nlo_ew}

%
\subfile{05numerical_result}

\section{Discussions and conclusions} \label{sec: conclusions}
We have discussed the impact of the NLO EW corrections on the decay rates of the CP-odd Higgs boson in the 2HDMs with the softly broken discrete $Z_{2}$ symmetry.
We have calculated the decay rates of the CP-odd Higgs boson based on the improved on-shell renormalization scheme~\cite{Kanemura:2017wtm}.
The decay rates of the CP-odd Higgs boson decays into a pair of quarks and leptons, $Z$ and the neutral Higgs bosons, and $W^{\pm}$ and the charged Higgs bosons have been calculated including the NLO EW corrections.
On the other hand, the one-loop induced decays, $A\to W^{\pm}W^{\mp},\, ZZ,\, \gamma\gamma,\, Z\gamma$ and $gg$, have been calculated at LO EW.
We have presented the explicit formulae for the decay rates including the NLO EW corrections and higher-order QCD corrections.
The NLO EW corrections have been written in terms of the renormalized vertex functions, and the analytical formulae have been presented.

The size of the NLO EW corrections on the partial decay widths, total decay widths and the decay branching ratios have been comprehensively studied in the near alignment scenario, where the couplings constants of the discovered Higgs boson are close to those in the SM.
We have found that the decay branching ratio of $A\to Zh$ can be negative in the near alignment scenario.
This is because the tree-level vertex becomes quite small, and the NLO correction can be larger than the LO contribution.
In this paper, we have solved this problem by introducing the square of the NLO amplitude.
However, we should include the contributions from the tree-level amplitude times the two-loop amplitude since the square of the one-loop amplitude corresponds to the NNLO.
In the alignment limit, this contribution vanishes due to the multiplication of the tree-level amplitude.
Therefore, we expect that the contribution from two-loop diagrams is also sub-leading in the nearly alignment scenario.
For other decay processes such as $A\to f\bar{f}$, we have IR divergences in the one-loop amplitudes.
Therefore, we need to take into account higher-order QED corrections in order to include partial NNLO contributions.
Since the partial decay widths do not become negative except for $A\to Zh$ in the nearly alignment case, we include the partial NNLO contribution only for $A\to Zh$ in this paper.

We have discussed the behavior of the NLO EW corrections to the partial decay widths in detail.
The NLO EW corrections to  $A\to b\bar{b}$ and the total decay width depend on the types of 2HDMs.
On the other hand, those for the other decay modes are almost independent of the types of 2HDMs.
We have also analyzed the correlations between the branching ratios of $A$ and the scaling factor of the $hZZ$ coupling including higher-order corrections.
We have found that the NLO EW corrections sizably modify theoretical predictions at LO.
For example, it is expected that $\mathrm{BR}(A\to Zh)$ and $\Delta\kappa_{Z}$ become large at the same time because both of them are promotional to $\abs{c_{\beta-\alpha}}$ at LO. 
However, $\Delta\kappa_{Z}$ can be sizable even when $\mathrm{BR}(A\to Zh)$ is small due to the non-decoupling effect.
We have shown that the NLO EW corrections sizably modify the decay branching ratio, and those behaviors have been discussed in detail.

We have shown that there are characteristic predictions in the decay pattern of the CP-odd Higgs boson depending on the types of the 2HDMs.
We can discriminate the types of the 2HDMs by the correlations between $A\to b\bar{b}$ and $A\to \tau\bar{\tau}$ if the size of these decay branching ratios is enough large.
When the decay branching ratios of $A\to b\bar{b}$ and $A\to \tau\bar{\tau}$ are small, the CP-odd Higgs boson mainly decays into $t\bar{t},\, Zh$ and $ZH$ in our scenario, and we would be able to determine the types of 2HDMs through the decay patterns of the additional CP-even Higgs boson and the deviations in the SM-like Higgs boson couplings.

\begin{acknowledgments}
We would like to thank Mariko Kikuchi and Kei Yagyu for their useful comments.
This work is supported in part by JSPS KAKENHI Grant No.~20H00160, No.~21F21324~[S.K.], No.~20H01894, No.~21K20363~[K.S.], and No.~22J01147~[M.A.].

\end{acknowledgments}
\clearpage

\appendix
\subfile{A01inputs}

\subfile{A02scalar_couplings}

\subfile{A03three_body}

\subfile{A04loop_induced}

\subfile{A05vertex_function}

\subfile{A06real_photon_emission}

\subfile{A07numerical_result}

\bibliographystyle{junsrt}
\bibliography{refs}

\end{document}

%% file: 01introduction.tex
\section{Introduction}
After the discovery of the new particle with a mass of 125 GeV at the LHC in 2012~\cite{Aad:2012tfa, Chatrchyan:2012ufa}, it has turned out that its properties are in agreement with those of the Higgs boson in the standard model (SM) under the current experimental and theoretical uncertainties.
While no signal for new physics beyond the SM has been observed yet, there are phenomena that cannot be explained within the SM such as dark matter, baryon asymmetry of the universe and tiny neutrino masses.
In addition to these phenomenological problems, there are theoretical problems in the SM such as the hierarchy problem, incomplete descriptions for the gauge coupling unification and the flavor structure, and so on.
Therefore, the SM must be replaced by a more fundamental theory.

While the Higgs boson was found, the structure of the Higgs sector remains unknown.
There is no theoretical principle to insist on the minimal structure of the Higgs sector as introduced in the SM.
The possibility that the Higgs sector takes a non-minimal form is not excluded experimentally at all.
Furthermore, such non-minimal Higgs sectors are often introduced in various new physics models, where the above-mentioned problems are tried to be solved.
Therefore, unraveling the structure of the Higgs sector is one of the central interests of current and future high-energy physics.
The direction of new physics can be determined by reconstructing the Higgs sector experimentally.
Among various extended Higgs models, the two Higgs doublet model (2HDM) is a representative model that contains two CP-even Higgs bosons $h/H$, a CP-odd Higgs boson $A$ and charged Higgs bosons $H^{\pm}$.

Current measurements of the discovered Higgs boson at the LHC show that its couplings with the SM particles are consistent with the SM~\cite{ATLAS:2019nkf, CMS:2020gsy}.
There are two distinctive scenarios to explain this observation in the 2HDM~\cite{Gunion:2002zf, Carena:2013ooa}.
The first one is the decoupling scenario, where masses of the additional Higgs bosons are sufficiently higher than the electroweak scale, and this leads to the SM-like Higgs boson couplings simultaneously.
The second one is the alignment-without-decoupling scenario, where the couplings of the SM-like Higgs boson approximately take their SM values keeping masses of the additional Higgs bosons to be at the electroweak (EW) scale.

In Ref.~\cite{Aiko:2020ksl}, we have studied the testability of such SM-like scenarios at the HL-LHC and future lepton colliders.
It was shown that the so-called Higgs-to-Higgs decays such as $A\to Zh$ and $H\to hh$ are quite important to investigate the 2HDM, especially in the nearly alignment scenario, where the SM-like Higgs boson couplings slightly deviate from those SM values.
In the nearly alignment scenario, there is an upper bound for the typical mass scale of the additional Higgs bosons due to the theoretical constraints such as perturbative unitarity and vacuum stability.
Thus, measuring deviations in the SM-like Higgs boson couplings is also useful to investigate the 2HDM.
Therefore, direct searches of the additional Higgs bosons and indirect studies of the property of the discovered Higgs boson are complementary, and the combined study is powerful to test the extended Higgs sector.

In the indirect study of extended Higgs sectors, the precision calculation for the decay branching ratios of the SM-like Higgs boson is important since the effect of higher-order corrections can be comparable with the precise measurements in the future collider experiments, such as 
the HL-LHC~\cite{ApollinariG.:2017ojx}, the International Linear Collider (ILC)~\cite{Baer:2013cma, Fujii:2017vwa, Asai:2017pwp, Fujii:2019zll}, 
the Future Circular Collider (FCC-ee)~\cite{Gomez-Ceballos:2013zzn} and the Circular Electron Positron Collider (CEPC)~\cite{CEPC-SPPCStudyGroup:2015csa}.
In Refs.~\cite{Aoki:2009ha, Kanemura:2014bqa}, it has been pointed out that various extended Higgs models can be discriminated from the SM by comparing patterns of deviations in Higgs boson couplings at tree level analysis. 
This study then has been extended including one-loop corrections~\cite{Kanemura:2004mg, Kanemura:2014dja, Kanemura:2015mxa, Kanemura:2015fra, Kanemura:2016lkz, Kanemura:2016sos, Kanemura:2017wtm, Aiko:2021nkb}. 
In the context of 2HDMs, many studies for EW corrections to the Higgs boson couplings and/or decays have been performed~\cite{Arhrib:2003ph, Arhrib:2016snv, Kanemura:2004mg, Kanemura:2014dja, Kanemura:2015mxa, Kanemura:2017wtm, Kanemura:2018yai, Kanemura:2019kjg, Gu:2017ckc, Chen:2018shg, Han:2020lta, LopezVal:2010vk, Castilla-Valdez:2015sng, Xie:2018yiv, Altenkamp:2017ldc, Altenkamp:2017kxk, Altenkamp:2018bcs, Kanemura:2016sos, Arhrib:2015hoa, Krause:2019qwe}. 
Several numerical computation tools have been published, e.g., \texttt{H-COUP}~\cite{Kanemura:2017gbi, Kanemura:2019slf}, \texttt{2HDECAY}~\cite{Krause:2018wmo} and \texttt{Prophecy4f}~\cite{Denner:2019fcr}. 

The study for direct searches of the additional Higgs bosons in Ref.~\cite{Aiko:2020ksl} is performed at leading order (LO).
This leads us to investigate the impact of higher-order corrections.
This is because radiative corrections would change the size of partial decay widths significantly since the Higgs-to-Higgs decays are sensitive to the magnitude of deviations in the SM-like Higgs boson couplings.
We have studied the radiative corrections for decays of the charged Higgs bosons in Ref.~\cite{Aiko:2021can}.
The authors in Ref.~\cite{Kanemura:2022ldq} have studied those for decays of the additional CP-even Higgs boson.
In this paper, we study decays of the CP-odd Higgs boson including the higher-order corrections.
The analytical results of decay rates of all the additional Higgs bosons will be implemented in a new version of our developing program \texttt{H-COUP~v3}~\cite{HCOUPv3}.
There are also important previous works done by several other groups for the higher-order corrections to decays of the additional Higgs bosons~\cite{Santos:1996hs, Akeroyd:1998uw, Akeroyd:2000xa, Krause:2016oke, Krause:2016xku, Krause:2019qwe, Su:2019dsf}.
For decays of the CP-odd Higgs boson, the possible size of the next-to-leading order (NLO) EW corrections are discussed in various renormalization schemes in Ref.~\cite{Krause:2019qwe}.
However, its dependence on the model parameters has not been exhibited.

In this paper, we calculate the full set of decay rates of the CP-odd Higgs boson including the higher-order corrections in the 2HDMs with the softly broken $Z_{2}$ symmetry.
We calculate NLO EW corrections to the decay rates of the CP-odd Higgs boson into a pair of quarks and leptons, $Z$ and the neutral Higgs bosons, and $W^{\pm}$ and the charged Higgs bosons.
The one-loop induced decays, $A\to W^{\pm}W^{\mp},\, ZZ,\, \gamma\gamma,\, Z\gamma$ and $gg$, are calculated at LO for EW corrections.
We present the explicit formulae for the decay rates with NLO EW corrections as well as QCD corrections.
The former is written by the renormalized vertex functions for the CP-odd Higgs boson, and the analytical formulae are given in Appendix~\ref{app: vertex}.
We comprehensively study the impact of the electroweak corrections on the partial decay widths and decay branching ratios.
We find that the radiative corrections can sizably modify the branching ratios, especially for the $A\to Zh$ decay mode in the nearly alignment scenario.
In addition, we have characteristic predictions on the decay pattern depending on the types of Yukawa interaction, by which we can discriminate the types of Yukawa interaction in future collider experiments.

What is new in this paper is the following.
First, we provide analytic formulae for the NLO EW corrections to the various decay modes of the CP-odd Higgs boson based on the improved on-shell renormalization scheme~\cite{Kanemura:2017wtm}.
Second, we have newly implemented these results in the \texttt{H-COUP} program~\cite{Kanemura:2017gbi, Kanemura:2019slf}, and behaviors of the higher-order corrections are studied in detail.
We clearly exhibit the dependence of the EW corrections for the decay rates as well as the decay branching ratios on the model parameters.
In addition, we analyze the correlations between the branching ratios of $A$ and the scaling factor of the $hZZ$ coupling including higher-order corrections.
We show that NLO EW corrections sizably modify theoretical predictions at LO.
Finally, we discuss the discrimination of the types of the 2HDM by studying the correlations of the decay branching ratios of the CP-odd Higgs boson.

This paper is organized as follows. In Sec.~\ref{sec: model}, we introduce Lagrangian of the 2HDM, and the constraints on model parameters are discussed. 
In Sec.~\ref{sec: decay_rate}, we give formulae for the decay rates of the CP-odd Higgs boson in terms of the renormalized vertex functions. 
In Sec.~\ref{sec: nlo_ew}, we examine the theoretical behaviors of NLO EW corrections to the decay rates and model parameter dependence on the branching ratios with NLO corrections. 
In Sec.~\ref{sec: numerical_result}, we discuss the impact of NLO EW corrections on the branching ratios and the discrimination of four types of 2HDMs.
Conclusions are given in \ref{sec: conclusions}.
In Appendices, we give analytic expressions for scalar couplings, self-energies and vertex functions of the CP-odd Higgs boson, and decay rates with real photon emission.

%% file: 02model.tex
\section{Two Higgs doublet model} \label{sec: model}
In this section, we define 2HDM Lagrangian to fix our notation.
There are two $SU(2)_{L}$ doublet Higgs fields $\Phi_{1}$ and $\Phi_{2}$ with the hypercharge $Y=1/2$.
We impose a softly-broken $Z_{2}$ symmetry to prohibit tree-level flavor changing neutral currents~\cite{Glashow:1976nt, Paschos:1976ay}.
The $Z_{2}$ charge assignment is shown in Table~\ref{tab: Z2_assignment}.
In addition, we assume CP conservation in the Higgs potential for simplicity.

\subsection{Lagrangian}
\begin{table}[t] 
\begin{center}
\begin{tabular}{l||ccccccc|ccc}\hline
& $\Phi_{1}$ & $\Phi_{2}$ &$Q_L$&$L_L$&$u_R$&$d_R$&$e_R$&$\zeta_u$ &$\zeta_d$&$\zeta_e$ \\\hline\hline
Type-I & $+$ & $-$ & $+$ & $+$ & $-$ & $-$ & $-$ 
& $\cot\beta$ & $\cot\beta$ & $\cot\beta$ \\ \hline
Type-II & $+$ & $-$ & $+$ & $+$ & $-$ & $+$ & $+$ & 
$\cot\beta$&$-\tan\beta$&$-\tan\beta$ \\ \hline
Type-X (lepton specific) & $+$ & $-$ & $+$ & $+$ & $-$ & $-$ & $+$ 
& $\cot\beta$ & $\cot\beta$ & $-\tan\beta$ \\ \hline
Type-Y (flipped) & $+$ & $-$ & $+$ & $+$ & $-$ & $+$ & $-$ & 
$\cot\beta$&$-\tan\beta$&$\cot\beta$ \\\hline
\end{tabular}
\caption{$Z_2$ charge assignments and $\zeta_f$ ($f=u, d, e$) factors in four types of 2HDMs~\cite{Aoki:2009ha}.}
\label{tab: Z2_assignment}
\end{center}
\end{table}

The Higgs potential under the softly-broken $Z_{2}$ symmetry is given by 
\begin{align}
V & = 
m_{1}^{2}\abs{\Phi_{1}}^{2}+m_{2}^{2}\abs{\Phi_{2}}^{2}-\qty(m_{3}^{2}\Phi_{1}^{\dagger} \Phi_{2}+\mathrm{h.c.}) \notag\\
&\quad
+\frac{\lambda_{1}}{2}\abs{\Phi_{1}}^{4}+\frac{\lambda_2}{2}\abs{\Phi_{2}}^{4}
+\lambda_{3}\abs{\Phi_{1}}^{2}\abs{\Phi_{2}}^{2}+\lambda_{4}\abs{\Phi_{1}^{\dagger}\Phi_{2}}^{2}
+\qty[\frac{\lambda_5}{2}\qty(\Phi_{1}^{\dagger}\Phi_{2})^{2}+\mathrm{h.c.}], \label{eq:pot-thdm}
\end{align}
where $m_{3}^{2}$ corresponds to the softly breaking parameter of the $Z_{2}$ symmetry.
Since we assume CP conservation in the Higgs potential, $m_{3}^{2}$ and $\lambda_{5}$ are real.
The Higgs doublets are parametrized as
\begin{align}
\Phi_{i}=\mqty(\omega^{+}_{i}\\ \frac{1}{\sqrt{2}}\qty(v_{i}+\varphi_{i}+i z_{i}))\qc (i=1,2),
\end{align}
where $v_{1}$ and $v_{2}$ are the vacuum expectation values (VEVs) of $\Phi_{1}$ and $\Phi_{2}$, and the electroweak VEV is given by $v=\sqrt{v_{1}^{2}+v_{2}^{2}}$.

We introduce the Higgs basis~\cite{Davidson:2005cw}, where only one of the Higgs doublets acquires its VEV,
\begin{align}
\mqty(\Phi_{1}\\ \Phi_{2}) = R(\beta)\mqty(H_{1}\\ H_{2})
\qq{with}
R(\theta)=\mqty(c_{\theta} & -s_{\theta} \\ s_{\theta} & c_{\theta}),
\end{align}
where $\tan\beta=v_{2}/v_{1}$ with $0\le \beta \le \pi/2$, and $s_{\theta}\ (c_{\theta})$ is abbreviation of $\sin{\theta}\ (\cos{\theta})$.
The Higgs potential can be expressed as
\begin{align}
V(H_{1}, H_{2})
&=
 Y_{1}^{2}H_{1}^{\dagger}H_{1}+Y_{2}^{2}H_{2}^{\dagger}H_{2} - Y_{3}^{2}(H_{1}^{\dagger}H_{2}+H_{2}^{\dagger}H_{1}) \notag \\
&\quad
+\frac{1}{2}Z_{1}(H_{1}^{\dagger}H_{1})^{2}+\frac{1}{2}Z_{2}(H_{2}^{\dagger}H_{2})^{2}
+Z_{3}(H_{1}^{\dagger}H_{1})(H_{2}^{\dagger}H_{2})+Z_{4}(H_{1}^{\dagger}H_{2})(H_{2}^{\dagger}H_{1}) \notag \\
&\quad
+\left\{\frac{1}{2}Z_{5}(H_{1}^{\dagger}H_{2})^{2}+
\left[Z_{6}H_{1}^{\dagger}H_{1} + Z_{7}H_{2}^{\dagger}H_{2}\right]H_{1}^{\dagger}H_{2} + h.c.\right\},
\label{eq: Higgs_potential_Higgs_basis}
\end{align}
where $Y_{i}^{2}$ and $Z_{i}$ are functions of $m_{i}^{2}$ and $\lambda_{i}$.
We give the explicit formulae of them in terms of the masses and mixing angles of the Higgs bosons in Appendix~\ref{Apendix: scalar_couplings}.

The Higgs doublets are parametrized as
\begin{align}
H_{1}=\mqty(G^{+}\\ \frac{1}{\sqrt{2}}\qty(v+h_{1}+i G^{0}))\qc
H_{2}=\mqty(H^{+}\\ \frac{1}{\sqrt{2}}\qty(h_{2}+iA)),
\end{align}
where $G^{\pm}$ and $G^{0}$ are the Nambu-Goldstone bosons while $H^{\pm}$ and $A$ are the physical charged and CP-odd Higgs bosons. 
In the Higgs basis, the mass matrices of the charged and CP-odd states are diagonalized after imposing the stationary conditions,
\begin{align}
Y_{1}^{2} = -\frac{1}{2}Z_{1}v^{2}\qc
Y_{3}^{2} = \frac{1}{2}Z_{6}v^{2}.
\end{align}
The masses of $H^{\pm}$ and $A$ are given by
\begin{align}
m_{H^{\pm}}^{2} &= Y_{2}^{2}+\frac{1}{2}Z_{3}v^{2}, \\
m_{A}^{2} &= Y_{2}^{2}+\frac{1}{2}(Z_{3}+Z_{4}-Z_{5})v^{2}.
\end{align}
In general, the mass matrix of the CP-even states is not diagonalized on the Higgs basis,
\begin{align}
{\cal M}^{2}
&=
\mqty(Z_{1}v^{2} & Z_{6}v^{2} \\
Z_{6}v^{2} & Y_{2}^{2}+\frac{1}{2}Z_{345}v^{2}).
\end{align}
We need further rotation to define the CP-even mass eigenstates $h$ and $H$,
\begin{align}
\mqty(h_{1}\\ h_{2})
=
R(\alpha-\beta)\mqty(H\\ h).
\end{align}
The masses of the CP-even Higgs bosons and the mixing angle $\beta-\alpha$ are given by
\begin{align}
m_{H}^{2} &= \mathcal{M}_{11}^{2}c^{2}_{\beta-\alpha}+\mathcal{M}_{22}^{2}s^{2}_{\beta-\alpha}
-\mathcal{M}_{12}^{2}s_{2(\beta-\alpha)}, \\
m_{h}^{2} &= \mathcal{M}_{11}^{2}s^{2}_{\beta-\alpha}+\mathcal{M}_{22}^{2}c^{2}_{\beta-\alpha}
+\mathcal{M}_{12}^{2}s_{2(\beta-\alpha)},\\
\tan&{2(\beta-\alpha)} = \frac{-2{\cal M}_{12}^{2}}{{\cal M}_{11}^{2}-{\cal M}_{22}^{2}}.
\label{eq:b-a}
\end{align}
In this paper, we identify $h$ and $H$ as the observed Higgs boson with the mass 125 {\rm GeV} and an additional CP-even Higgs boson, respectively. 

The original eight parameters in the Higgs potential can be replaced by the physical parameters. 
While the mass of the discovered Higgs boson and the electroweak VEV are fixed, we have the following six parameters as inputs,
\begin{align}
m_{H}\qc m_{A}\qc m_{H^\pm}\qc M^2\qc \tan{\beta}\qc s_{\beta-\alpha}, \label{eq:thdminput}
\end{align}
where we have introduced a softly $Z_{2}$ breaking scale $\sqrt{M^{2}}=\sqrt{m_{12}^{2}/s_{\beta}c_{\beta}}$.
These parameters are constrained by theoretical arguments such as perturbative unitarity~\cite{Kanemura:1993hm, Akeroyd:2000wc, Ginzburg:2005dt, Kanemura:2015ska}, tree-level vacuum stability~\cite{Deshpande:1977rw, Klimenko:1984qx, Sher:1988mj, Nie:1998yn, Kanemura:1999xf} and the true vacuum condition~\cite{Barroso:2013awa, Branchina:2018qlf}.
In this paper, we take $M^{2}\ge 0$ which ensures the true vacuum condition.

Under the $Z_{2}$ symmetry, 2HDM can be classified into four independent types of Yukawa interactions as given in Table.~\ref{tab: Z2_assignment}~\cite{Barger:1989fj, Aoki:2009ha}.
We call them Type-I, Type-II, Type-X and Type-Y, respectively.
Yukawa Lagrangian is generally written in terms of $\Phi_{1}$ and $\Phi_{2}$ by 
\begin{align}
\mathcal{L}_Y =
&-Y_{u}\bar{Q}_L\tilde{\Phi}_{u} u_{R}
-Y_{d}\bar{Q}_L\Phi_d d_{R}
-Y_{e}\bar{L}_L\Phi_e e_{R}+\mathrm{h.c.}, \label{eq:Yukawa}
\end{align}
where $\tilde{\Phi}_{u}= i\sigma_2 \Phi^{*}_{u}$, and $\Phi_{u,d,e}$ are $\Phi_1$ or $\Phi_2$ depending on the types of 2HDMs. 

In the mass eigenstate, the interaction terms among the gauge bosons and the CP-even scalars are given by,
\begin{align}
\mathcal{L}_{\mathrm{kin}}
&=
\abs{D_{\mu}\Phi_{1}}^{2}+\abs{D_{\mu}\Phi_{2}}^{2} \notag \\
&\supset
[s_{\beta-\alpha}h+c_{\beta-\alpha}H]\left(\frac{2m_{W}^{2}}{v}W^{+\mu}W_{\mu}^{-}+\frac{m_{Z}^{2}}{v}Z^{\mu}Z_{\mu}\right). \label{int_gauge}
\end{align}
The Yukawa interaction terms among the fermions and the CP-even scalars are given by \begin{align}
{\cal L}_{int}
=
-\sum_{f=u, d, e}\frac{m_{f}}{v}\left(\zeta_{h}^{f}\overline{f}fh+\zeta_{H}^{f}\overline{f}fH\right),
\label{int_fermion}
\end{align}
where
\begin{align}
\zeta_{h}^{f}&=s_{\beta-\alpha}+\zeta_{f}c_{\beta-\alpha}, \\
\zeta_{H}^{f}&=c_{\beta-\alpha}-\zeta_{f}s_{\beta-\alpha},
\end{align}
and $\zeta_{f}$ is the type-dependent parameter given in Table \ref{tab: Z2_assignment}.
When $s_{\beta-\alpha}=1$, the couplings of $h$ with various SM particles become SM like. We call this SM-like limit as the alignment limit in this paper.

The alignment limit can be achieved in different two ways \cite{Gunion:2002zf, Kanemura:2004mg}; (i) decoupling of the additional Higgs bosons, and (ii) alignment without decoupling.
In the scenario (i), we take the decoupling limit, $M^{2} \gg f(\lambda_{i})v^{2}$. Then, we have
\begin{align}
\tan{2(\beta-\alpha)}
\simeq \frac{-2Z_{6}v^{2}}{-M^{2}} \simeq 0, \label{decouple}
\end{align}
where we have used that $Z_{6}$ does not depend on $\sqrt{M^{2}}$.
Eq.~\eqref{decouple} implies $s_{\beta-\alpha}=1$, and the couplings of $h$ become SM like.
In the decoupling scenario, masses of the additional Higgs bosons are close to $\sqrt{M^{2}}$, decoupling from electroweak physics.
In scenario (ii), the off-diagonal component of the mass matrix for the CP-even states is equal to zero, $Z_{6}=0$.
In this scenario, the additional Higgs bosons need not be decoupled, and their masses can be taken around the electroweak scale.

\subsection{Constraints from experimental data} \label{sec: constraint}
We here discuss experimental constraints on the 2HDMs.
Although these experimental constraints have been discussed in Ref.~\cite{Aiko:2021can}, we dare to explain them here again for completeness.

\noindent
$\bullet$ Electroweak precision tests

The constraint from the electroweak precision tests is imposed by the $S$ and $T$ parameters~\cite{Peskin:1990zt, Peskin:1991sw}.
The new physics contributions in the 2HDMs are defined by $\Delta S=S_{\rm 2HDM}-S_{\rm SM}$ and $\Delta T= T_{\rm 2HDM}-T_{\rm SM}$. 
The analytical formulae for $\Delta S$ and $\Delta T$ are given in Refs.~\cite{Bertolini:1985ia, Grimus:2008nb, Kanemura:2011sj, Kanemura:2015mxa}.
The experimental data are given in Ref.~\cite{Haller:2018nnx},
\begin{align}
\Delta S=0.04\pm 0.08\qc \Delta T=0.08\pm 0.07,
\end{align}
where the $U$ parameter is fixed to zero.
The reference values of the masses of the SM Higgs boson and the top quark are $m_{h, \mathrm{ref}}=125$ GeV and $m_{t, \mathrm{ref}}=172.5$ GeV, respectively.
The correlation coefficient in $\chi^{2}$ analysis is $+0.92$.
We require $\Delta S$ and $\Delta T$ to be within 95\% CL.

\noindent
$\bullet$ Signal strengths of the SM-like Higgs boson

Measurements of signal strengths for the SM-like Higgs boson constrain the parameter space of the 2HDMs.
We evaluate the decay rates of the SM-like Higgs boson, $\Gamma(h\to XY)$, including the NLO EW and higher-order QCD corrections by using \texttt{H-COUP v2}~\cite{Kanemura:2019slf}.
The analytic expressions for $\Gamma(h\to XY)$ are given in Ref.~\cite{Kanemura:2019kjg}.
We define the scaling factors at the one-loop level,
\begin{align}
\kappa_{X}=\sqrt{\frac{\Gamma^{\mathrm{2HDM}}_{\mathrm{LO+EW+QCD}}(h\to XY)}{\Gamma^{\mathrm{SM}}_{\mathrm{LO+QCD}}(h\to XY)}}.
\label{eq: scaling-factor}
\end{align}
We require that the scaling factors for $XY=bb,\, \tau\tau,\, \gamma\gamma,\, gg$ and $ZZ^{*}$ to be consistent with the values presented in Table 11 (a) of Ref.~\cite{ATLAS:2019nkf} at 95~\% CL.

\noindent
$\bullet$ Direct searches of the additional Higgs bosons

In Ref.~\cite{Aiko:2020ksl}, constraints from direct searches of the additional Higgs bosons at the LHC 13 TeV with 36 ${\rm fb^{-1}}$ have been evaluated for the alignment limit and nearly alignment scenario for all the types of 2HDMs.
The excluded regions at 95\% CL are shown in Figs.~10 and 11 in Ref.~\cite{Aiko:2020ksl}.

In the alignment limit, $A\to \tau\tau$ gives the lower bound, $m_{A} \gtrsim 350~\mathrm{GeV}$, depending on the value of $\tan{\beta}$ and the types of 2HDMs.
For Type-I and Type-Y, $\tan\beta\lesssim 2$ and $\tan\beta\lesssim 1.2$ is excluded, respectively.
For Type-II, $\tan\beta\lesssim 2.3$ and $\tan\beta\gtrsim 8$ are excluded, while $\tan\beta\lesssim 8$ is excluded for Type-X.
$H^{+}\to t\bar{b}$ gives the lower bound, $\mHp \gtrsim 650~\mathrm{GeV}$, for $\tan\beta \lesssim 1$ for the all types of 2HDMs.

In the nearly alignment scenarios with $s_{\beta-\alpha}=0.99$, $A\to Zh$ gives the lower bound on $m_{A}$, e.g., $m_{A}\gtrsim 950~(1100)~{\rm GeV}$ for $\tan\beta=1$ for Type-I (Type-II) and Type-X (Type-Y).
For $s_{\beta-\alpha}=0.995$, the lower bound on $m_{A}$ is relaxed, and $m_{A} \gtrsim 350~\mathrm{GeV}$ for $\tan{\beta}=5 (2)$ for Type-I (Type-Y).
For Type-II and Type-X with $s_{\beta-\alpha}=0.995$, $A\to \tau\tau$ gives the almost same bound as in the alignment limit, and it is stronger than the bound by $A\to Zh$.
For $s_{\beta-\alpha}=0.995$, $H\to hh$ gives the lower bound on $m_{H}$, e.g., $m_{H}\gtrsim 950~(1100)~{\rm GeV}$ for $\tan\beta=1$ with $m_{H}=\sqrt{M^{2}}$ for Type-I (Type-II) and Type-X (Type-Y).
We note that the partial decay width of $H\to hh$ depends on the value of $\sqrt{M^{2}}$, while the above-mentioned decay processes are independent of it at LO.

\noindent
$\bullet$ Flavor constraints

The mass of the charged Higgs bosons is constrained by the $B$ meson flavor violating decay, $B\to X_{s}\gamma$~\cite{Misiak:2017bgg, Misiak:2020vlo}.
For Type-II and Type-Y, $m_{H^{\pm}}\lesssim$ 800{\rm GeV} is excleded for $\tan\beta>1$~\cite{Misiak:2017bgg,Misiak:2020vlo}.
For Type-I and Type-X, the constraint is weaker than that for Type-II and Type-Y.
The excluded regions are given in lower $\tan\beta$ regions, e.g., $m_{H^{\pm}}\lesssim 400~(180)~{\rm GeV}$ for $\tan\beta=1~(1.5)$~\cite{Misiak:2017bgg}.
For Type-II, high $\tan\beta$ regions are also constrained by the $B$ meson rare leptonic decay, $B_{s}\to \mu^{+}\mu^{-}$~\cite{Cheng:2015yfu}.
For $\mHp=800~{\rm GeV}$, the region with $\tan\beta\gtrsim 10$ is excluded~\cite{Haller:2018nnx}. 
Comprehensive studies for constraints from various flavor observables such as $B$ meson decays, $D$ meson decays and $B_{0}-\bar{B}_{0}$ mixing, are performed in Refs.~\cite{Haller:2018nnx, Enomoto:2015wbn}.

%% file: 03decay_rate.tex
\section{Decay rates with higher-order corrections} \label{sec: decay_rate}
In this section, we first define the renormalized vertex functions of the CP-odd Higgs boson.
The decay rates of $A\to f\bar{f},\, Zh/H$ and $W^{\pm}H^{\mp}$ including the NLO EW corrections are given in terms of the form factors of the vertex functions.
We also give the decay rates of the loop-induced decays, $A\to W^{+}W^{-}, ZZ, Z\gamma, \gamma\gamma$ and $gg$ at LO EW.
We include higher-order QCD corrections for the decays into a pair of quarks, photons and gluons.

We here give an overview of our calculation scheme of NLO EW corrections.
In order to treat UV divergences, we employ the on-shell renormalization scheme, where external-leg corrections are absorbed into the field renormalization constant.
For the gauge sector and the fermion sector, we follow the on-shell renormalization scheme developed in Ref.~\cite{Bohm:1986rj, Hollik:1988ii}.
For the renormalization of the Higgs sector, we employ the improved on-shell renormalization scheme developed in Ref.~\cite{Kanemura:2017wtm}.
In the improved on-shell renormalization scheme, the masses of the additional Higgs bosons, the mixing angles and the wave function renormalization constants are renormalized by imposing the on-shell conditions for the Higgs bosons in the mass eigenstates.
The gauge dependences, which appear in the renormalization of mixing angles~\cite{Yamada:2001px}, are removed by applying the pinch technique~\cite{Krause:2016oke, Kanemura:2017wtm}.
For the calculation of the decays of the CP-odd Higgs boson, we do not need to renormalize 
$M^{2}$, while it can be renormalized by the minimal subtraction~\cite{Kanemura:2004mg}.
Among the various renormalization scheme discussed in Ref.~\cite{Krause:2019qwe}, our renormalization scheme is essentially the same as the $\mathrm{pOS}^{o}$ scheme, where $\delta \beta$ is determined by the pinched self-energies of the CP-odd scalars~\cite{Kanemura:2004mg, Kanemura:2017wtm}.
This scheme provides relatively stable numerical results compared with other renormalization schemes such as the $\overline{\mathrm{MS}}$ scheme~\cite{Krause:2016oke, Krause:2019qwe}.
We perform the calculation in the 't Hooft-Feynman gauge, i.e., $\xi=1$, and implement the analytical formulae into \texttt{H-COUP} program~\cite{Kanemura:2017gbi, Kanemura:2019slf}.
For numerical evaluations of the Passarino-Veltman functions~\cite{Passarino:1978jh}, we use  
\texttt{LoopTools}~\cite{Hahn:1998yk}.

We regularize the IR divergences, which appear in one-loop diagrams containing a virtual photon, by introducing the finite photon mass.
The photon-mass dependence is canceled by adding the decay rate of the real-photon emission process.
We confirm the cancellation of the photon-mass dependence numerically.

In the following calculation, the contribution of $AZ$ mixing, which can arises in the amplitudes for $A$ decays at loop level, vanishes due to the Slavnov-Taylor identity~\cite{Williams:2011bu},
\begin{align}
\widehat{\Pi}_{AG^{0}}(m_{A}^{2})+i\frac{m_{A}^{2}}{m_{Z}}\widehat{\Pi}_{AZ}(m_{A}^{2}) = 0,
\end{align}
where $\widehat{\Pi}_{AG^{0}\, (AZ)}(q^{2})$ denotes the renormalized self-energy for $AG^{0}\, (AZ)$ mixing.
Since we impose the on-shell renormalization condition, $\widehat{\Pi}_{AG^{0}}(m_{A}^{2})=0$, the contribution of $AZ$ mixing vanishes in the decay amplitude.

\subsection{Form factors for vertex functions of the CP-odd Higgs boson}

\subsubsection{$Af\bar{f}$ vertex} \label{sec:Aff_vetex}
\begin{figure}[t]
	\centering
	\includegraphics[scale=0.45]{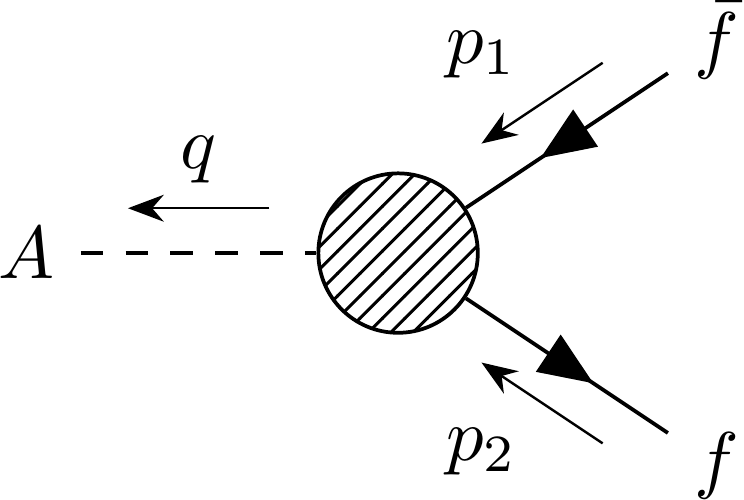}
	\caption{Momentum assignment for the renormalized $Af\bar{f}$ vertex.}
	\label{Mom_A_to_ff}
\end{figure}

The renormalized $Af\bar{f}$ vertex functions can be expressed as
\begin{align}
\widehat{\Gamma}_{Af\bar{f}}(p_{1}, p_{2}, q) &=
\widehat{\Gamma}_{Af\bar{f}}^{S}+\gamma_{5}\widehat{\Gamma}_{Af\bar{f}}^{P}
+\slashed{p}_{1}\widehat{\Gamma}_{Af\bar{f}}^{V_{1}}
+\slashed{p}_{2}\widehat{\Gamma}_{Af\bar{f}}^{V_{2}}
+\slashed{p}_{1}\gamma_{5}\widehat{\Gamma}_{Af\bar{f}}^{A_{1}}
+\slashed{p}_{2}\gamma_{5}\widehat{\Gamma}_{Af\bar{f}}^{A_{2}} \notag \\
&\quad
+\slashed{p}_{1}\slashed{p}_{2}\widehat{\Gamma}_{Af\bar{f}}^{T}
+\slashed{p}_{1}\slashed{p}_{2}\gamma_{5}\widehat{\Gamma}_{Af\bar{f}}^{PT},
\end{align}
where $p_{1}\ (p_{2})$ is the incoming four-momentum of the anti-fermion (fermion), and $q^{\mu}\, (=p_{1}+p_{2})$ is the outgoing four-momentum of the CP-odd Higgs boson (see Fig.~\ref{Mom_A_to_ff}).

The renormalized form factors are composed of the tree-level and the one-loop parts as
\begin{align}
\widehat{\Gamma}_{Af\bar{f}}^{X} = \Gamma_{Af\bar{f}}^{X, \mathrm{tree}}+\Gamma_{Af\bar{f}}^{X, \mathrm{loop}}\qc
(X=S,\, P,\, V_{1},\, V_{2},\, A_{1},\, A_{2},\, T,\, PT),
\end{align}
where the tree-level couplings for $Af\bar{f}$ vertex are given by
\begin{align}
\Gamma_{Af\bar{f}}^{P, \mathrm{tree}} = i2I_{f}\frac{m_{f}\zeta_{f}}{v}\qc
\Gamma_{Af\bar{f}}^{X, \mathrm{tree}} = 0\qc (X\neq P),
\end{align}
where $I_{f}$ is the third component of the iso-spin of fermions.

The one-loop parts are further decomposed into contributions from 1PI diagrams and counterterms,
\begin{align}
\Gamma_{Af\bar{f}}^{X, \mathrm{loop}} = \Gamma_{Af\bar{f}}^{X, \mathrm{1PI}} + \delta \Gamma_{Af\bar{f}}^{X}.
\end{align}
The 1PI diagrams contributions $\Gamma_{Af\bar{f}}^{X, \mathrm{1PI}}$ are given in Appendix~\ref{app: vertex}.
The counterterms $\delta \Gamma_{Af\bar{f}}^{X}$ are given by
\begin{align}
\delta \Gamma_{Af\bar{f}}^{P} &= \Gamma_{Af\bar{f}}^{P, \mathrm{tree}}\qty[\frac{\delta m_{f}}{m_{f}}-\frac{\delta v}{v}-\zeta_{f}\delta\beta_{f}+\delta Z_{V}^{f}+\frac{\delta Z_{A}}{2}+\frac{\delta C_{A}}{\zeta_{f}}-\qty(\frac{1}{\zeta_{f}}+\zeta_{f})\delta\beta^{\mathrm{PT}}], \label{eq: del_Gam_P} \\
\delta \Gamma_{Af\bar{f}}^{X} &= 0\qc (X\neq P).
\end{align}
The analytic expressions for the counterterms in Eq.~\eqref{eq: del_Gam_P} are given in Refs.~\cite{Kanemura:2017wtm, Aiko:2021can}.

When we neglect the effects of the CP violation, the renormalized $Af\bar{f}$ vertex functions satisfy the following relations,
\begin{align}
\widehat{\Gamma}_{Af\bar{f}}^{S} = \widehat{\Gamma}_{Af\bar{f}}^{T} = 0\qc
\widehat{\Gamma}_{Af\bar{f}}^{V_{1}} = \widehat{\Gamma}_{Af\bar{f}}^{V_{2}}\qc
\widehat{\Gamma}_{Af\bar{f}}^{A_{1}} = \widehat{\Gamma}_{Af\bar{f}}^{A_{2}}\qc (q^{2}=m_{A}^{2},\, p_{1}^{2}=p_{2}^{2}=m_{f}^{2}). \label{eq: CP_conditions}
\end{align}

\subsubsection{$AV\phi$ vertex} \label{sec:AVphi_vetex}
\begin{figure}[t]
	\centering
	\includegraphics[scale=0.45]{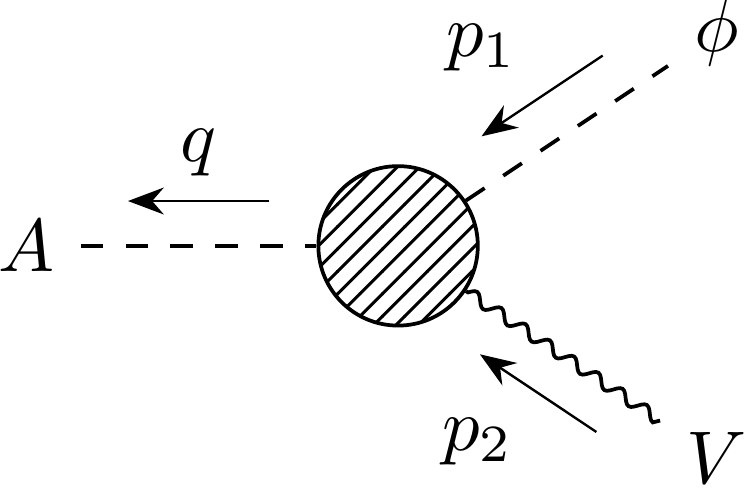}
	\caption{Momentum assignment for the renormalized $AV\phi$ vertex.}
	\label{Mom_A_to_Vphi}
\end{figure}

The renormalized $AV\phi$ vertex functions $(V, \phi)=(Z, h),\, (Z, H)$ and $(W^{\pm}, H^{\mp})$ can be expressed as
\begin{align}
\widehat{\Gamma}_{AV\phi}^{\mu}(p_{1}, p_{2}, q) &= 
(p_{1}+q)^{\mu}\widehat{\Gamma}_{AV\phi},
\end{align}
where $p_{1}$ and $p_{2}$ denote the incoming four-momentum of the scalar boson $\phi$ and the gauge boson $V$, respectively. The momentum $q$ is the outgoing four-momentum of the CP-odd Higgs boson (see Fig.~\ref{Mom_A_to_Vphi}).
Since we assume that the external gauge boson is on-shell, the term proportional to $p_{2}$ vanishes due to the orthogonality of the polarization vector.
For the $A\to W^{\pm}H^{\mp}$ decays, we use the renormalized $H^{\pm}W^{\mp}A$ vertex function given in Ref.~\cite{Aiko:2021can},
\begin{align}
\widehat{\Gamma}_{AW^{\pm}H^{\mp}}^{\mu}(p_{1}, p_{2}, q)
&= -(q+p_{1})^{\mu}\widehat{\Gamma}_{H^{\pm}W^{\pm}A}(q^{2}, p_{2}^{2}, p_{1}^{2}),
\end{align}
where the additional minus sign comes from the change of momentum assignment.

The renormalized form factors are composed of the tree-level and the one-loop parts as
\begin{align}
\widehat{\Gamma}_{AV\phi} = \Gamma_{AV\phi}^{\mathrm{tree}}+\Gamma_{AV\phi}^{\mathrm{loop}},
\end{align}
where $\Gamma_{AZ\phi}^{\mathrm{tree}}$ are given by
\begin{align}
\Gamma_{AZh}^{\mathrm{tree}} = i\frac{m_{Z}}{v}c_{\beta-\alpha}\qc
\Gamma_{AZH}^{\mathrm{tree}} = -i\frac{m_{Z}}{v}s_{\beta-\alpha}\qc
\Gamma_{AW^{\mp}H^{\pm}}^{\mathrm{tree}} = -i\frac{m_{W}}{v}.
\label{eq: Gam_AVphi_LO}
\end{align}
The one-loop parts are further decomposed into contributions from 1PI diagrams and counterterms,
\begin{align}
\Gamma_{AV\phi}^{\mathrm{loop}} = \Gamma_{AV\phi}^{\mathrm{1PI}} + \delta \Gamma_{AV\phi}.
\end{align}
The 1PI diagrams contributions $\Gamma_{AV\phi}^{\mathrm{1PI}}$ are given in Appendix~\ref{app: vertex}.
The counterterms $\delta \Gamma_{AZ\phi}$ vertices are given by
\begin{align}
\delta \Gamma_{AZh} &= \Gamma_{AZh}^{\mathrm{tree}}\bigg[\frac{\delta m_{Z}^{2}}{2m_{Z}^{2}}-\frac{\delta v}{v}+\frac{1}{2}\qty(\delta Z_{h}+\delta Z_{A}+\delta Z_{Z}) \notag \\
&\quad
+\tan{(\beta-\alpha)}\qty(\delta C_{A}-\delta C_{h}-\delta\beta^{\mathrm{PT}}+\delta\alpha^{\mathrm{PT}})\bigg], \\
\delta \Gamma_{AZH} &= \Gamma_{AZH}^{\mathrm{tree}}\bigg[\frac{\delta m_{Z}^{2}}{2m_{Z}^{2}}-\frac{\delta v}{v}+\frac{1}{2}\qty(\delta Z_{H}+\delta Z_{A}+\delta Z_{Z}) \notag \\
&\quad
-\cot{(\beta-\alpha)}\qty(\delta C_{A}+\delta C_{h}-\delta\beta^{\mathrm{PT}}+\delta\alpha^{\mathrm{PT}})\bigg], \\
\delta \Gamma_{AW^{\mp}H^{\pm}} &= \Gamma_{AW^{\mp}H^{\pm}}^{\mathrm{tree}}\bigg[\frac{\delta m_{W}^{2}}{2m_{W}^{2}}-\frac{\delta v}{v}+\frac{1}{2}\qty(\delta Z_{H^{\pm}}+\delta Z_{A}+\delta Z_{W})\bigg].
\end{align}
The analytic expressions for the counterterms are given in Refs.~\cite{Kanemura:2017wtm, Aiko:2021can}.

\subsubsection{$AV_{1}V_{2}$ vertex} \label{sec: AVV_vetex}
\begin{figure}[t]
	\centering
	\includegraphics[scale=0.45]{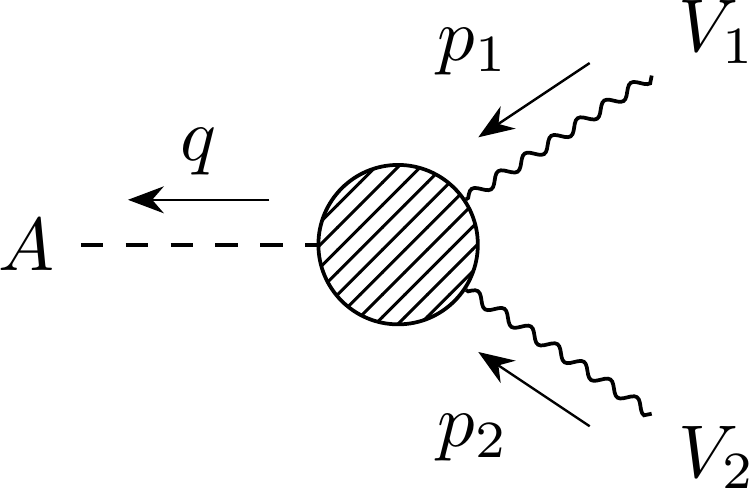}
	\caption{Momentum assignment for the renormalized $AV_{1}V_{2}$ vertex.}
	\label{Mom_A_to_VV}
\end{figure}

The $AV_{1}V_{2}$ vertex functions $(V_{1}, V_{2}) = (W^{\pm}, W^{\mp}),\, (Z, Z),\, (Z, \gamma)$ and $(\gamma, \gamma)$ appear at one-loop level.
They can be expressed as
\begin{align}
\widehat{\Gamma}_{AV_{1}V_{2}}^{\mu\nu}(p_{1}, p_{2}, q) &= 
g^{\mu\nu}\Gamma_{AV_{1}V_{2}}^{1, \mathrm{1PI}}
+p^{\nu}_{1}p^{\mu}_{2}\Gamma_{AV_{1}V_{2}}^{2, \mathrm{1PI}}
+i\epsilon^{\mu\nu\rho\sigma}p_{1\rho}p_{2\sigma}\Gamma_{AV_{1}V_{2}}^{3, \mathrm{1PI}},
\end{align}
where $p_{1}$ and $p_{2}$ denote the incoming four-momentum of the gauge bosons $V_{i}\, (i=1,2)$, and $q$ is the outgoing four-momentum of the CP-odd Higgs boson (see Fig.~\ref{Mom_A_to_VV}).
We take $\epsilon^{0123}=+1$ and assume that the external gauge bosons are on-shell.
When we neglect the effects of the CP violation, the 1PI vertices satisfy
\begin{align}
\Gamma_{AV_{1}V_{2}}^{1, \mathrm{1PI}}=\Gamma_{AV_{1}V_{2}}^{2, \mathrm{1PI}}=0.
\end{align}

In the CP-conserving 2HDMs, radiative corrections consist of fermion loops only~\cite{Gunion:1991cw}.
This is because the bosonic sector separately conserves $P$ and $C$ symmetries~\cite{Gunion:1986nh}, and $AV_{1}V_{2}$ couplings are prohibited.
On the other hand, the fermion sector breaks them, and $\Gamma_{AV_{1}V_{2}}^{3, \mathrm{1PI}}$ is induced at the one-loop level.
The 1PI diagrams contributions $\Gamma_{AV_{1}V_{2}}^{3, \mathrm{1PI}}$ are given in Appendix~\ref{app: vertex}.

\subsection{Decay rates of $A\to f\bar{f}$}
The decay rates of the CP-odd Higgs boson into a pair of fermions with NLO EW and QCD corrections are given by
\begin{align}
\Gamma(A\to f\bar{f}) &=
\Gamma_{\mathrm{LO}}(A\to f\bar{f})\qty(1+\Delta_{\mathrm{EW}}^{f}+\Delta_{\mathrm{QCD}}^{f})+\Gamma(A\to f\bar{f}\gamma).
\end{align}
The decay rate at LO is given by
\begin{align}
\Gamma_{\mathrm{LO}}(A\to f\bar{f}) &= N_{c}^{f}\frac{m_{A}}{8\pi}\abs{\Gamma_{Af\bar{f}}^{P, \mathrm{tree}}}^{2}
\lambda^{1/2}\qty(\frac{m_{f}^{2}}{m_{A}^{2}}, \frac{m_{f}^{2}}{m_{A}^{2}}),
\end{align}
where $N_{c}^{f}$ is the color factor, and the kinematical factor $\lambda(x,y)$ is given by
\begin{align}
\lambda(x,y) = (1-x-y)^{2}-4xy. \label{eq: lam_kin}
\end{align}
For the decays into a pair of light quarks, $A\to q\bar{q}\, (q\neq t)$, we replace the quark mass in the Yukawa coupling to the running mass $\overline{m}_{q}(\mu)$ evaluated at $\mu=m_{A}$~\cite{Braaten:1980yq, Sakai:1980fa, Inami:1980qp, Drees:1990dq}.
Factors $\Delta_{\mathrm{EW}}^{f}$ and $\Delta_{\mathrm{QCD}}^{f}$ denote the EW and QCD corrections, respectively.
Since the NLO EW correction includes the IR divergences, we regularize them by introducing the finite photon mass.
The photon mass dependence is canceled by adding the decay rates of real photon emission $\Gamma(A\to f\bar{f}\gamma)$.
The analytic expression of $\Gamma(A\to f\bar{f}\gamma)$ is given in Appendix~\ref{sec:real_emission}.

The EW correction $\Delta_{\mathrm{EW}}^{f}$ is given by
\begin{align}
\Delta_{\mathrm{EW}}^{f} &=
\frac{2}{\abs{\Gamma_{Af\bar{f}}^{P, \mathrm{tree}}}^{2}}\Re\qty(\Gamma^{P, \mathrm{tree}}_{Af\bar{f}}G^{P, \mathrm{loop}*}_{Af\bar{f}})
-\Delta r,
\end{align}
with
\begin{align}
G^{P, \mathrm{loop}}_{Af\bar{f}} 
&=
\Gamma_{Af\bar{f}}^{P, \mathrm{loop}}-2m_{f}\Gamma_{Af\bar{f}}^{A_{1}, \mathrm{loop}}
+m_{A}^{2}\qty(1-\frac{3m_{f}^{2}}{m_{A}^{2}})\Gamma_{Af\bar{f}}^{PT, \mathrm{loop}},
\end{align}
where we have used Eq.~\eqref{eq: CP_conditions}.
The one-loop weak correction to the muon decay, $\Delta r$, is introduced by the resummation of universal leading higher-order corrections~\cite{Sirlin:1980nh, Denner:1991kt}.

For $A\to q\bar{q}$, we apply the QCD corrections at NNLO in the $\overline{\mathrm{MS}}$ scheme.
The QCD correction is given by
\begin{align}
\Delta_{\mathrm{QCD}}^{q} &= \Delta_{qq}+\Delta_{A}.
\end{align}
The correction $\Delta_{qq}$ is evaluated in the chiral limit, $m_{A}\gg m_{q}$, and it is the same for CP-odd and CP-even particles.
The NLO correction is given in Refs.~\cite{Braaten:1980yq, Sakai:1980fa, Inami:1980qp, Drees:1990dq}, while the NNLO correction is given in Refs.~\cite{Gorishnii:1990zu, Gorishnii:1991zr}.
\begin{align}
\Delta_{qq}
&=
\frac{\alpha_{s}(m_{A})}{\pi}\frac{17}{4}C_{F} \notag \\
&\quad
+\qty(\frac{\alpha_{s}(m_{A})}{\pi})^{2}\qty[\frac{10801}{144}-\frac{39}{2}\zeta(3)-\qty(\frac{65}{24}-\frac{2}{3}\zeta(3))N_{f}-\pi^{2}\qty(\frac{19}{12}-\frac{1}{18}N_{f})],
\end{align}
with the color factor $C_{F}=4/3$ and the Riemann zeta function $\zeta(n)$.
We take the number of active flavors $N_{f}$ as $N_{f}=5$ for $m_{A}\leq m_{t}$, while $N_{f}=6$ for $m_{A}> m_{t}$. 
The correction $\Delta_{A}$ includes the logarithms of the light-quarks and top-quark masses.
In the heavy top-mass limit, $m_{t}\gg m_{A}$, and $\mu=m_{A}$, it is given by~\cite{Chetyrkin:1995pd, Larin:1995sq}
\begin{align}
\Delta_{A} = \qty(\frac{\alpha_{s}(m_{A})}{\pi})^{2}\qty(3.83+\ln{\frac{m_{t}^{2}}{m_{A}^{2}}}+\frac{1}{6}\ln^{2}{\frac{\overline{m}_{q}^{2}(m_{A})}{m_{A}^{2}}}).
\end{align}

For the decay into the top-quark pair, the effects of the top-quark mass in the QCD corrections are significant near the threshold region.
The QCD correction in the on-shell scheme is given by~\cite{Drees:1989du, Djouadi:1994gf}
\begin{align}
\Delta_{\mathrm{QCD}}^{t} = \frac{\alpha_{s}(\mu)}{\pi}C_{F}
\qty[\frac{L(\beta_{t})}{\beta_{t}}-\frac{1}{16\beta_{t}}(19+2\beta_{t}^{2}+3\beta_{t}^{4})\ln{\rho_{t}}+\frac{3}{8}(7-\beta_{t}^{2})],
\end{align}
with $\beta_{t}=\lambda^{1/2}(m_{t}^{2}/m_{A}^{2},m_{t}^{2}/m_{A}^{2})$ and $\rho_{t}=(1-\beta_{t})/(1+\beta_{t})$.
The function $L(\beta_{t})$ is given by
\begin{align}
L(\beta_{t}) &= (1+\beta_{t}^{2})\qty[4\mathrm{Li}_{2}(\rho_{t})+2\mathrm{Li}_{2}(-\rho_{t})
+3\ln{\rho_{t}}\ln{\frac{2}{1+\beta_{t}}}+2\ln{\rho_{t}}\ln{\beta_{t}}]  \notag \\
&\quad
-3\beta_{t}\ln{\frac{4}{1-\beta_{t}^{2}}}-4\beta_{t}\ln{\beta_{t}},
\end{align}
where $\mathrm{Li}_{2}(x)$ is the dilog function.
When $m_{A}\gg m_{t}$, we can treat the top-quark as massless, and apply the QCD corrections in the $\overline{\mathrm{MS}}$ scheme.
In order to treat the transition between the two regions, we use linear interpolation for the QCD corrections to $A\to t\bar{t}$ as discussed in Ref.~\cite{Djouadi:1997yw}.
\begin{align}
\Gamma(A\to t\bar{t}) &= 
R^{2}\Gamma_{\mathrm{LO}}^{\mathrm{pole}}(A\to t\bar{t})\qty(1+\Delta_{\mathrm{QCD}}^{t})
+(1-R^{2})\Gamma_{\mathrm{LO}}^{\overline{\mathrm{MS}}}(A\to t\bar{t})\qty(1+\Delta_{\mathrm{QCD}}^{t, \overline{\mathrm{MS}}}) \notag \\
&\quad
+\Gamma_{\mathrm{LO}}^{\mathrm{pole}}(A\to t\bar{t})\Delta_{\mathrm{EW}}^{t}+\Gamma(A\to t\bar{t}\gamma),
\end{align}
with $R=2m_{t}/m_{A}$.
In the evaluation of $\Gamma_{\mathrm{LO}}^{\mathrm{pole}}(A\to t\bar{t})$, we use the top-quark pole mass for the Yukawa coupling, while the running mass is used for $\Gamma_{\mathrm{LO}}^{\overline{\mathrm{MS}}}(A\to t\bar{t})$.
The QCD correction in the $\overline{\mathrm{MS}}$ scheme $\Delta_{\mathrm{QCD}}^{t, \overline{\mathrm{MS}}}$ is evaluated as similar to the decays into the light quarks.

When $m_{t}+m_{W}\leq m_{A} < 2m_{t}$, the CP-odd Higgs boson decays into a pair of on-shell and off-shell top quarks, $A\to tt^{*} \to tbW$.
We calculate the decay rate at the tree level and give the analytic expressions in Appendix~\ref{sec: three_body}.

\subsection{Decay rates of $A\to V\phi$}
The decay rate for the CP-odd Higgs boson decays into a $Z$ boson and a CP-even Higgs boson $\phi\ (\phi=h, H)$ with NLO EW corrections is given by
\begin{align}
\Gamma(A\to Z\phi) &=
\Gamma_{\mathrm{LO}}(A\to Z\phi)\qty(1+\Delta_{\mathrm{EW}}^{\phi}).
\end{align}
The decay rate at LO is given by
\begin{align}
\Gamma_{\mathrm{LO}}(A\to Z\phi) &= \frac{\abs{\Gamma_{AZ\phi}^{\mathrm{tree}}}^{2}}{16\pi}\frac{m_{A}^{3}}{m_{Z}^{2}}\lambda^{3/2}\qty(\frac{m_{\phi}^{2}}{m_{A}^{2}}, \frac{m_{Z}^{2}}{m_{A}^{2}}).
\end{align}
The EW correction $\Delta_{\mathrm{EW}}^{\phi}$ is given by
\begin{align}
\Delta_{\mathrm{EW}}^{\phi} = \frac{2\Re\qty(\Gamma^{\mathrm{tree}}_{AZ\phi}\Gamma^{\mathrm{loop}*}_{AZ\phi})}{\abs{\Gamma_{AZ\phi}^{\mathrm{tree}}}^{2}}
-\Delta r-\Re\widehat{\Pi}'_{ZZ}(m_{Z}^{2}).
\end{align}
The term $\widehat{\Pi}'_{ZZ}(m_{Z}^{2})$ is included since the residue of renormalized $Z$ boson's propagator is not unity~\cite{Bohm:1986rj, Hollik:1988ii}.

The decay rate for the CP-odd Higgs boson decays into a $W^{\pm}$ boson and a charged Higgs boson $H^{\pm}$ with NLO EW corrections is given by
\begin{align}
\Gamma(A\to W^{\pm}H^{\mp}) &=
\Gamma_{\mathrm{LO}}(A\to W^{\pm}H^{\mp})\qty(1+\Delta_{\mathrm{EW}}^{H^{\pm}})
+\Gamma(A\to H^{\pm}W^{\mp}\gamma).
\end{align}
The decay rate at LO is given by
\begin{align}
\Gamma_{\mathrm{LO}}(A\to W^{\pm}H^{\mp}) &= \frac{\abs{\Gamma_{AW^{\mp}H^{\pm}}^{\mathrm{tree}}}^{2}}{8\pi}\frac{m_{A}^{3}}{m_{W}^{2}}\lambda^{3/2}\qty(\frac{m_{H^{\pm}}^{2}}{m_{A}^{2}}, \frac{m_{W}^{2}}{m_{A}^{2}}).
\end{align}
The EW correction $\Delta_{\mathrm{EW}}^{H^{\pm}}$ is given by
\begin{align}
\Delta_{\mathrm{EW}}^{H^{\pm}} = \frac{2\Re\qty(\Gamma^{\mathrm{tree}}_{AW^{\mp}H^{\pm}}\Gamma^{\mathrm{loop}*}_{AW^{\mp}H^{\pm}})}{\abs{\Gamma_{AW^{\mp}H^{\pm}}^{\mathrm{tree}}}^{2}}
-\Delta r-\Re\widehat{\Pi}'_{WW}(m_{W}^{2}).
\end{align}
The term $\widehat{\Pi}'_{WW}(m_{W}^{2})$ is included since the residue of renormalized $W^{\pm}$ bosons propagator is not unity.
The IR divergences in the NLO EW correction are regularized by introducing the finite photon mass.
The photon mass dependence is removed by including the decay rates of real photon emission $A\to H^{\pm}W^{\mp}\gamma$ as similar to $A\to f\bar{f}$.
The analytic expression of $\Gamma(A\to H^{\pm}W^{\mp}\gamma)$ is given in Appendix~\ref{sec:real_emission}.

When $m_{\phi}\leq m_{A} < m_{\phi} +m_{Z}$, the CP-odd Higgs boson decays into a pair of on-shell CP-even Higgs boson $\phi$ and off-shell $Z$ boson, $A\to \phi Z^{*} \to \phi f\bar{f}$.
In addition, when $m_{H^{\pm}}\leq m_{A} < m_{H^{\pm}} +m_{W}$, the CP-odd Higgs boson decays into a pair of on-shell charged Higgs bosons $H^{\pm}$ and off-shell $W^{\pm}$ bosons, $A\to H^{\pm} W^{\mp*} \to H^{\pm}ff'$.
We calculate these decay rates at the tree level and give the analytic expressions in Appendix~\ref{sec: three_body}.

\subsection{Decay rates of loop-induced processes}
We calculate the decay rates of $A\to W^{+}W^{-},\, ZZ$ and $Z\gamma$ at LO, while higher-order QCD corrections are included for the decay rates of $\gamma\gamma$ and $gg$.
The decay rates for the $A\to W^{+}W^{-},\, ZZ$ and $Z\gamma$ are given by
\begin{align}
\Gamma(A\to V_{i}V_{j}) = \frac{m_{A}^{3}}{32\pi(1+\delta_{ij})}\lambda^{3/2}\qty(\frac{m_{V_{i}}^{2}}{m_{A}^{2}}, \frac{m_{V_{j}}^{2}}{m_{A}^{2}})\abs{\Gamma^{3}_{AV_{i}V_{j}}}^{2}.
\end{align}

The decay rate for the CP-odd Higgs boson into a pair of photons is given by
\begin{align}
\Gamma(A\to \gamma\gamma) &= \frac{G_{F}\alpha_{\mathrm{em}}^{2}m_{A}^{3}}{128\sqrt{2}\pi^{3}}\Bigg|
\sum_{\ell}Q_{\ell}^{2}\kappa_{\ell}^{A}I_{F}^{A}(\tau_{\ell}) \notag \\
&\quad
+\sum_{q}N_{c}^{q}Q_{q}^{2}\kappa_{q}^{A}I_{F}^{A}(\tau_{q})\qty[1+\frac{\alpha_{s}(m_{A})}{\pi}\qty(C_{1}^{A}(\tau_{q})+C_{2}^{A}(\tau_{q})\ln{\frac{4\tau_{q}\mu^{2}}{m_{A}^{2}}})]
\Bigg|^{2}, \label{eq: Gam_Agamgam}
\end{align}
where $\kappa_{f}^{A} = -2iI_{f}\zeta_{f}$, $\tau_{f} = m_{A}^{2}/(4m_{f}^{2})$ and $Q_{f}$ is the electric charge of the fermion.
The terms in the first line correspond to the contributions from the charged leptons, while the terms in the second line are those from the quarks.
The loop function $I_{F}^{A}(\tau_{f})$ is given by
\begin{align}
I_{F}^{A}(\tau_{f})
&= \frac{2}{\tau_{f}}f(\tau_{f}), \label{eq: IFA}
\end{align}
with
\begin{align}
f(\tau) =
\begin{dcases}
{\arcsin^{2}(\sqrt{\tau}) \qquad (\tau \leq 1)}, \\
{-\frac{1}{4}\qty[\ln{\frac{1+\sqrt{1-\tau^{-1}}}{1-\sqrt{1-\tau^{-1}}}}-i\pi]^{2} \qquad (\tau>1)}.
\end{dcases}
\label{eq: f_tau}
\end{align}
The NLO QCD correction in the $\overline{\mathrm{MS}}$ scheme is given in Refs.~\cite{Spira:1995rr, Harlander:2005rq}.
The analytic expressions for QCD corrections are given in Appendix~\ref{sec: loop_induced}.

The decay rate of the CP-odd Higgs boson into a pair of gluons is given by
\begin{align}
\Gamma(A\to gg) = \Gamma_{\mathrm{LO}}(A\to gg)
\qty[1+\frac{\alpha_{s}(m_{A})}{\pi}E_{A}^{(1)}+\qty(\frac{\alpha_{s}(m_{A})}{\pi})^{2}E_{A}^{(2)}],
\label{eq: Gam_Agg}
\end{align}
where the decay rate at LO is given by
\begin{align}
\Gamma_{\mathrm{LO}}(A\to gg) =
\frac{\sqrt{2}G_{F}\alpha_{s}^{2}m_{A}^{3}}{128\pi^{3}}\abs{\sum_{q}\kappa_{q}^{A}I_{F}^{A}(\tau_{q})}^{2},
\end{align}
with the loop function $I_{F}^{A}(\tau)$ defined in Eq.~\eqref{eq: IFA}.
The NLO QCD correction $E_{A}^{(1)}$ is given in Refs.~\cite{Spira:1995rr, Harlander:2005rq}, while the NNLO QCD correction $E_{A}^{(2)}$ in heavy top-mass limit is given in Ref.~\cite{Chetyrkin:1998mw}.
The analytic expressions for QCD corrections are given in Appendix~\ref{sec: loop_induced}.

%% file: 04nlo_ew.tex
\section{Next-to-leading-order electroweak corrections} \label{sec: nlo_ew}

\subsection{Branching ratios}
In this subsection, we examine the decay branching ratios of the CP-odd Higgs boson including the higher-order corrections in the four types of 2HDMs.
The qualitative behavior does not change from those at LO, while the size of higher-order corrections can reach several dozens of percent. 
Therefore, we first discuss their behavior and summarize the dominant decay modes in each type of 2HDMs.
Since $\mathrm{BR}(A\to Zh)$ can be negative near $s_{\beta-\alpha}=1$, we include the contribution of the square of the NLO amplitude.
We discuss this point at the end of this subsection.
The magnitude of the higher-order corrections will be examined in the next subsection.

\begin{figure}[t]
	\centering
	\includegraphics[scale=0.75]{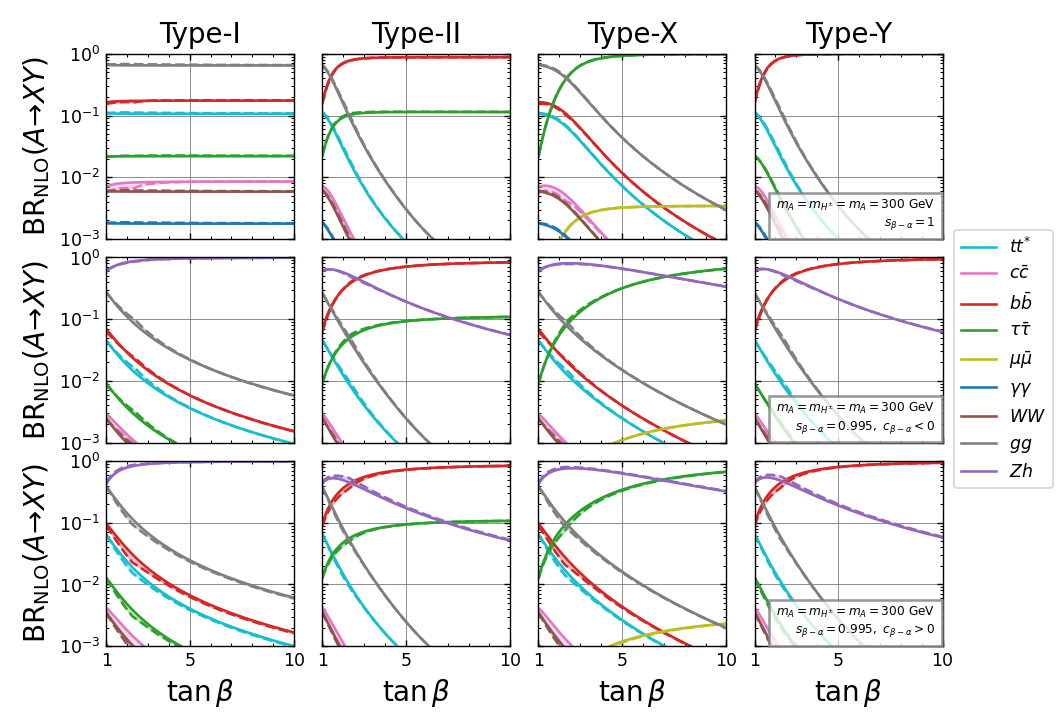}
	\caption{
BR($A\to XY$) as a function of $\tan{\beta}$ in the alignment limit $s_{\beta-\alpha}=1$ (first row) and in the nearly alignment case $s_{\beta-\alpha}=0.995$ with $\cba<0$ (second row) and $\cba>0$ (third row) including the higher-order EW and QCD corrections.
The masses of the additional Higgs bosons are degenerate and taken to be $m_{\Phi}=300$ GeV.
Each decay mode is specified by color as given in the legend. 
The solid and dashed lines correspond to the results with $\sqrt{M^{2}}= M_{\rm max}$ and $M_{\rm min}$, respectively, where $M_{\rm max}$ ($M_{\rm min}$) is a maximum (minimum) value of $\sqrt{M^{2}}$ satisfying the theoretical constraints and $S$ and $T$ parameters.
}
	\label{BrA300_NLO}
\end{figure}

In Fig.~\ref{BrA300_NLO}, we show BR$(A\to XY)$ including the higher-order corrections as a function of $\tan{\beta}$.
We have assumed that the masses of additional Higgs bosons are degenerate, i.e., $m_{\Phi}\equiv m_{A}=m_{H}=m_{H^{\pm}}$ and $m_{\Phi}=300$ GeV, where the CP-odd Higgs boson cannot decay into a pair of top-quarks.
However, the magnitude of the three-body decay width of $A\to tt^{*}$ is not negligible, and we include it at LO.
We have scanned $\sqrt{M^{2}}$ for $0<\sqrt{M^{2}}<m_{A}+500~{\rm GeV}$ under the theoretical constraints and the $S$ and $T$ parameters.
$M_{max}\, (M_{min})$ denotes the maximum (minimum) value of $\sqrt{M^{2}}$.
In order to discuss the theoretical behavior of NLO EW corrections, we dare to omit the constraint from the direct and indirect search and flavor experiment.
The results in Type-I, Type-II, Type-X and Type-Y are shown from the first to fourth columns in order.
The results in the alignment limit are shown in the first row, while those in the nearly alignment scenario with $c_{\beta-\alpha}<0$ and $c_{\beta-\alpha}>0$ are shown in the second and third rows, respectively.

In the alignment limit, the CP-odd Higgs boson mainly decays into a pair of fermions and gluons.
Since $\kappa_{f}=\cot{\beta}$ in Type-I, the partial decay widths of $A\to f\bar{f}$ and $A\to gg$ monotonically decrease as $\tan{\beta}$ increases.
Therefore, the decay branching ratios are almost constant in Type-I.
On the other hand, there is $\tan{\beta}$ enhancement in either or both decays into a pair of down-type quarks and leptons in the other types of 2HDMs.
In Type-II, both $A\to b\bar{b}$ and $A\to \tau\bar{\tau}$ can be dominant with large $\tan{\beta}$, while $A\to \tau\bar{\tau}$ ($A\to b\bar{b}$) can be dominant in Type-X (Type-Y).

In the nearly alignment case with $c_{\beta-\alpha}<0$, the CP-odd Higgs boson decays into the $Z$ boson and the SM-like Higgs boson.
Since the tree-level $hAZ$ vertex is independent of $\tan{\beta}$, the partial decay width of $A\to hZ$ is almost constant in all types of 2HDMs.
In Type-I, the other decay modes monotonically decrease as $\tan{\beta}$ becomes large, and $A\to Zh$ can be the dominant decay mode despite the tree-level vertex being small in the nearly alignment case.
In the other types of 2HDMs, $A\to b\bar{b}$ or $A\to \tau\bar{\tau}$ is enhanced with large $\tan{\beta}$, but the decay branching ratio of $A\to Zh$ is still several percent with $\tan{\beta}=10$.
The results with $c_{\beta-\alpha}>0$ are similar to those with $c_{\beta-\alpha}<0$.

\begin{figure}[t]
	\centering
	\includegraphics[scale=0.75]{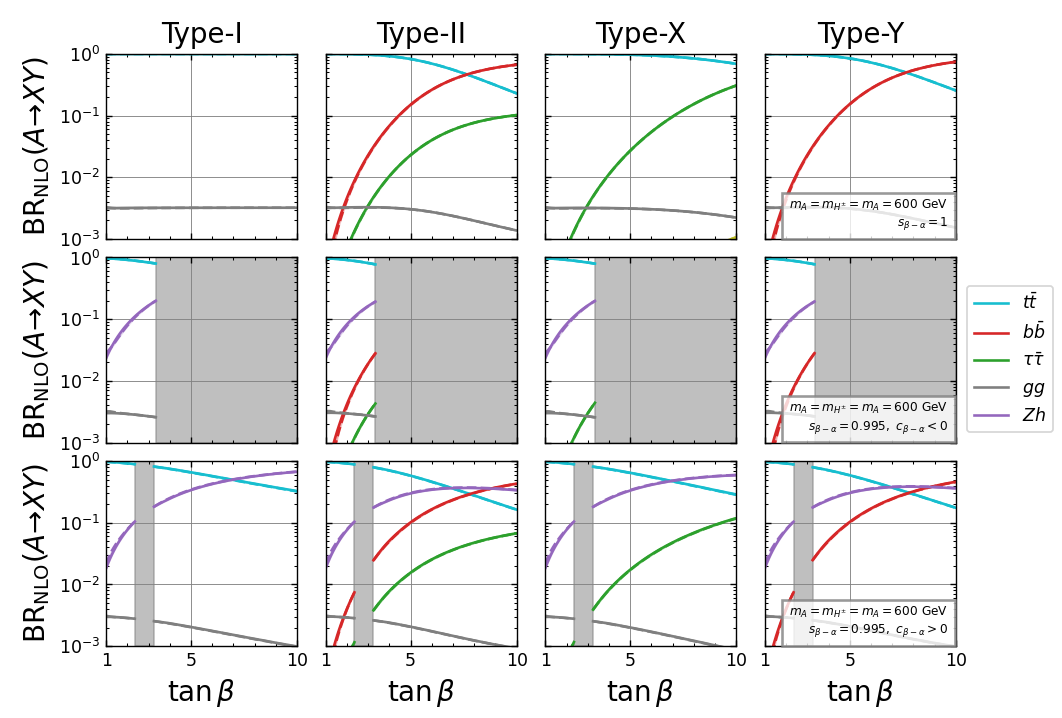}
	\caption{BR($A\to XY$) as a function of $\tan{\beta}$ in the alignment limit $s_{\beta-\alpha}=1$ (first row) and in the nearly alignment case $s_{\beta-\alpha}=0.995$ with $\cba<0$ (second row) and $\cba>0$ (third row) including the higher-order EW and QCD corrections.
The masses of the additional Higgs bosons are degenerate and taken to be $m_{\Phi}=600$ GeV.
Each decay mode is specified by color as given in the legend. 
The solid and dashed lines correspond to the results with $\sqrt{M^{2}}= M_{\rm max}$ and $M_{\rm min}$, respectively.
The gray-shaded region is excluded by the theoretical constraints.}
	\label{BrA600_NLO}
\end{figure}

In Fig.~\ref{BrA600_NLO}, we show BR$(A\to XY)$ with $m_{\Phi}=600$ GeV, where the CP-odd Higgs boson can decay into a pair of top quarks.
While BR$(A\to t\bar{t})$ monotonically decreases as $\tan{\beta}$ becomes large, it is the dominant decay mode for all of the types of 2HDMs due to the large top Yukawa coupling.
In the alignment limit, the CP-odd Higgs boson mainly decays into a pair of fermions.
In Type-I, $A\to t\bar{t}$ is dominant, and the decay branching ratios are almost constant similar to the case with $m_{\Phi}=300$ GeV.
In Type-II, both $A\to b\bar{b}$ and $A\to \tau\bar{\tau}$ are enhanced with large $\tan{\beta}$, and they are comparable with $A\to t\bar{t}$.
In Type-X and Type-Y, $A\to \tau\bar{\tau}$ and $A\to b\bar{b}$ are enhanced, respectively, and they are comparable with $A\to t\bar{t}$.

In the nearly alignment case with $s_{\beta-\alpha}=0.995$ and $m_{\Phi}=600$ GeV, the unitarity and vacuum stability bounds exclude the shaded regions.
In the case with $c_{\beta-\alpha}<0$, $A\to t\bar{t}$ is dominant in all types of 2HDMs since the possible value of the $\tan{\beta}$ is not enough large.
On the other hand, in the case with $c_{\beta-\alpha}<0$, there are parameter regions with large $\tan{\beta}$, where $A\to Zh$ and $A\to b\bar{b}$ can be dominant.

\begin{figure}[t]
	\begin{minipage}{0.45\linewidth}
		\centering
		\includegraphics[scale=0.5]{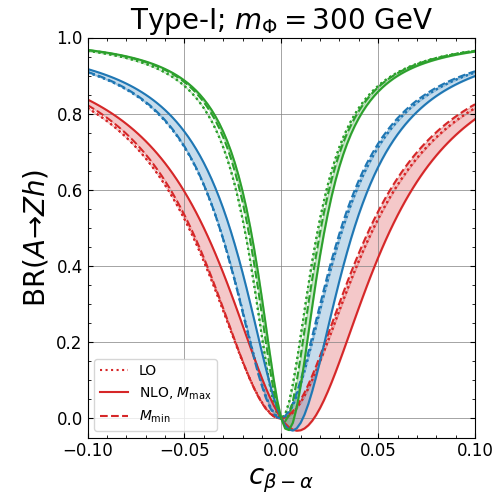}
	\end{minipage}
	\begin{minipage}{0.45\linewidth}
		\centering
		\includegraphics[scale=0.5]{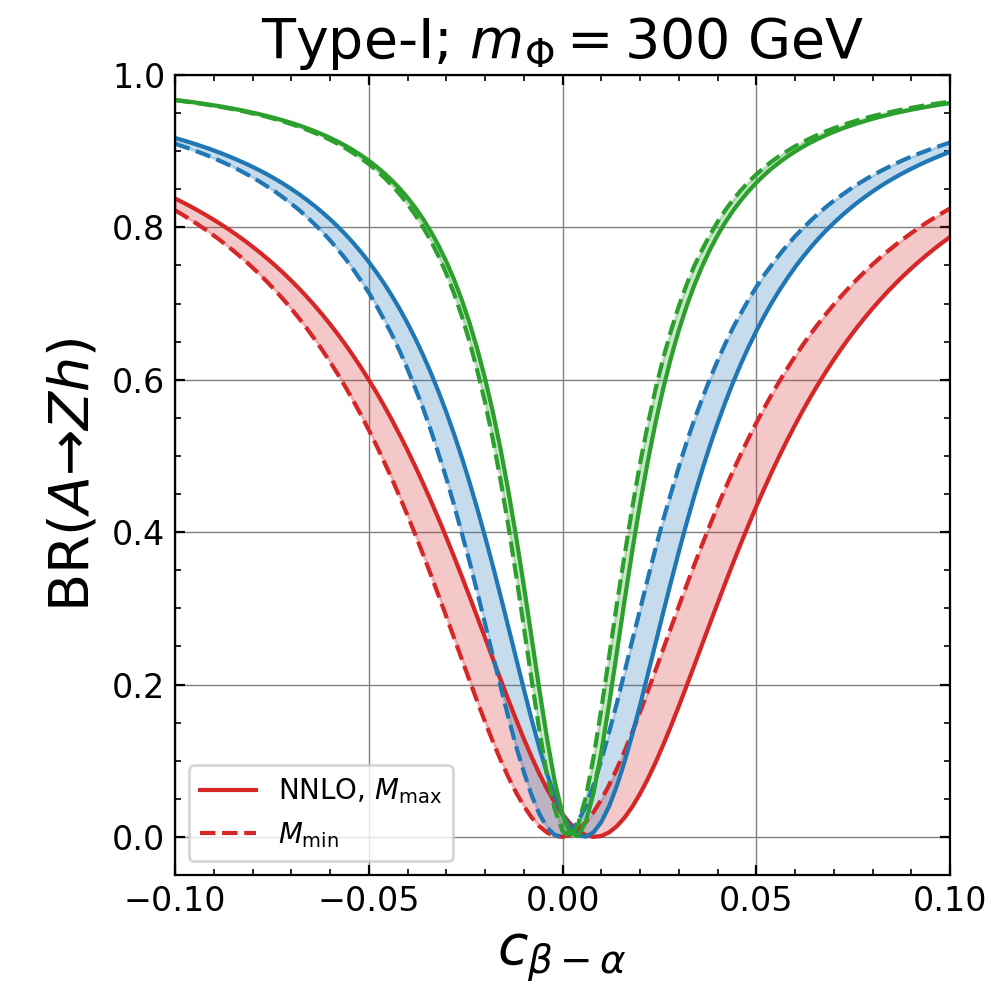}
	\end{minipage} \\
	\begin{minipage}{0.45\linewidth}
		\centering
		\includegraphics[scale=0.5]{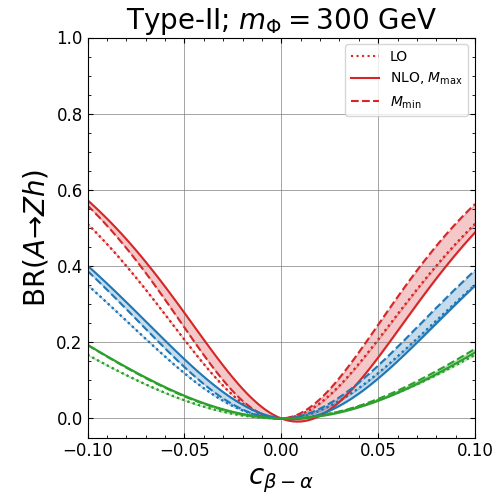}
	\end{minipage}
	\begin{minipage}{0.45\linewidth}
		\centering
		\includegraphics[scale=0.5]{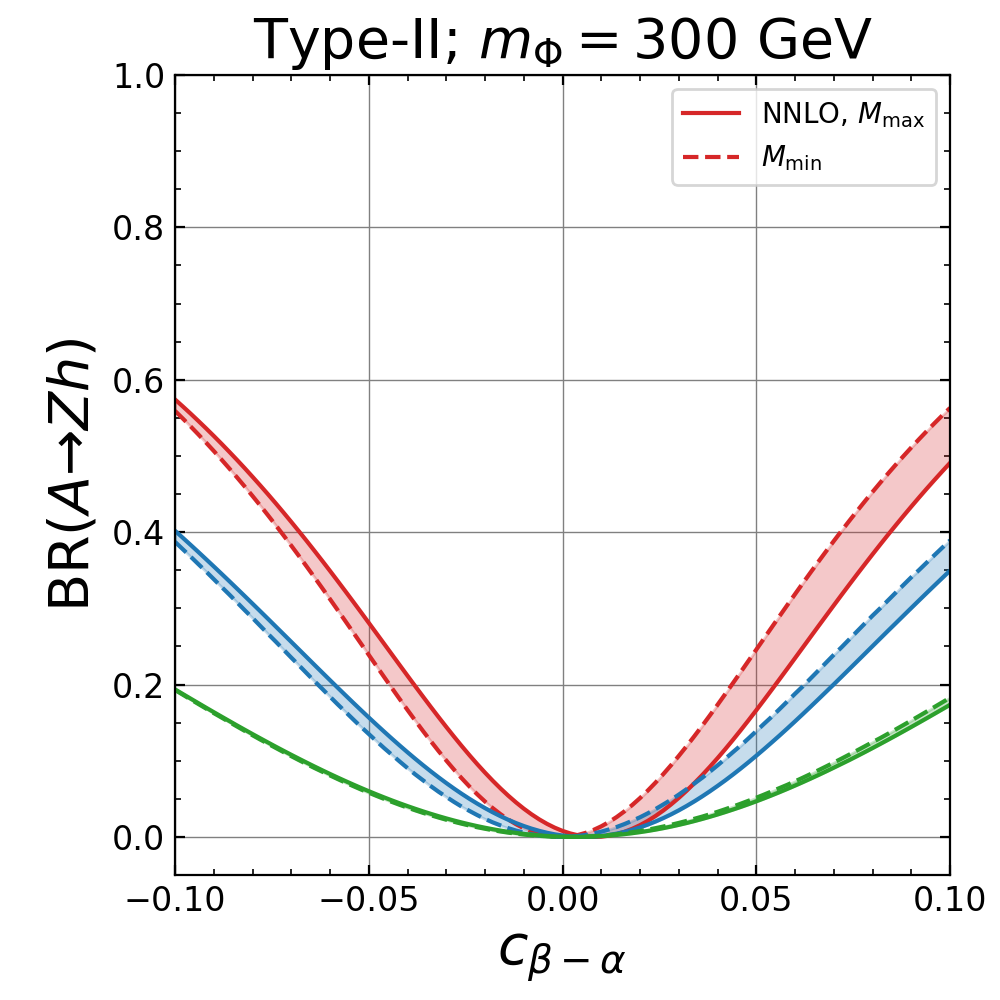}
	\end{minipage}
	\caption{BR($A\to Zh$) including the higher-order corrections with $m_{\Phi}=300$ GeV, and $\tan{\beta}$=2 (red), 3 (blue) and 5 (green).
	We take the masses of additional Higgs bosons to be degenerate.
	The plots in the left (right) column show the results without (with) partial NNLO EW corrections.
	The first (second) row shows the results in Type-I (Type-II) 2HDM.
	$M_{max}\, (M_{min})$ denotes the maximum (minimum) value of $\sqrt{M^{2}}$.}
	\label{BrA_Zh_NLO_and_NNLO}
\end{figure}

As we have mentioned, the partial decay width and the decay branching ratio of $A\to Zh$ can be negative if we truncate the perturbation up to NLO.
In Fig.~\ref{BrA_Zh_NLO_and_NNLO}, we show BR$(A\to Zh)$ including the higher-order corrections with $\tan{\beta}$=2 (red), 3 (blue) and 5 (green).
We assume that the masses of additional Higgs bosons are degenerate and $m_{\Phi}=300$ GeV.
The plots in the left (right) column show the results without (with) partial NNLO EW corrections.
The first and second rows correspond to the results in Type-I and Type-II 2HDMs, respectively.
When we truncate the square of the one-loop amplitude, which corresponds to the two-loop order, the decay branching ratio becomes negative near $c_{\beta-\alpha}=0$.
This is because the tree-level amplitude is proportional to $c_{\beta-\alpha}$, and the LO contribution can be smaller than the NLO contribution.
Since the partial decay width must be positive, the calculation up to NLO is not valid near $c_{\beta-\alpha}=0$.

This problem can be solved by including the square of the NLO amplitude.
As shown in the right column of Fig.~\ref{BrA_Zh_NLO_and_NNLO}, the decay branching ratio becomes positive even with $c_{\beta-\alpha}\simeq 0$, and it gives a physically meaningful result.
Since we have no infrared divergence in the NLO amplitude of $A\to Zh$, we can safely include the square of the NLO amplitude\footnote{In the Higgs-to-Higgs decays of the other additional Higgs bosons, such as $H^{\pm}\to W^{\pm}h$ and $H\to W^{\pm}W^{\mp}$, there are infrared divergences in the NLO amplitude. In order to make the partial decay widths of these decay processes positive even when $c_{\beta-\alpha}$ closes to zero, we need to take into account higher-order corrections to real photon emission.}.

\subsection{Total decay width}
\begin{figure}[t]
	\centering
	\includegraphics[scale=0.75]{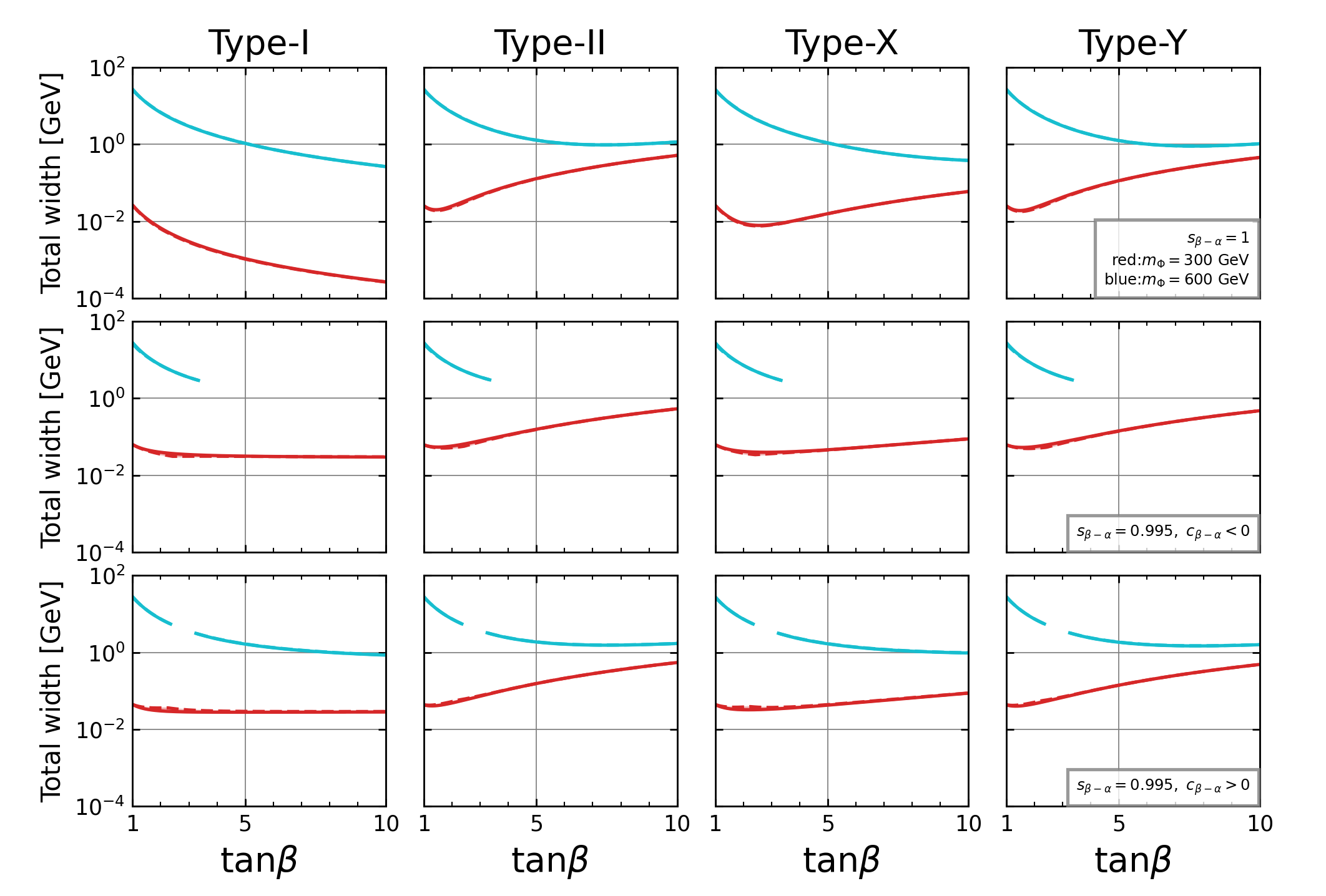}
	\caption{Total decay width as a function of $\tan{\beta}$ in the alignment limit $s_{\beta-\alpha}=1$ (first row) and in the nearly alignment case $s_{\beta-\alpha}=0.995$ with $\cba<0$ (second row) and $\cba>0$ (third row) including the higher-order EW and QCD corrections.
The masses of the additional Higgs bosons are degenerate and taken to be $m_{\Phi}=300$ GeV (red) and 600 GeV (blue).
The solid and dashed lines correspond to the results with $\sqrt{M^{2}}= M_{\rm max}$ and $M_{\rm min}$, respectively.
The lines with $m_{\Phi}=600$ GeV are terminated when $c_{\beta-\alpha}<0$ and they are discontinuous when $c_{\beta-\alpha}>0$ due to the unitarity and vacuum stability bounds.}
	\label{tGamA_NLO}
\end{figure}

In this subsection, we discuss the total decay width of the CP-odd Higgs boson including the higher-order corrections in the four types of 2HDMs.
Similar to the decay branching ratio, the qualitative behavior does not change from those at LO.

In Fig.~\ref{tGamA_NLO}, we show the total decay width of the CP-odd Higgs boson including the higher-order corrections as a function of $\tan{\beta}$.
We assume that the masses of additional Higgs bosons are degenerate, and the red and blue lines correspond to $m_{\Phi}=300$ and 600 GeV, respectively.
As similar to the Figs.~\ref{BrA300_NLO} and \ref{BrA600_NLO}, we have scanned $\sqrt{M^{2}}$ under the theoretical constraints and $S$ and $T$ parameters.

In Type-I with the alignment limit, the total decay width monotonically decreases as $\tan{\beta}$ becomes large because the partial decay widths of $A\to f\bar{f}$ and $A\to gg$ are proportional to $\cot^{2}{\beta}$.
On the other hand, the total decay width increases as $\tan{\beta}$ becomes large in the other types of 2HDMs since either or both $A\to b\bar{b}$ and $A\to \tau\bar{\tau}$ are enhanced, especially with $m_{\Phi}=300$ GeV.
Above the top-quark threshold, $A\to t\bar{t}$ is dominant, and we cannot see sizable enhancement in the total decay width below $\tan{\beta}=10$.

In the nearly alignment scenario with $m_{\Phi}=300$ GeV, the total decay width behaves as almost constant in Type-I, since the dominant $A\to Zh$ is independent of $\tan{\beta}$ at the LO.
In the other types of 2HDMs, $A\to b\bar{b}$ or $A\to \tau\bar{\tau}$ is enhanced as $\tan{\beta}$ becomes large, and they increase the total decay width.
Above the top-quark thresholds, $A\to t\bar{t}$ is dominant, and the behavior is almost the same as that in the alignment limit.
Since the unitarity and vacuum stability bounds exclude the parameter regions as shown in Fig.~\ref{BrA600_NLO}, the lines with $m_{\Phi}=600$ GeV are terminated when $c_{\beta-\alpha}<0$ and they are discontinuous when $c_{\beta-\alpha}>0$.

\subsection{Impact of NLO EW corrections to the decay rates} \label{subsec: Del_EW}

In this subsection, we examine the impact of NLO EW corrections on the decay rates of the CP-odd Higgs boson in Type-I and Type-II.
We do not show the results in Type-X and Type-Y since they are almost similar to those in Type-I or Type-II.
We introduce the following quantity to parametrize the NLO EW corrections.
\begin{align}
\Delta_{\rm EW}(A \to XY)
=\frac{\Gamma_{\rm LO+EW}(A \to XY)}{\Gamma_{\rm LO}(A \to XY)}-1,
\label{eq: dGamA}
\end{align}
where $\Gamma_{\mathrm{LO+EW}}(A \to XY)$ is the decay rate of $A\to XY$ including the NLO EW correction but no QCD correction. 
For the calculation of decay rates at LO, $\Gamma_{\rm LO}(A\to XY)$, we employ the quark pole masses, instead of the running masses.
The NLO EW corrections on the total decay width are parameterized by
\begin{align}
\Delta_{\mathrm{EW}}^{\mathrm{tot}}
&=
\frac{\Gamma^{\mathrm{tot}}_{\mathrm{LO+EW}}}{\Gamma^{\mathrm{tot}}_{\mathrm{LO}}}-1,
\label{eq: dGamAtot}
\end{align}
where $\Gamma_{\mathrm{LO+EW}}^{\mathrm{tot}}$ is the total decay width of the CP-odd Higgs boson including the NLO EW corrections but no QCD correction.
As similar to $\Delta_{\rm EW}(A \to XY)$, we use the quark pole masses in the calculation of total decay width at LO, $\Gamma_{\rm LO}^{\mathrm{tot}}$.

We evaluate $\Delta_{\rm EW}(A \to XY)$ and $\Delta_{\mathrm{EW}}^{\mathrm{tot}}$ in the alignment limit, $\sba=1$, and the nearly alignment scenario, $\sba=0.995$ with $c_{\beta-\alpha}<0$ and $c_{\beta-\alpha}>0$.
We assume that the masses of additional Higgs bosons are degenerate.
For each scenario, we take $\tan{\beta}=1, 3,$ and 10, and scan $\sqrt{M^{2}}$ under the constraints of perturbative unitarity, vacuum stability and $S$ and $T$ parameters.
In order to discuss the theoretical behavior of NLO EW corrections, we dare to omit the constraint from the direct and indirect search and flavor experiment.

\begin{figure}[t]
	\centering
	\includegraphics[scale=0.525]{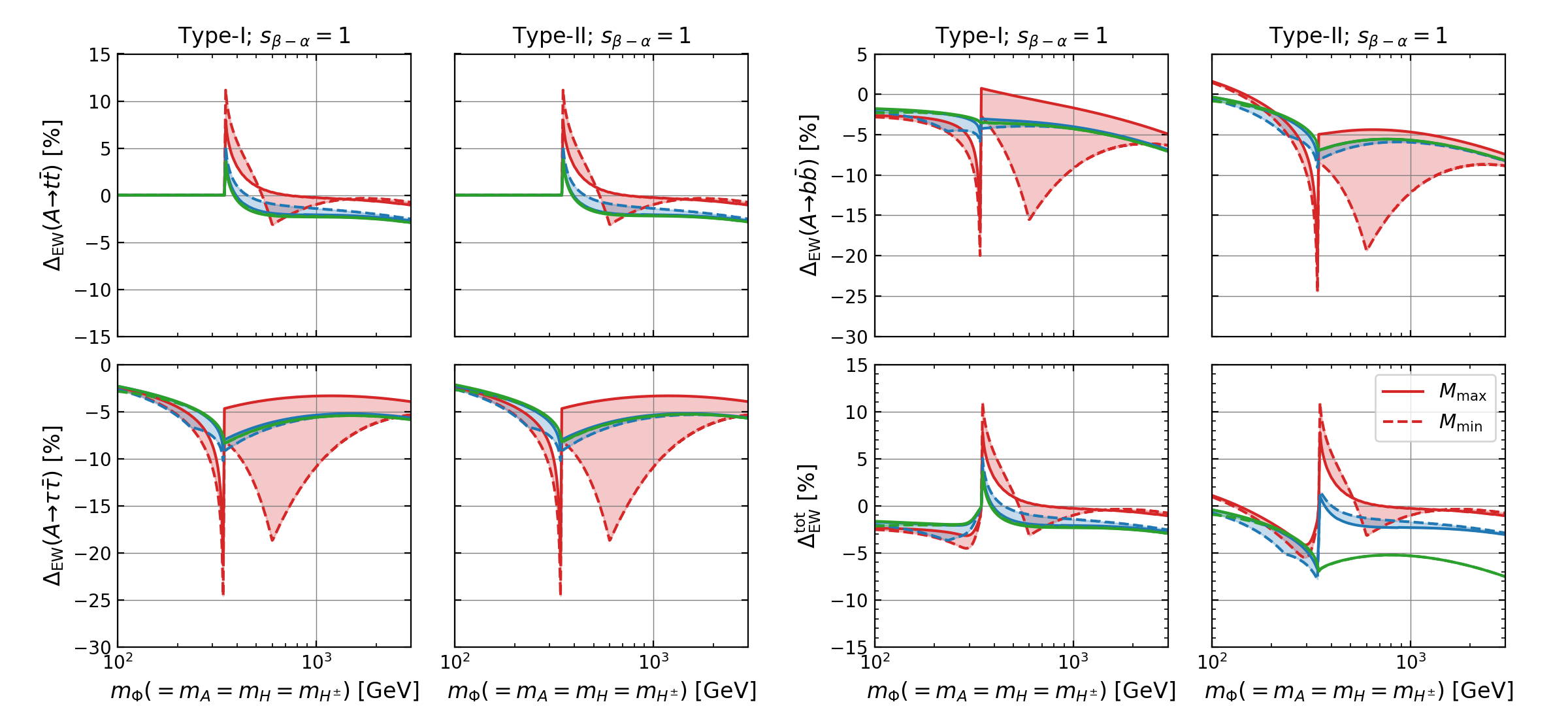}
\caption{NLO EW corrections to the partial decay widths of the CP-odd Higgs boson with $s_{\beta-\alpha}=1$ and $\tan{\beta}$=1 (red), 3 (blue), 10 (green).
	Masses of additional Higgs bosons are degenerate, $m_{\Phi} = m_{H^{\pm}} = m_{A} = m_{H}$. 
	$M_{max} (M_{min})$ denotes the maximum (minimum) value of $\sqrt{M^{2}}$ satisfying the theoretical constraints and $S,\,T$ parameters.}
	\label{del_EW_sba1}
\end{figure}

In Fig.~\ref{del_EW_sba1}, we show $\Delta_{\rm EW}(A \to XY)$ as a function of $m_{\Phi}$ with $s_{\beta-\alpha}=1$.
The red, blue, and green colored regions correspond to $\tan{\beta}$=1, 3, and 10, respectively.
The solid and dashed lines correspond to the results with maximum and minimum values of $\sqrt{M^{2}}$ under the constraints, respectively.

For $\tb=1$, there are two kinks at $m_{\Phi}\simeq 350~{\rm GeV}$ and $m_{\Phi}\simeq 600~{\rm GeV}$. 
The kink at $m_{\Phi}\simeq 350~{\rm GeV}$ is the threshold of the top quark.
For $A\to t\bar{t}$, threshold effects appear both in 1PI triangle diagrams and top-quark loop diagrams in the two-point function of the neutral Higgs bosons.
The latter contributions also apper in $A\to b\bar{b}$ and $A\to \tau\bar{\tau}$.
The kink at $m_{\Phi}\simeq 600~{\rm GeV}$ corresponds to the point where the minimum value of $\sqrt{M^{2}}$ changes from zero to non-zero due to the theoretical bounds.
At this point, the scalar couplings take maximal value under the constraints, especially from perturbative unitarity, and the non-decoupling effects of scalar loops become dominant.
The behaviors of $\Delta_{\rm EW}(A \to t\bar{t})$ and $\Delta_{\rm EW}(A \to \tau\bar{\tau})$ in Type-I are almost the same as those in Type-II.
The size of non-decoupling effects for $\Delta_{\rm EW}(A \to t\bar{t})$ reaches about $-3\%$, while that for $\Delta_{\rm EW}(A \to \tau\bar{\tau})$ reaches about $-19\%$.
On the other hand, the behavior of $\Delta_{\rm EW}(A \to b\bar{b})$ in Type-I is different from that in Type-II.
This is because of the contributions from 1PI diagrams including the virtual $G^{\pm}, H^{\pm}$ and $W^{\pm}$ bosons, that proportional to $\zeta_{t}\zeta_{b}$.
Due to the large top Yukawa coupling and the type dependence of $\zeta_{t}\zeta_{b}$, the behavior of $\Delta_{\rm EW}(A \to b\bar{b})$ depends on the types of 2HDMs.
The size of non-decoupling effects for $\Delta_{\rm EW}(A \to b\bar{b})$ reaches about $-15\%$ and $-20\%$ in Type-I and Type-II, respectively.

For $\tb=3$ and 10, $\sqrt{M^{2}}$ is almost degenerate with $m_{\Phi}$ due to the theoretical constraints.
$\Delta_{\rm EW}(A \to t\bar{t})$ is almost constant above the top threshold, and it is about $-2\%$ for $m_{\Phi}=1$ TeV.
The size of $\Delta_{\rm EW}(A \to b\bar{b})$ is about $-4\%$ ($-6\%$) for Type-I (Type-II) with $m_{\Phi}=1$ TeV, while that of $\Delta_{\rm EW}(A \to \tau\bar{\tau})$ is about $-5\%$ both in Type-I and Type-II.

The behavior of $\Delta^{\mathrm{tot}}_{\mathrm{EW}}$ in Type-I is different from that in Type-II, especially with $\tan{\beta}=10$.
In Type-I, it is almost the same as $\Delta_{\rm EW}(A \to t\bar{t})$, and $\Delta^{\mathrm{tot}}_{\mathrm{EW}}$ is about $-2\%$ for $m_{\Phi}=1$ TeV with $\tan{\beta}=10$.
On the other hand, $\Delta^{\mathrm{tot}}_{\mathrm{EW}}$ is about $-5\%$ for $m_{\Phi}=1$ TeV with $\tan{\beta}=10$ in Type-II.
This is because of $\tan{\beta}$ enhancement of $\mathrm{BR}(A\to b\bar{b})$ in Type-II.
Above the top-quark threshold, $\Delta^{\mathrm{tot}}_{\mathrm{EW}}$ can be expanded as
\begin{align}
\Delta^{\mathrm{tot}}_{\mathrm{EW}}
&=
\Delta_{\mathrm{EW}}(A\to t\bar{t})\mathrm{BR}_{\mathrm{LO}}(A\to t\bar{t})
+\Delta_{\mathrm{EW}}(A\to b\bar{b})\mathrm{BR}_{\mathrm{LO}}(A\to b\bar{b})
+ \text{other channels}.
\label{eq: dGamAtot_decompose}
\end{align}
As we can see from Fig.~\ref{BrA600_NLO}, $\mathrm{BR}(A\to t\bar{t})\approx1$ in Type-I, and $\Delta^{\mathrm{tot}}_{\mathrm{EW}}\approx \Delta_{\mathrm{EW}}(A\to t\bar{t})$.
On the other hand, $\mathrm{BR}(A\to b\bar{b})$ is also dominant in Type-II with large $\tan{\beta}$.
Therefore, $\Delta_{\mathrm{EW}}(A\to b\bar{b})$ also contributes to $\Delta^{\mathrm{tot}}_{\mathrm{EW}}$.

\begin{figure}[t]
	\centering
	\includegraphics[scale=0.6]{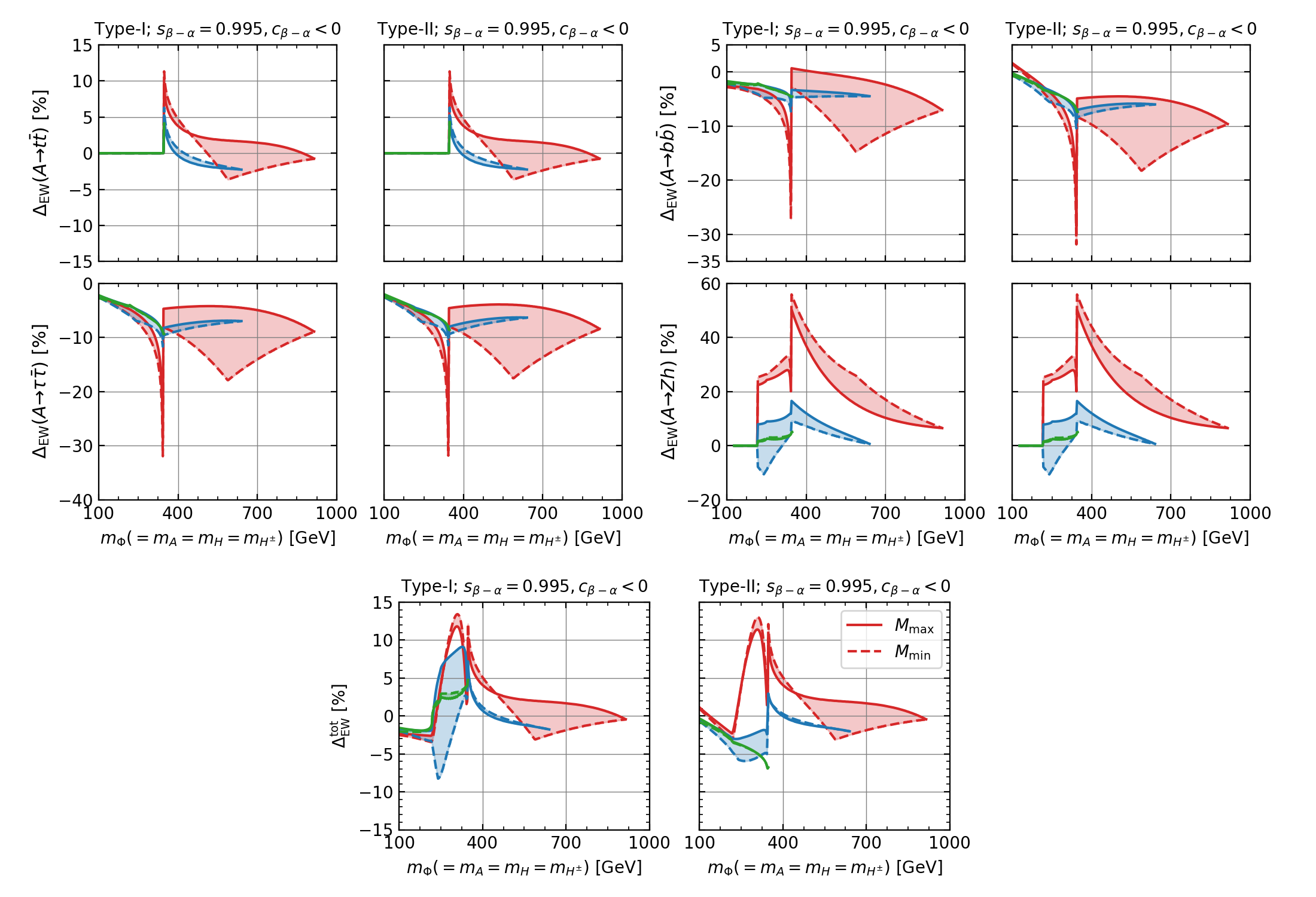}
	\caption{NLO corrections for the decay widths of the CP-odd Higgs boson with $s_{\beta-\alpha}=0.995$, $c_{\beta-\alpha}<0$ and $\tan{\beta}$=1 (red), 3 (blue), 10 (green).}
	\label{del_EW_sba995m}
\end{figure}

In Fig.~\ref{del_EW_sba995m}, we show $\Delta_{\rm EW}(A \to XY)$ and $\Delta^{\mathrm{tot}}_{\mathrm{EW}}$ as a function of $m_{\Phi}$ with $\sba=0.995$ and $\cba<0$.
The main differences between the case with $s_{\beta-\alpha}=1$ and $s_{\beta-\alpha}\neq1$ are upper bound on $m_{\Phi}$ and the new decay mode $A\to Zh$.
Since we cannot take the decoupling limit keeping $s_{\beta-\alpha}\neq1$, there is an upper bound on $m_{\Phi}$, and it should be below 1~TeV with $s_{\beta-\alpha}=0.995$ and $\tan{\beta}=1$ under the theoretical constraints, especially perturbative unitarity.
Since the deviation from the alignment limit is small, the qualitative behavior of $\Delta_{\rm EW}(A \to f\bar{f})$ is almost unchanged, except for the region with large $m_{\Phi}$.

The behavior of $\Delta_{\rm EW}(A\to Zh)$ in Type-I is almost the same as that in Type-II.
There are two kinks at $m_{\Phi}\simeq 200~{\rm GeV}$ and $350~{\rm GeV}$. 
The first one corresponds to the threshold effects of $Zh$, and the second one corresponds to those of the top quark.
Near the top-quark threshold, the magnitude of the NLO EW correction is above $50\%$, while it decreases as $m_{\Phi}$ becomes large.
When $\tan{\beta}=1$, $\Delta_{\rm EW}(A\to Zh)$ is positive, and non-decoupling effects of scalar loops enhance its magnitude.
On the other hand, non-decoupling effects decrease the size of $\Delta_{\rm EW}(A\to Zh)$ when $\tan{\beta}=3$, and it can be negative for $m_{\Phi}\leq 300~\mathrm{GeV}$.

The behavior of $\Delta^{\mathrm{tot}}_{\mathrm{EW}}$ can be understood by Eq.~\eqref{eq: dGamAtot_decompose}, while we have an additional contribution from $A\to Zh$.
When $\tan{\beta}=1$, the behavior of $\Delta^{\mathrm{tot}}_{\mathrm{EW}}$ in Type-I is almost the same as that in Type-II, except for below the $Zh$ threshold.
Below the $Zh$ threshold, $A\to b\bar{b}$ and $A\to gg$ are the main decay modes, and $\Delta_{\mathrm{EW}}(A\to b\bar{b})$ causes the difference between Type-I and Type-II. 
Above the $Zh$ threshold, $A\to Zh$ and $A\to gg$ are the main decay modes, while $A\to tt^{*}$ also contribute to the decay branching ratio, especially for $m_{\Phi}\lesssim 2m_{t}$.
Above the top-quark threshold, $A\to t\bar{t}$ mainly determine $\Delta^{\mathrm{tot}}_{\mathrm{EW}}$.
When $\tan{\beta} = 10$, $A\to t\bar{t}$ is suppressed by $\cot{\beta}$, and $A\to Zh$ can be dominant, while $A\to b\bar{b}$ also contributes to $\Delta^{\mathrm{tot}}_{\mathrm{EW}}$ in Type-II.

\begin{figure}[t]
	\centering
	\includegraphics[scale=0.6]{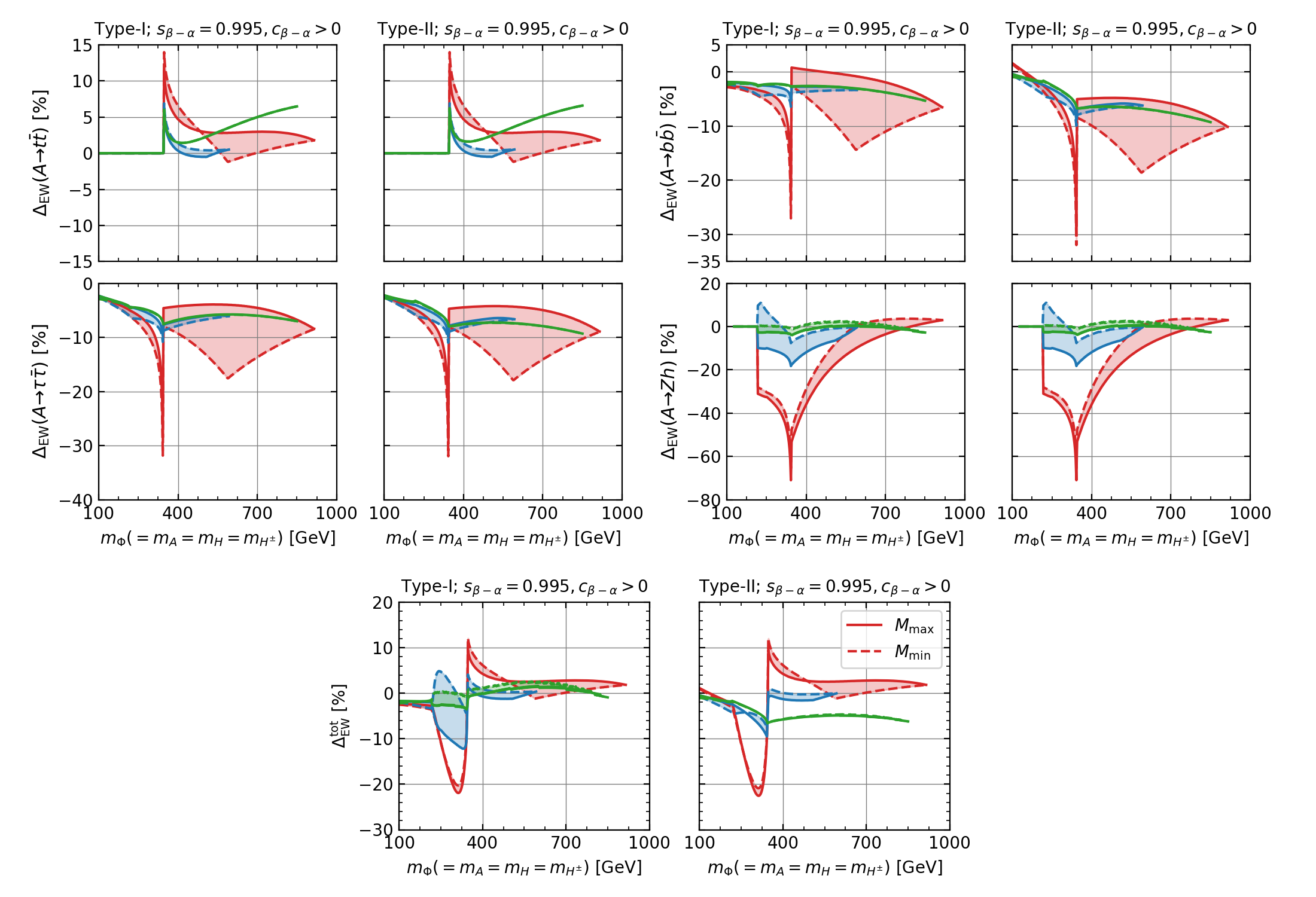}
	\caption{NLO corrections for the decay widths of the CP-odd Higgs boson with $s_{\beta-\alpha}=0.995$, $c_{\beta-\alpha}>0$ and $\tan{\beta}$=1 (red), 3 (blue), 10 (green).}
	\label{del_EW_sba995p}
\end{figure}

In Fig.~\ref{del_EW_sba995p}, we show $\Delta_{\rm EW}(A \to XY)$ and $\Delta^{\mathrm{tot}}_{\mathrm{EW}}$ as a function of $m_{\Phi}$ with $\sba=0.995$ and $\cba>0$.
The main differences from the case with $\cba<0$ are the possible value of $\tan{\beta}$ and the sign of $\Delta_{\rm EW}(A \to Zh)$.
The allowed regions for $\tb=10$ are broader than those in $c_{\beta-\alpha}<0$, and the upper bound on $m_{\Phi}$ is about 850~GeV.
The sign of the NLO EW corrections to $A\to Zh$ is opposite to those in $c_{\beta-\alpha}<0$, and higher-order corrections mainly decrease the partial decay width.
Above the top-threshold, the behavior of $\Delta^{\mathrm{tot}}_{\mathrm{EW}}$ is mainly determined by $A\to t\bar{t}$ and $A\to Zh$ in Type-I similar to the case with $c_{\beta-\alpha}<0$, while $A\to b\bar{b}$ also contributes to $\Delta^{\mathrm{tot}}_{\mathrm{EW}}$ when $\tan{\beta}$ is large in Type-II.

%% file: 05numerical_result.tex
\section{Decay pattern of the CP-odd Higgs boson in the nearly alignment scenario} \label{sec: numerical_result}

In this section, we discuss the decay pattern of the CP-odd Higgs boson in the nearly alignment scenario including the higher-order corrections.
In addition to the theoretical constraints, we take into account the constraint from the $S$ and $T$ parameters, the signal strength of the SM-like Higgs boson, the direct searches of the additional Higgs bosons, and flavor experiments explained in Sec.~\ref{sec: constraint}.
Although there are constraints from the direct searches for $A\to Zh$ and $H\to hh$ at the LHC, they are quite sensitive to the value of $s_{\beta-\alpha}$, and the excluded region would be changed when we include the higher-order corrections.
Therefore, we dare to omit the constraint from the Higgs-to-Higgs decay modes in this paper.
We will study how the higher-order corrections modify the constraints from the direct searches for the additional Higgs bosons elsewhere.

We consider the following two scenarios for the mass spectrum of the additional Higgs bosons.
\begin{align}
\text{Scenario~A}&:\quad m_{A} = m_{H^{\pm}} = 300~{\rm GeV}\qc m_{A}\leq m_{H},\\
\text{Scenario~B}&:\quad m_{A} = m_{H^{\pm}} = 800~{\rm GeV}\qc m_{H}\leq m_{A}-m_{Z}. 
\end{align}
In order to avoid the constraint from the $T$ parameter, we assume that the Higgs potential respects the custodial symmetry\footnote{We can also consider the scenario with the twisted-custodial symmetry~\cite{Gerard:2007kn, Haber:2010bw, Aiko:2020atr}, which requires $m_{H}=m_{H^{\pm}}=\sqrt{M^{2}}$ and $s_{\beta-\alpha}=1$. In this scenario, the Higgs-to-Higgs decays are prohibited.}~\cite{Pomarol:1993mu}, and the masses of the CP-odd and charged Higgs bosons are degenerate, $m_{A}=m_{H^{\pm}}$.
In Scenario~A, $A\to t\bar{t}$ does not open, and the CP-odd Higgs boson mainly decays into $Zh,\,tt^{*},\, b\bar{b},\, \tau\bar{\tau}$ and $gg$ as shown in Fig.~\ref{BrA300_NLO}.
In Scenario~B, the CP-odd Higgs boson mainly decays into $Zh,\, ZH,\, t\bar{t},\, b\bar{b}$ and $\tau\bar{\tau}$ as shown in Fig.~\ref{BrA600_NLO}.

For Scenario~A, we study Type-I because Type-II and Type-Y are excluded by the flavor constraints~\cite{Haller:2018nnx}, and Type-X is also excluded by the constraint from $A\to \tau\bar{\tau}$ \cite{Aiko:2020ksl}.
We take $\tan{\beta}=2$ and $5$.
The mass of the additional CP-even Higgs boson is scanned as
\begin{align}
m_{A} \leq m_{H} \leq m_{A}+500~{\rm GeV}. \label{eq: scan_mbH_A}
\end{align}
The remaining parameters are scanned as
\begin{align}
0.995 \leq \sba \leq 1\qc 0 \leq \sqrt{M^{2}} \leq m_{A}+500{\rm GeV}, \label{eq: scan_others_A}
\end{align}
with both of $\cba<0$ and $\cba>0$.
The lower bound on $\tan{\beta}$ comes from the constraint of $B_{d}\to\mu\mu$~\cite{Haller:2018nnx}.
In addition, the region with $\tb\lesssim 2$ is excluded by the direct searches for $A\to \tau\bar{\tau}$ and $H^{\pm}\to tb$~\cite{Aiko:2020ksl}.

For Scenario~B, we study all types of 2HDMs.
We scan $\tan{\beta}$ as
\begin{align}
2<\tb<10. \label{eq: scan_others_B}
\end{align}
For $m_{H^{\pm}}=800~\mathrm{GeV}$, the region with $\tan{\beta}\lesssim 1.2$ is excluded for all types of 2HDM by $B_{d}\to \mu\mu$~\cite{Haller:2018nnx}.
While there are allowed parameter regions with $\tan{\beta}\leq 2$, we take $\tan{\beta}=2$ as a minimum value conservatively.
We also scan the mass of the additional CP-even Higgs bosons as
\begin{align}
m_{A}-500~{\rm GeV} < m_{H} < m_{A}-m_{Z}.
\end{align}
The scan regions of the remaining parameters are the same as those in Eqs.~\eqref{eq: scan_mbH_A} and \eqref{eq: scan_others_A}.

In order to study the size of the NLO EW corrections on the decay branching ratios, we define $\Delta_{\mathrm{EW}}^{\mathrm{BR}}(A\to XY)$ as
\begin{align}
\Delta_{\mathrm{EW}}^{\mathrm{BR}}(A\to XY)
&=
\frac{\mathrm{BR}_{\mathrm{LO+EW+QCD}}(A\to XY)}{\mathrm{BR}_{\mathrm{LO+QCD}}(A\to XY)}-1,
\label{eq: dBRA}
\end{align}
where $\mathrm{BR}_{\mathrm{LO+EW+QCD}}(A\to XY)$ includes the NLO EW and the higher-order QCD corrections, while $\mathrm{BR}_{\mathrm{LO+QCD}}(A\to XY)$ includes only the higher-order QCD corrections.
We employ the running quark masses for the evaluation of the LO parts.
At the one-loop level, we can approximate $\Delta_{\mathrm{EW}}^{\mathrm{BR}}(A\to XY)$ as
\begin{align}
\Delta_{\mathrm{EW}}^{\mathrm{BR}}(A\to XY)
&\approx
\overline{\Delta}_{\mathrm{EW}}(A\to XY)-\overline{\Delta}_{\mathrm{EW}}^{\mathrm{tot}},
\label{eq: dBRA decomposed}
\end{align}
where $\overline{\Delta}_{\mathrm{EW}}(A\to XY)$ and $\overline{\Delta}_{\mathrm{EW}}^{\mathrm{tot}}$ are given by
\begin{align}
\overline{\Delta}_{\mathrm{EW}}(A\to XY)
&=
\frac{\Gamma_{\mathrm{LO+EW+QCD}}(A\to XY)}{\Gamma_{\mathrm{LO+QCD}}(A\to XY)}-1, \label{eq: dEWA} \\
\overline{\Delta}_{\mathrm{EW}}^{\mathrm{tot}}
&=
\frac{\Gamma^{\mathrm{tot}}_{\mathrm{LO+EW+QCD}}}{\Gamma^{\mathrm{tot}}_{\mathrm{LO+QCD}}}-1.
\label{eq: dGamAbartot}
\end{align}
For the decays into a pair of quarks, we can approximate $\overline{\Delta}_{\mathrm{EW}}(A\to Q\bar{Q})\, (Q=q, t)$ as
\begin{align}
\overline{\Delta}_{\mathrm{EW}}(A\to Q\bar{Q})
&\approx
\frac{m_{Q}^{2}}{\overline{m}_{Q}^{2}(m_{A})}\Delta_{\mathrm{EW}}(A\to Q\bar{Q}),
\label{eq: dEWbar_AQQ}
\end{align}
where $\Delta_{\mathrm{EW}}(A\to XY)$ given in Eq.~\eqref{eq: dGamA}.
The ratio of the pole and running mass of the quarks enhances the size of  $\overline{\Delta}_{\mathrm{EW}}(A\to Q\bar{Q})$.
For instance, $m_{b}/\overline{m}_{b}(m_{A})\approx 2$ for $m_{A}=500~\mathrm{GeV}$, and the size of $\overline{\Delta}_{\mathrm{EW}}(A\to b\bar{b})$ becomes four times larger than $\Delta_{\mathrm{EW}}(A\to b\bar{b})$.
For $A\to \ell\bar{\ell}$ and $A\to V\phi$, $\overline{\Delta}_{\mathrm{EW}}(A\to XY)=\Delta_{\mathrm{EW}}(A\to XY)$ because we have no QCD correction at the one-loop level.

In order to study the deviation in the SM-like Higgs boson coupling, we define $\Delta\kappa_{Z}$ as
\begin{align}
\Delta\kappa_{Z} = \kappa_{Z}-1,
\end{align}
where $\kappa_{Z}$ is defined in Eq.~\eqref{eq: scaling-factor} with the partial decay rate of $h\to ZZ^{*}$.
In the 2HDM, $\Delta\kappa_{Z}$ is mostly negative independently of the sign of $c_{\beta-\alpha}$~\cite{Kanemura:2018yai}.
On the other hand, allowed parameter regions depend on the sign of $c_{\beta-\alpha}$.
Therefore, we take $\mathrm{Sign}(c_{\beta-\alpha})\abs{\Delta\kappa_{Z}}$ as the horizontal axis in some figures shown below to clearly exhibit the difference due to the sign of the $c_{\beta-\alpha}$.

\subsection{Scenario~A}
\begin{figure}[t]
	\centering
	\includegraphics[scale=0.7]{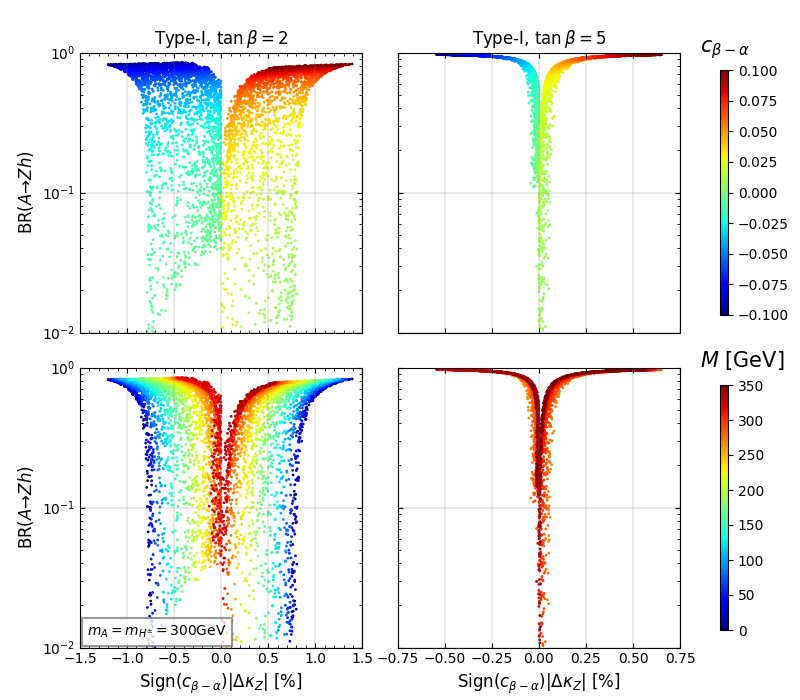}
	\caption{Decay branching ratios of $A\to Zh$ in Type-I 2HDM in Scenario~A.
	The color differences correspond to the values of $c_{\beta-\alpha}$ and $\sqrt{M^{2}}$, in the top and bottom panels, respectively.}
	\label{BrA300_NLO_scan}
\end{figure}

In Fig.~\ref{BrA300_NLO_scan}, we show the branching ratio of $A\to Zh$ including the NLO EW corrections in Type-I for Scenario~A.
The plots in the first column are the results with $\tan{\beta}=2$, while those in the second column are the results with $\tan{\beta}=5$.
The color differences correspond to the values of $c_{\beta-\alpha}$ and $\sqrt{M^{2}}$ in the top and bottom panels, respectively.
When $c_{\beta-\alpha}\simeq 0.1$, $\BR(A\to Zh)$ reaches almost 80\% (100\%) for $\tan{\beta}=2\, (5)$.
Since $\Delta\kappa_{Z}$ is proportional to $\abs{c_{\beta-\alpha}}$ at LO, we expect that $\BR(A\to Zh)$ becomes large when $\Delta\kappa_{Z}$ is large.
However, this is not necessarily true when we include NLO corrections.
When $\tan{\beta}=2$, we have the parameter points, where $\Delta\kappa_{Z}$ sizably deviates from the SM value, while the $\mathrm{BR}(A\to Zh)$ is small.
This is because of the non-decoupling scalar-loop effects in $\Delta\kappa_{Z}$.
When $\sqrt{M^{2}}\approx 0$, the non-decoupling effects make $\Delta\kappa_{Z}$ large even with $c_{\beta-\alpha}\approx 0$.
When $\tan{\beta}=5$, $\sqrt{M^{2}}$ is almost degenerate with $m_{A}$ due to perturbative unitarity and vacuum stability, and the size of the higher-order corrections is smaller than the tree-level contributions.
Therefore, the correlation between $\mathrm{BR}(A\to Zh)$ and $\Delta\kappa_{Z}$ follows the expectation at LO.
We note that $\BR(A\to Zh)$ can reach about 100\% even with $\Delta \kappa_{Z}\leq 0.6$, where it is difficult to observe the deviation in the $hZZ$ coupling at the ILC250~\cite{Barklow:2017suo}.
In this sense, $A\to Zh$ is useful to explore the nearly alignment scenario.

\begin{figure}[t]
	\centering
	\includegraphics[scale=0.5]{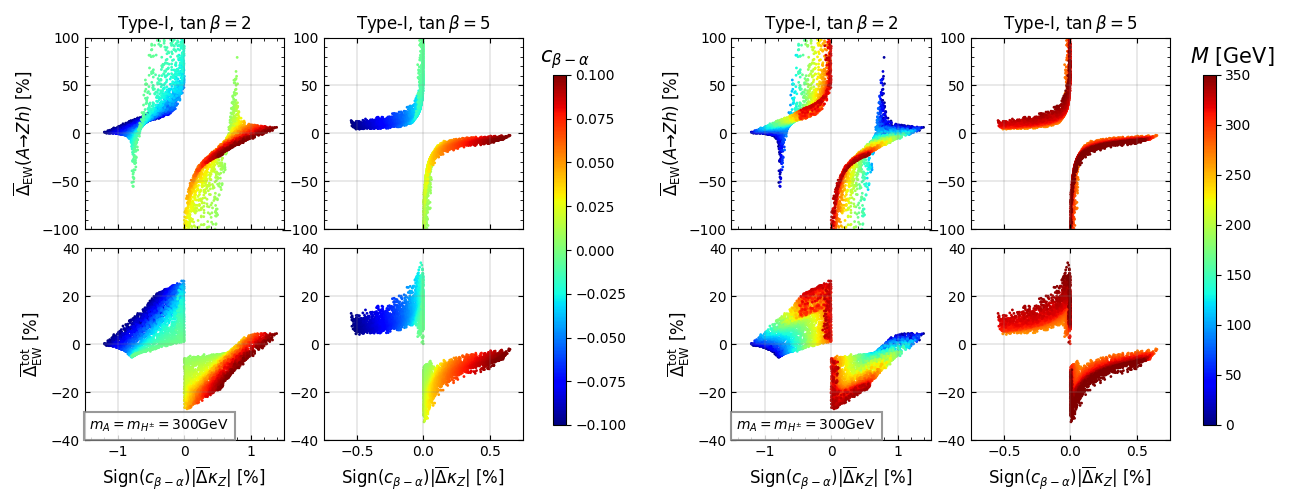}
	\caption{NLO corrections to the decay rates of $A\to Zh$ and the total decay width of the CP-odd Higgs boson in Type-I 2HDM in Scenario~A.
	The color differences correspond to the values of $c_{\beta-\alpha}$ and $\sqrt{M^{2}}$, in the left and right panels, respectively.}
	\label{dGamA300_NLO_scan}
\end{figure}

In Fig.~\ref{dGamA300_NLO_scan}, we show the size of NLO EW corrections to the decay rates $\overline{\Delta}_{\mathrm{EW}}(A\to Zh)$ and the total decay width $\overline{\Delta}^{\mathrm{tot}}_{\mathrm{EW}}$.
The plots in the first and third columns are the results with $\tan{\beta}=2$, while those in the second and fourth columns are the results with $\tan{\beta}=5$.
The color differences correspond to the values of $c_{\beta-\alpha}$ and $\sqrt{M^{2}}$ in the left and right panels, respectively.
The sign of $\overline{\Delta}_{\mathrm{EW}}(A\to Zh)$ is opposite with the sign of $c_{\beta-\alpha}$, except for $\sqrt{M^{2}}\lesssim 150$ GeV with $\tan{\beta}=2$.
The size of $\overline{\Delta}_{\mathrm{EW}}(A\to Zh)$ becomes quite large when $c_{\beta-\alpha}\simeq 0$ since the LO contribution is close to zero.
We note that such large $\overline{\Delta}_{\mathrm{EW}}(A\to Zh)$ does not mean a breakdown of the perturbation, since it is caused by the smallness of the tree-level coupling.
When $\sqrt{M^{2}}\lesssim 150$ GeV with $\tan{\beta}=2$, the non-decoupling effects of scalar loops change the sign of $\overline{\Delta}_{\mathrm{EW}}(A\to Zh)$ as shown Figs.~\ref{del_EW_sba995m} and \ref{del_EW_sba995p}.
In addition, $\overline{\Delta}_{\mathrm{EW}}(A\to Zh)$ becomes large due to the smallness of the LO contribution, while $\Delta\kappa_{Z}$ takes large value even if $c_{\beta-\alpha}\simeq 0$ due to the non-decoupling effects.
Therefore, we have two kinks around $\abs{\kappa_{Z}}\simeq 0.8$ with $c_{\beta-\alpha}\simeq 0$.

As similar to Eq.~\eqref{eq: dGamAtot_decompose}, we can expand $\overline{\Delta}^{\mathrm{tot}}_{\mathrm{EW}}$ as
\begin{align}
\overline{\Delta}^{\mathrm{tot}}_{\mathrm{EW}}
&=
\overline{\Delta}_{\mathrm{EW}}(A\to b\bar{b})\mathrm{BR}_{\mathrm{LO}}(A\to b\bar{b})+\overline{\Delta}_{\mathrm{EW}}(A\to Zh)\mathrm{BR}_{\mathrm{LO}}(A\to Zh) \notag \\
&\quad
+ \text{other channels}.
\label{eq: dGamAbartot_decompose}
\end{align}
When $c_{\beta-\alpha}\simeq 0.1$, the behavior of $\overline{\Delta}^{\mathrm{tot}}_{\mathrm{EW}}$ is mainly determined by $\overline{\Delta}_{\mathrm{EW}}(A\to Zh)$ since $\mathrm{BR}_{\mathrm{LO}}(A\to Zh)$ is dominant.
On the other hand, when $c_{\beta-\alpha}\simeq 0$, both $\overline{\Delta}_{\mathrm{EW}}(A\to b\bar{b})$ and $\overline{\Delta}_{\mathrm{EW}}(A\to Zh)$ contribute to $\overline{\Delta}^{\mathrm{tot}}_{\mathrm{EW}}$.
While $\overline{\Delta}_{\mathrm{EW}}(A\to Zh)$ takes quite large value with $c_{\beta-\alpha}\simeq0$, $\mathrm{BR}_{\mathrm{LO}}(A\to Zh)$ is close to zero.
Therefore, we have no singular behavior on $\overline{\Delta}^{\mathrm{tot}}_{\mathrm{EW}}$ even at $\abs{\Delta \kappa_{Z}}\simeq 0$.

\begin{figure}[t]
	\centering
	\includegraphics[scale=0.7]{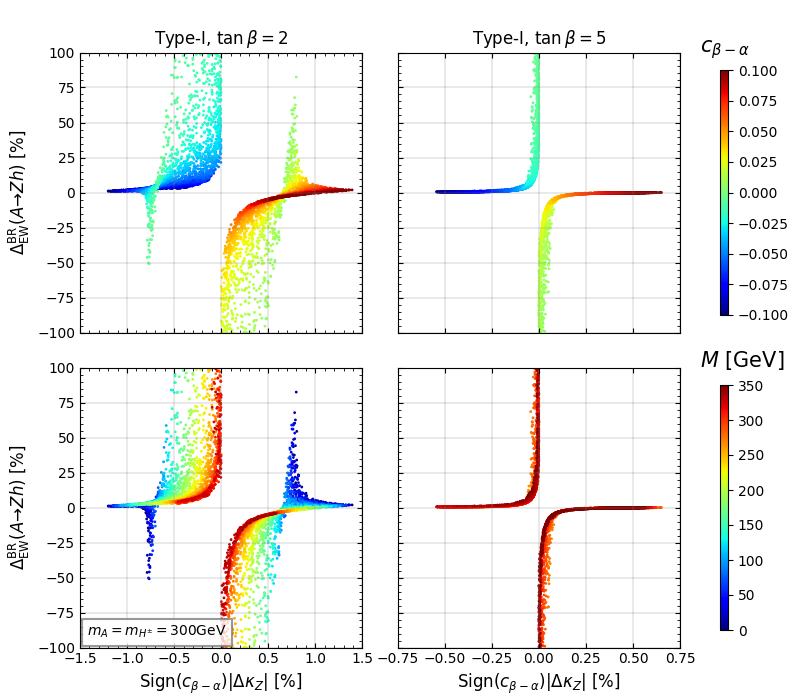}
	\caption{NLO corrections to the decay branching ratios of the CP-odd Higgs boson in Type-I 2HDM in Scenario~A.
	The color differences correspond to the values of $c_{\beta-\alpha}$ and $\sqrt{M^{2}}$, in the top and bottom panels, respectively.}
	\label{dBrA300_NLO_scan}
\end{figure}

In Fig.~\ref{dBrA300_NLO_scan}, we show the size of NLO EW corrections to the decay branching ratios $\Delta_{\mathrm{EW}}^{\mathrm{BR}}(A\to Zh)$.
The behavior of $\Delta_{\mathrm{EW}}^{\mathrm{BR}}(A\to Zh)$ can be understood from $\overline{\Delta}_{\mathrm{EW}}(A\to Zh)$ and $\overline{\Delta}^{\mathrm{tot}}_{\mathrm{EW}}$ as shown in Eq.~\eqref{eq: dBRA decomposed}.
When $c_{\beta-\alpha}\simeq 0.1$, $\mathrm{BR}_{\mathrm{LO}}(A\to Zh)\simeq1$ and $\Delta^{\mathrm{tot}}_{\mathrm{EW}}\simeq \Delta_{\mathrm{EW}}(A\to Zh)$.
Therefore, $\overline{\Delta}_{\mathrm{EW}}(A\to Zh)$ and $\overline{\Delta}^{\mathrm{tot}}_{\mathrm{EW}}$ are canceled out, and $\Delta_{\mathrm{EW}}^{\mathrm{BR}}(A\to XY)$ is close to zero.
When $c_{\beta-\alpha}\simeq 0$, $\mathrm{BR}_{\mathrm{LO}}(A\to Zh)\simeq0$ and $\Delta_{\mathrm{EW}}^{\mathrm{BR}}(A\to Zh)$ can be larger than 100\%.
\subsection{Scenario~B} \label{subsec: scenarioB}
\begin{figure}[t]
	\centering
	\includegraphics[scale=0.7]{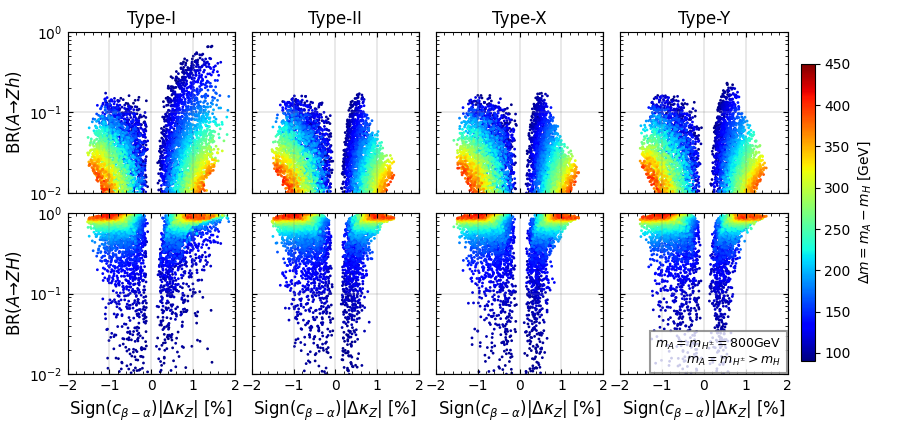}
	\caption{Decay branching ratios of the CP-odd Higgs boson in Scenario~B.
	Predictions on Type-I, Type-II, Type-X and Type-Y are shown from the left to the right panels in order.
	The color differences correspond to the values of $\Delta m = m_{A}-m_{H}$.}
	\label{BrA800_NLO_B}
\end{figure}

In Fig.~\ref{BrA800_NLO_B}, we show the branching ratios of the CP-odd Higgs boson including the  NLO EW corrections in Scenario~B.
The results in Type-I, Type-II, Type-X and Type-Y are shown from the left to the right panels in order.
The color differences correspond to the values of $\Delta m = m_{A}-m_{H}$.
In Scenario~B, $A\to ZH$ is open in addition to $A\to Zh,\, t\bar{t},\, b\bar{b}$ and $\tau\bar{\tau}$, and it can be the dominant decay mode depending on $\Delta m$.
The value of $\tan{\beta}$ is approximately 2 independently of the types of 2HDM for both $c_{\beta-\alpha}>0$ and $c_{\beta-\alpha}<0$.
The exceptional regions, where $\tan{\beta}$ can be large are $c_{\beta-\alpha}\simeq 0$ in all types of 2HDMs and $c_{\beta-\alpha}>0$ in Type-I.
The value of $\sqrt{M^{2}}$ is almost degenerate with $m_{H}$.
As similar to Scenario~A with $\tan{\beta}=2$, we have parameter regions where non-decoupling effects enlarge $\Delta\kappa_{Z}$ even with $c_{\beta-\alpha}\simeq 0$ in all types of 2HDMs.
The maximal size of $\Delta\kappa_{Z}$ with $c_{\beta-\alpha}\simeq 0$ is about $1\%$.
On the other hand, $\Delta\kappa_{Z}$ can be larger than $1\%$ if $c_{\beta-\alpha}\neq 0$, and $\Delta\kappa_{Z}$ reaches about $1.5\%$ independently of the types of 2HDM for both $c_{\beta-\alpha}>0$ and $c_{\beta-\alpha}<0$.

For $A\to Zh$, the branching ratio reaches about $80\%$ ($20\%$) in Type-I with $c_{\beta-\alpha}>0$ ($c_{\beta-\alpha}<0$).
In Type-II, Type-X, and Type-Y, it reaches about $20\%$ independently of the sign of $c_{\beta-\alpha}$.
The branching ratio of $A\to ZH$ reaches about 100\% independently of the types of 2HDMs when $\Delta m\gtrsim 300$ GeV.
Since the tree-level $AZH$ coupling is proportional to $s_{\beta-\alpha}$, we expect that $\mathrm{BR}(A\to ZH)$ can be large with $\Delta\kappa_{Z}\simeq 0$.
However, $\mathrm{BR}(A\to ZH)$ tends to be small when $\Delta\kappa_{Z}\simeq 0$.
This is because of the theoretical bounds such as perturbative unitary and vacuum stability.
If $\Delta m$ is so large that $\mathrm{BR}(A\to ZH)\simeq 1$, $c_{\beta-\alpha}\neq 0$ is favored by the theoretical bounds, and $\Delta\kappa_{Z}$ becomes large due to the both of tree-level mixing and the non-decoupling effects.

\begin{figure}[t]
	\centering
	\includegraphics[scale=0.7]{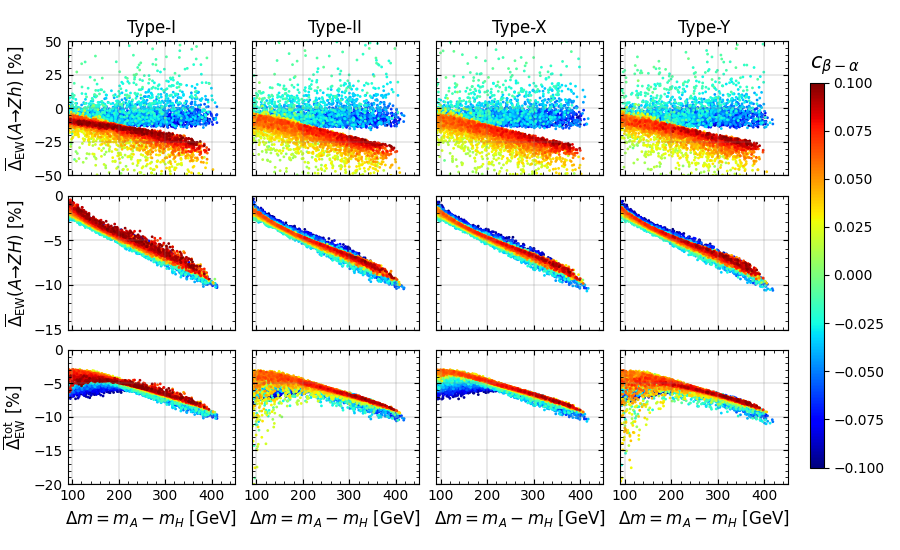}
	\caption{NLO corrections to the decay rates and the total decay width of the CP-odd Higgs boson as a function of $m_{A}-m_{H}$ in Scenario~B.
	Predictions on Type-I, Type-II, Type-X and Type-Y are shown from the left to the right panels in order.
	The color differences correspond to the values of $c_{\beta-\alpha}$.}
	\label{dGamA800_NLO_B}
\end{figure}

\begin{figure}[t]
	\centering
	\includegraphics[scale=0.7]{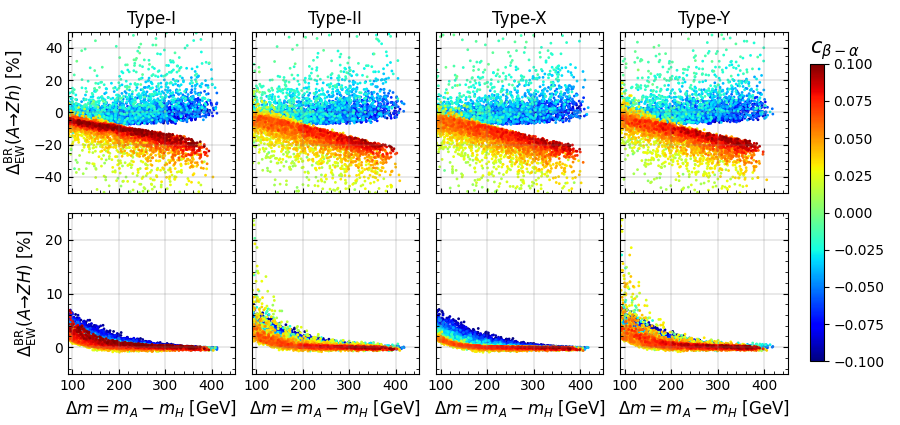}
	\caption{NLO corrections to the decay branching ratios for the CP-odd Higgs boson as a function of $m_{A}-m_{H}$ in Scenario~B.
	Predictions on Type-I, Type-II, Type-X and Type-Y are shown from the left to the right panels in order.
	The color differences correspond to the values of $c_{\beta-\alpha}$.}
	\label{dBrA800_NLO_B}
\end{figure}

In Fig.~\ref{dGamA800_NLO_B}, we show the size of $\overline{\Delta}_{\mathrm{EW}}(A\to XY)$ and $\overline{\Delta}^{\mathrm{tot}}_{\mathrm{EW}}$ as a function of $\Delta m$.
The color differences correspond to the values of $c_{\beta-\alpha}$.
We find that both $\overline{\Delta}_{\mathrm{EW}}(A\to Zh)$ and $\overline{\Delta}_{\mathrm{EW}}(A\to ZH)$ are the almost same among all types of 2HDMs, while $\overline{\Delta}^{\mathrm{tot}}_{\mathrm{EW}}$ shows type-dependent behavior, especially for $\Delta m\lesssim 200$~GeV. 
When $c_{\beta-\alpha}<0$, the magnitude of $\overline{\Delta}_{\mathrm{EW}}(A\to Zh)$ is almost independent of $\Delta m$.
On the other hand, it monotonically increases as $\Delta m$ becomes large when $c_{\beta-\alpha}>0$.
We note that $\overline{\Delta}_{\mathrm{EW}}(A\to Zh)$ can be larger than $100\%$ if $c_{\beta-\alpha}$ is quite small, while we take the range of $\overline{\Delta}_{\mathrm{EW}}(A\to Zh)$ as $[-50,\, 50]\%$ for the illustration purpose.
The magnitude of $\overline{\Delta}_{\mathrm{EW}}(A\to ZH)$ increases as $\Delta m$ becomes large due to the non-decoupling effects of scalar loops.
When $\Delta m \gtrsim 200~\mathrm{GeV}$, $\overline{\Delta}^{\mathrm{tot}}_{\mathrm{EW}}$ is mainly determined by $A\to ZH$, and its magnitude increases as $\Delta m$ becomes large in all types of 2HDMs.
In Type-II and Type-Y, $A\to b\bar{b}$ contributes to $\overline{\Delta}^{\mathrm{tot}}_{\mathrm{EW}}$ when $\Delta m \lesssim 200~\mathrm{GeV}$ with large $\tan{\beta}$, and $\overline{\Delta}^{\mathrm{tot}}_{\mathrm{EW}}$ reaches about $-20\%$.


In Fig.~\ref{dBrA800_NLO_B}, we show the size of $\Delta_{\mathrm{EW}}^{\mathrm{BR}}(A\to XY)$ as a function of $\Delta m$.
The color differences correspond to the values of $c_{\beta-\alpha}$.
$\Delta_{\mathrm{EW}}^{\mathrm{BR}}(A\to Zh)$ is negative when $c_{\beta-\alpha}>0$, while it can be positive when $c_{\beta-\alpha}<0$ due to the contribution of $\overline{\Delta}^{\mathrm{tot}}_{\mathrm{EW}}$.
$\Delta_{\mathrm{EW}}^{\mathrm{BR}}(A\to ZH)$ is close to zero in large $\Delta m$ region due to the cancellation between $\overline{\Delta}_{\mathrm{EW}}(A\to ZH)$ and $\Delta_{\mathrm{EW}}^{\mathrm{tot}}$.
When $\Delta m\lesssim 150~\mathrm{GeV}$ and $c_{\beta-\alpha}\simeq 0$, $\Delta_{\mathrm{EW}}^{\mathrm{BR}}(A\to ZH)$ becomes large in Type-II and Type-Y, where $A\to b\bar{b}$ contributes to $\overline{\Delta}^{\mathrm{tot}}_{\mathrm{EW}}$.

Finally, we would like to comment on the difference between our results and those in Ref.~\cite{Krause:2019qwe}.
We have compared the maximum value of $\Delta_{\mathrm{EW}}^{\mathrm{BR}}(A\to XY)$ with those in Table~13 in Ref.~\cite{Krause:2019qwe}, and it was found that our results for $\Delta_{\mathrm{EW}}^{\mathrm{BR}}(A\to t\bar{t})$ and $\Delta_{\mathrm{EW}}^{\mathrm{BR}}(A\to b\bar{b})$ take larger values.
For example, $\Delta_{\mathrm{EW}}^{\mathrm{BR}}(A\to b\bar{b})$ reaches $-40\%$ in Scenario~B, while the maximum value is about $20\%$ in Ref.~\cite{Krause:2019qwe} (See the discussion in Appendix~\ref{app: numerical_result}).
This difference might be understood by the treatment of the running quark masses because the sizes of the EW corrections are enhanced by the ratio of the pole and running quark masses as shown in Eq.~\eqref{eq: dEWbar_AQQ}.
Although the size of the EW corrections for the total decay width also depends on the treatment of the quark masses, we have found that the results for $A\to \tau\bar{\tau}$, $A\to Zh$ and $A\to ZH$ are consistent with Ref.~\cite{Krause:2019qwe}.

\subsection{Discrimination of Types of the Yukawa interaction in the 2HDMs}
In this subsection, we discuss the discrimination of the types of 2HDMs by the decays of the CP-odd Higgs boson.
Since $\Gamma_{Af\bar{f}}^{P, \mathrm{tree}}$ is proportional to $\zeta_{f}$, we can discriminate the types of 2HDMs by the correlations among $\BR(A\to f\bar{f})$.
We study the correlation between $\BR(A\to \tau\bar{\tau})$ and $\BR(A\to b\bar{b})$ including the NLO EW and higher-order QCD corrections in Type-I, Type-II, Type-X and Type-Y in Scenario~B.

\begin{figure}[t]
	\begin{minipage}{0.45\linewidth}
		\centering
		\includegraphics[scale=0.5]{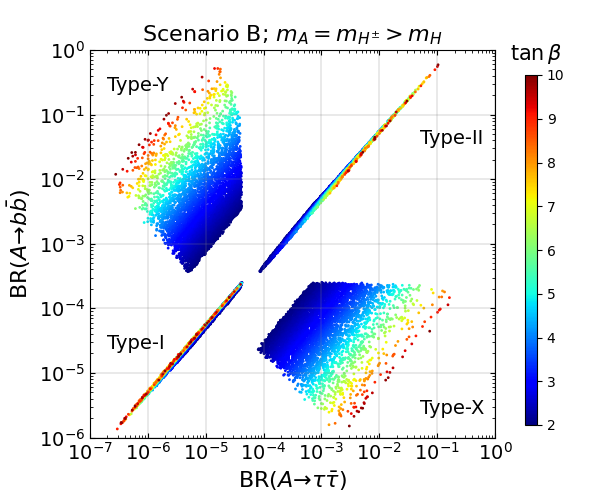}
	\end{minipage}
	\begin{minipage}{0.45\linewidth}
		\centering
		\includegraphics[scale=0.5]{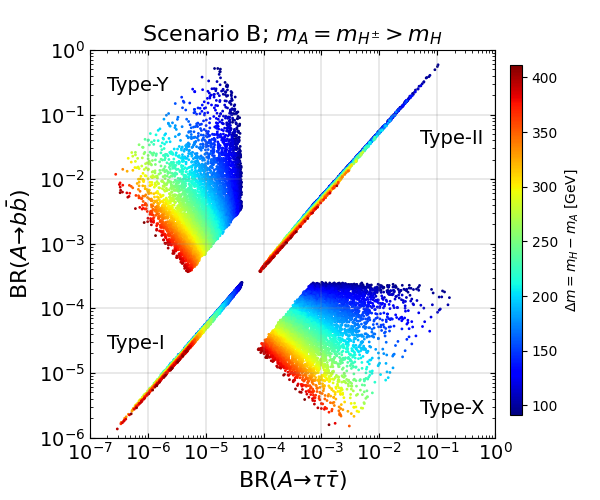}
	\end{minipage}	
	\caption{Correlation between $\BR(A\to \tau\bar{\tau})$ and $\BR(A\to b\bar{b})$ in Type-I, Type-II, Type-X and Type-Y for Scenario~B.
	The color differences correspond to the values of $\tb$ and $\Delta m$ in the left and the right panels, respectively.}
	\label{corr_BrA800_B}
\end{figure}

In Fig.~\ref{corr_BrA800_B}, we show the correlation between $\BR(A\to \tau\bar{\tau})$ and $\BR(A\to b\bar{b})$ in Scenario~B.
The color differences correspond to the values of $\tan{\beta}$ and $\Delta m$ in the left and right panels, respectively.
In Scenario~B, there is $A\to ZH$, and it can be dominant depending on $\Delta m$.
From the left panel of Fig.~\ref{corr_BrA800_B}, we can see that both $\BR(A\to \tau\bar{\tau})$ and $\BR(A\to b\bar{b})$ become small in Type-I, while both of them can take several dozens of percent due to the $\tan{\beta}$ enhancement in Type-II.
On the other hand, $\BR(A\to \tau\bar{\tau})$ can be large in Type-X, while $\BR(A\to b\bar{b})$ becomes large in Type-Y.
From the right panel of Fig.~\ref{corr_BrA800_B}, we can see that both of $\BR(A\to \tau\bar{\tau})$ and $\BR(A\to b\bar{b})$ can be several dozens of percent if $\Delta m\lesssim 150$ GeV, while they are below 1\% when $\Delta m\simeq 400$ GeV.
Therefore, we can discriminate the types of 2HDMs by examining the decay pattern of the CP-odd Higgs boson if $\Delta m$ is not so large.
We note that $A\to b\bar{b}$ and $A\to \tau\bar{\tau}$ can be dominant when $c_{\beta-\alpha}\approx0$.
Therefore, the types of Yukawa interactions can be discriminated by examining the decay pattern of the CP-odd Higgs boson even in the case with $c_{\beta-\alpha}\simeq0$, where it would be difficult to measure the deviation in the SM-like Higgs boson couplings. 
When $\Delta m$ is large, we have sizable deviations in the SM-like Higgs boson couplings as discussed in Sec.~\ref{subsec: scenarioB}, and the measurement of SM-like Higgs boson couplings can be used to study the types of 2HDMs~\cite{Kanemura:2014bqa, Kanemura:2019kjg}.
Therefore, the decay pattern of the CP-odd Higgs boson and the deviation in the SM-like Higgs boson couplings play a comprehensive role, and we can determine the types of 2HDMs by using them.
%

%% file: A01inputs.tex
\section{Input parameters} \label{sec: SM_inputs}
\begin{table}[t]
\centering
\begin{tabular}{c|c|c} \hline
Input parameter & Symbol & Value \\ \hline
fine-structure constant at the Thomson limit & $\alpha^{-1}_{\mathrm{em}}(0)$ & 137.035999084 \\
$Z$ boson mass & $m_{Z}\ \qty[\text{GeV}]$ & 91.1876 \\
Fermi constant & $G_{F}\ \qty[\text{GeV}^{-2}]$ & $1.1663787\times 10^{-5}$ \\
strong coupling constant at $m_{Z}$ & $\alpha_{s}(m_{Z})$ & 0.1179 \\
Higgs boson mass & $m_{h}\ \qty[\text{GeV}]$ & 125.1 \\
top-quark pole mass & $m_{t}\ \qty[\text{GeV}]$ & 172.5 \\
bottom-quark pole mass & $m_{b}\ \qty[\text{GeV}]$ & 4.78 \\
charm-quark pole mass &  $m_{c}\ \qty[\text{GeV}]$ & 1.67 \\
bottom-quark running mass & $\bar{m}_{b}(\bar{m}_{b})\ \qty[\text{GeV}]$ & 4.18 \\
charm-quark running mass &  $\bar{m}_{c}(\bar{m}_{c})\ \qty[\text{GeV}]$ & 1.27 \\
tauon mass & $m_{\tau}\ \qty[\text{GeV}]$ & 1.77686 \\
muon mass & $m_{\mu}\ \qty[\text{GeV}]$ & 0.1056583745 \\
electron mass & $m_{e}\ \qty[\text{GeV}]$ & $5.109989461\times 10^{-4}$ \\
Total decay width of the top quark & $\Gamma_{t}\ \qty[\text{GeV}]$ & $1.42$ \\
Total decay width of the $W^{\pm}$ bosons & $\Gamma_{W}\ \qty[\text{GeV}]$ & $2.085$ \\
Total decay width of the $Z$ boson & $\Gamma_{Z}\ \qty[\text{GeV}]$ & $2.4952$ \\ \hline
\end{tabular}
\caption{SM input parameters. The values are taken from Ref.~\cite{Zyla:2020zbs}.}
\label{Table: inputs}
\end{table}

We work in the scheme where $\alpha_{\mathrm{em}}(0), G_{F}$ and $m_{Z}$ are input parameters following the \texttt{H-COUP} program~\cite{Kanemura:2017gbi,Kanemura:2019slf}.
In Table~\ref{Table: inputs}, we list the SM input parameters.
The values of input parameters are taken from Ref.~\cite{Zyla:2020zbs}.
The expressions for the running parameters such as $\alpha_{s}(\mu)$ and $\overline{m}_{q}(\mu)$ are given in Ref.~\cite{Aiko:2020ksl}.

%% file: A02scalar_couplings.tex
\section{Scalar couplings} \label{Apendix: scalar_couplings}
We list the expression of the scalar trilinear couplings in terms of the coefficients of the Higgs potential on the Higgs basis given in Eq.~\eqref{eq: Higgs_potential_Higgs_basis}.
We use the following notation for these couplings,
\begin{align}
\mathcal{L} = +\lambda_{\phi_{i}\phi_{j}\phi_{k}}\phi_{i}\phi_{j}\phi_{k}+\cdots.
\end{align}
In terms of the masses of Higgs bosons and mixing angles, the coefficients $Z_{1},..., Z_{7}$ are given as~\cite{Davidson:2005cw, Bernon:2015qea}
\begin{align}
Z_{1}v^{2} &= m_{H}^{2}\cos^{2}{(\beta-\alpha)} + m_{h}^{2}\sin^{2}{(\beta-\alpha)}, \\
Z_{2}v^{2} &= m_{H}^{2}\cos^{2}{(\beta-\alpha)} + m_{h}^{2}\sin^{2}{(\beta-\alpha)} + 4(m_{h}^{2}-m_{H}^{2})\cos{(\beta-\alpha)}\sin{(\beta-\alpha)}\cot{2\beta} \notag \\
&\quad
+ 4\left[(m_{H}^{2}-M^{2})\sin^{2}{(\beta-\alpha)} + (m_{h}^{2}-M^{2})\cos^{2}{(\beta-\alpha)}\right]\cot^{2}{2\beta}, \\
Z_{3}v^{2} &= m_{H}^{2}\cos^{2}{(\beta-\alpha)} + m_{h}^{2}\sin^{2}{(\beta-\alpha)}+ 2(m_{h}^{2}-m_{H}^{2})\cos{(\beta-\alpha)}\sin{(\beta-\alpha)}\cot{2\beta} \notag \\
&\quad
+ 2(m_{H^{\pm}}^{2} - M^{2}), \\
Z_{4}v^{2} &= m_{H}^{2}\sin^{2}{(\beta-\alpha)} + m_{h}^{2}\cos^{2}{(\beta-\alpha)} + m_{A}^{2} - 2m_{H^{\pm}}^{2}, \\
Z_{5}v^{2} &= m_{H}^{2}\sin^{2}{(\beta-\alpha)} + m_{h}^{2}\cos^{2}{(\beta-\alpha)} - m_{A}^{2}, \\
Z_{6}v^{2} &= (m_{h}^{2} - m_{H}^{2})\cos{(\beta-\alpha)}\sin{(\beta-\alpha)}, \\
Z_{7}v^{2} &= 2\left[(m_{H}^{2}-M^{2})\sin^{2}{(\beta-\alpha)} + (m_{h}^{2}-M^{2})\cos^{2}{(\beta-\alpha)}\right]\cot{2\beta} \notag \\
&\quad
+(m_{h}^{2}-m_{H}^{2})\cos{(\beta-\alpha)}\sin{(\beta-\alpha)}.
\end{align}

The Higgs trilinear couplings relevant for the decays of the CP-odd Higgs boson are given by
\begin{align}
\lambda_{AG^{0}h} &= -\qty[Z_{6}s_{\beta-\alpha}+Z_{5}c_{\beta-\alpha}]v, \\
\lambda_{AG^{0}H} &= -\qty[Z_{6}c_{\beta-\alpha}-Z_{5}s_{\beta-\alpha}]v, \\
\lambda_{hG^{0}G^{0}} &= -\frac{1}{2}(Z_{1}s_{\beta-\alpha}+Z_{6}c_{\beta-\alpha})v, \\
\lambda_{HG^{0}G^{0}} &= -\frac{1}{2}(Z_{1}c_{\beta-\alpha}-Z_{6}s_{\beta-\alpha})v, \\
\lambda_{AAh} &= -\frac{1}{2}\qty[Z_{7}c_{\beta-\alpha}+(Z_{3}+Z_{4}-Z_{5})s_{\beta-\alpha}]v, \\
\lambda_{AAH} &= -\frac{1}{2}\qty[-Z_{7}s_{\beta-\alpha}+(Z_{3}+Z_{4}-Z_{5})c_{\beta-\alpha}]v, \\
\lambda_{hhh} &= -\frac{1}{2}\qty[Z_{1}s_{\beta-\alpha}^{3}+Z_{345}c_{\beta-\alpha}^{2}s_{\beta-\alpha}+3Z_{6}s_{\beta-\alpha}^{2}c_{\beta-\alpha}+Z_{7}c_{\beta-\alpha}^{3}]v, \\
\lambda_{hhH} &= -\frac{1}{2}[3Z_{1}s_{\beta-\alpha}^{2}c_{\beta-\alpha}+Z_{345}(c_{\beta-\alpha}^{3}-2s_{\beta-\alpha}^{2}c_{\beta-\alpha}) \notag \\
&\qquad
-3Z_{6}(s_{\beta-\alpha}^{3}-2s_{\beta-\alpha}c_{\beta-\alpha}^{2})-3Z_{7}c_{\beta-\alpha}^{2}s_{\beta-\alpha}]v, \\
\lambda_{hHH} &= -\frac{1}{2}[3Z_{1}c_{\beta-\alpha}^{2}s_{\beta-\alpha}+Z_{345}(s_{\beta-\alpha}^{3}-2c_{\beta-\alpha}^{2}s_{\beta-\alpha}) \notag \\
&\qquad
+3Z_{6}(c_{\beta-\alpha}^{3}-2c_{\beta-\alpha}s_{\beta-\alpha}^{2})+3Z_{7}c_{\beta-\alpha}s_{\beta-\alpha}^{2}]v, \\
\lambda_{HHH} &= -\frac{1}{2}\qty[Z_{1}c_{\beta-\alpha}^{3}+Z_{345}s_{\beta-\alpha}^{2}c_{\beta-\alpha}-3Z_{6}c_{\beta-\alpha}^{2}s_{\beta-\alpha}-Z_{7}s_{\beta-\alpha}^{3}]v, \\
\lambda_{H^{\pm}G^{\mp}A} &= \pm\frac{i}{2}(Z_{4}-Z_{5})v, \\
\lambda_{H^{\pm}G^{\mp}h} &= -\qty[Z_{6}s_{\beta-\alpha}+\frac{1}{2}(Z_{4}+Z_{5})c_{\beta-\alpha}]v, \\
\lambda_{H^{\pm}G^{\mp}H} &= -\qty[Z_{6}c_{\beta-\alpha}-\frac{1}{2}(Z_{4}+Z_{5})s_{\beta-\alpha}]v,
\end{align}
where $Z_{345}=Z_{3}+Z_{4}+Z_{5}$.

%% file: A03three_body.tex
\section{Three-body decays of the CP-odd Higgs boson} \label{sec: three_body}

\subsection{$A\to tt^{*} \to tbW$}
When $m_{t}+m_{W} \leq m_{A} < 2m_{t}$, the CP-odd Higgs boson can decay into a pair of on-shell and off-shell top quarks, $A\to tt^{*} \to tbW$.
The decay rate is given by~\cite{Djouadi:1995gv, Barradas:1996xb}
\begin{align}
\Gamma(A\to tt^{*} \to tbW)
=
2\int_{y_{t}^{-}}^{y_{t}^{+}}\dd y_{t}
\int_{y_{b}^{-}}^{y_{b}^{+}}\dd y_{b}
\frac{\dd \Gamma}{\dd y_{t}\dd y_{b}}(A\to t\bar{b}W^{-}),
\end{align}
with the scaling variables $y_{t\, (b)} = 1-2E_{t\, (b)}/m_{A}$.
Since the charge conjugate $\bar{t}bW^{+}$ final state doubles the partial decay width, we multiply factor two.
The kinematic boundaries are given by
\begin{align}
y_{t}^{+} &= (1-x_{t})^{2}-x_{t}^{2}, \\
y_{t}^{-} &= (x_{b}+x_{W})^{2}-x_{t}^{2}, \\
y_{b}^{\pm} &= x_{t}^{2}-x_{b}^{2}+x_{W}^{2}+\frac{(1-2x_{t}^{2}-y_{t})(y_{t}+x_{t}^{2}-x_{b}^{2}+x_{W}^{2})}{2(y_{t}+x_{t}^{2})} \notag \\
&\quad
\pm \frac{1}{2}\lambda^{1/2}(y_{t}+x_{t}^{2}, x_{t}^{2})\lambda^{1/2}\qty(\frac{x_{b}^{2}}{y_{t}+x_{t}^{2}}, \frac{x_{W}^{2}}{y_{t}+x_{t}^{2}}),
\end{align}
where $x_{i} = m_{i}/m_{A}\, (i=t, b, W)$, and the kinematical factor $\lambda(x,y)$ is given in Eq.~\eqref{eq: lam_kin}.
The Dalitz density is given by
\begin{align}
\frac{\dd \Gamma}{\dd y_{t}\dd y_{b}}(A\to t\bar{b}W^{-})
=
3\frac{G_{F}^{2}m_{t}^{2}\zeta_{u}^{2}m_{A}^{3}}{64\pi^{3}}
\frac{\Gamma_{0}}{y_{t}^{2}+\kappa_{t}\gamma_{t}},
\end{align}
with 
\begin{align}
\Gamma_{0} = -y_{t}^{3}+(1-\kappa_{t}+\kappa_{W}-y_{b})y_{t}^{2}+[2\kappa_{t}-y_{b}(\kappa_{t}-2\kappa_{W})]y_{t}+(\kappa_{t}-\kappa_{W})(\kappa_{t}+2\kappa_{W}),
\end{align}
and $\kappa_{i} = m_{i}^{2}/m_{A}^{2}\, (i=t, b, W)$.
The reduced decay width for the virtual top quark is defined by $\gamma_{t}=\Gamma_{t}^{2}/m_{A}^{2}$ with the total decay width of the top quark $\Gamma_{t}$.
In the numerical evaluation, we use the on-shell mass of the top quark in the Yukawa interaction.

\subsection{$A\to \phi Z^{*} \to \phi f\bar{f}$}
When $m_{\phi} \leq m_{A} < m_{Z}+m_{\phi}$, the CP-odd Higgs boson can decay into a pair of on-shell CP-even Higgs boson $\phi\, (=h, H)$ and off-shell $Z$ boson, $A\to \phi Z^{*} \to \phi f\bar{f}$.
The decay rate is given by~\cite{Djouadi:1995gv}
\begin{align}
\Gamma(A\to \phi Z^{*})
=
\int_{x_{f}^{-}}^{x_{f}^{+}}\dd x_{f}
\int_{x_{\bar{f}}^{-}}^{x_{\bar{f}}^{+}}\dd x_{\bar{f}}
\frac{\dd \Gamma}{\dd x_{f}\dd x_{\bar{f}}}(A\to \phi Z^{*} \to \phi f\bar{f}),
\end{align}
with the scaling variables $x_{f\, (\bar{f})} = 2E_{f\, (\bar{f})}/m_{A}$.
We neglect the masses of the final-state fermions.
The kinematic boundaries are given by
\begin{align}
0 &< x_{f} < 1-\kappa_{\phi}, \\
1-x_{f}-\kappa_{\phi} &< x_{\bar{f}} < 1-\frac{\kappa_{\phi}}{1-x_{f}},
\end{align}
with $\kappa_{\phi} = m_{\phi}^{2}/m_{A}^{2}$.
The Dalitz density is given by
\begin{align}
\frac{\dd \Gamma}{\dd x_{f}\dd x_{\bar{f}}}(A\to \phi Z^{*} \to \phi f\bar{f})
&=
\frac{9G_{F}}{16\sqrt{2}\pi^{3}}\abs{\Gamma_{AZ\phi}^{\mathrm{tree}}}^{2}m_{Z}^{2}m_{A}
\qty(\frac{7}{12}-\frac{10}{9}s_{W}^{2}+\frac{40}{27}s_{W}^{4}) \notag \\
&\quad\times
\frac{(1-x_{f})(1-x_{\bar{f}})-\kappa_{\phi}}{(1-x_{f}-x_{\bar{f}}-\kappa_{\phi}+\kappa_{Z})^{2}+\kappa_{Z}\gamma_{Z}},
\end{align}
where $\kappa_{Z} = m_{Z}^{2}/m_{A}^{2}$, and the tree-level couplings $\Gamma_{AZ\phi}^{\mathrm{tree}}$ are given in Eq.~\eqref{eq: Gam_AVphi_LO}.
The reduced decay width for the virtual $Z$ boson is defined by $\gamma_{Z}=\Gamma_{Z}^{2}/m_{A}^{2}$ with the total decay width of the $Z$ boson $\Gamma_{Z}$.

\subsection{$A\to H^{\pm}W^{\mp *} \to H^{\pm}f\bar{f}'$}
When $m_{H^{\pm}} \leq m_{A} < m_{W}+m_{H^{\pm}}$, the CP-odd Higgs boson can decay into a pair of on-shell charged Higgs bosons $H^{\pm}$ and off-shell $W^{\pm}$ boson, $A\to H^{\pm}W^{\mp *} \to H^{\pm}f\bar{f}'$.
The decay rate is given by
\begin{align}
\Gamma(A\to H^{\pm} W^{\mp*})
=
2\int_{x_{f}^{-}}^{x_{f}^{+}}\dd x_{f}
\int_{x_{\bar{f}'}^{-}}^{x_{\bar{f}'}^{+}}\dd x_{\bar{f}'}
\frac{\dd \Gamma}{\dd x_{f}\dd x_{\bar{f}'}}(A\to H^{+} W^{-*}\to H^{+}f\bar{f}'),
\end{align}
with the scaling variables $x_{f\, (\bar{f}')} = 2E_{f\, (\bar{f}')}/m_{A}$.
Since the charge conjugate $H^{-}\bar{f}f'$ final state doubles the partial decay width, we multiply factor two.
We neglect the masses of the final state fermions.
The kinematic boundaries are given by
\begin{align}
0 &< x_{f} < 1-\kappa_{H^{\pm}}, \\
1-x_{f}-\kappa_{H^{\pm}} &< x_{\bar{f}'} < 1-\frac{\kappa_{H^{\pm}}}{1-x_{f}},
\end{align}
with $\kappa_{H^{\pm}} = m_{H^{\pm}}^{2}/m_{A}^{2}$.
The Dalitz density is given by
\begin{align}
\frac{\dd \Gamma}{\dd x_{f}\dd x_{\bar{f}'}}(A\to H^{+} W^{-*}\to H^{+}f\bar{f}')
&=
\frac{9G_{F}}{16\sqrt{2}\pi^{3}}\abs{\Gamma_{AW^{\mp}H^{\pm}}^{\mathrm{tree}}}^{2}m_{W}^{2}m_{A} \notag \\
&\quad\times
\frac{(1-x_{f})(1-x_{\bar{f}'})-\kappa_{H^{\pm}}}{(1-x_{f}-x_{\bar{f}'}-\kappa_{H^{\pm}}+\kappa_{W})^{2}+\kappa_{W}\gamma_{W}},
\end{align}
where $\kappa_{W} = m_{W}^{2}/m_{A}^{2}$, and the tree-level coupling $\Gamma_{AW^{\mp}H^{\pm}}^{\mathrm{tree}}$ is given in Eq.~\eqref{eq: Gam_AVphi_LO}.
The reduced decay width for the virtual $W$ boson is defined by $\gamma_{W}=\Gamma_{W}^{2}/m_{A}^{2}$ with the decay width of the $W^{\pm}$ boson $\Gamma_{W}$.

%% file: A04loop_induced.tex
\section{QCD corrections for $A\to \gamma\gamma$ and $A\to gg$} \label{sec: loop_induced}

\subsection{$A\to\gamma\gamma$}
The analytic expression for $I_{F}^{A}C_{1}^{A}(\tau)$ in Eq.~\eqref{eq: Gam_Agamgam} is given by~\cite{Harlander:2005rq}
\begin{align}
I_{F}^{A}(\tau)C_{1}^{A}(\tau) &= -\frac{\theta(1+\theta^{2})}{(1-\theta)^{3}(1+\theta)}\bigg[
72\mathrm{Li}_{4}(\theta)+96\mathrm{Li}_{4}(-\theta)-\frac{128}{3}\qty[\mathrm{Li}_{3}(\theta)-\mathrm{Li}_{3}(-\theta)]\ln{\theta} \notag \\
&\quad
+\frac{28}{3}\mathrm{Li}_{2}(\theta)\ln^{2}{\theta}+\frac{16}{3}\mathrm{Li}_{2}(-\theta)\ln^{2}{\theta}+\frac{1}{18}\ln^{4}{\theta}
+\frac{8}{3}\zeta(2)\ln^{2}{\theta}+\frac{32}{3}\zeta(3)\ln{\theta}+12\zeta(4)\bigg] \notag \\
&\quad
+\frac{\theta}{(1-\theta)^{2}}\bigg[-\frac{56}{3}\mathrm{Li}_{3}(\theta)-\frac{64}{3}\mathrm{Li}_{3}(-\theta)+16\mathrm{Li}_{2}(\theta)\ln{\theta}+\frac{32}{3}\mathrm{Li}_{2}(-\theta)\ln{\theta} \notag \\
&\quad
+\frac{20}{3}\ln{(1-\theta)}\ln^{2}{\theta}-\frac{8}{3}\zeta(2)\ln{\theta}+\frac{8}{3}\zeta(3)\bigg] \notag \\
&\quad
+\frac{2\theta(1+\theta)}{3(1-\theta)^{3}}\ln^{3}{\theta},
\end{align}
with
\begin{align}
\theta \equiv \theta(\tau) = \frac{\sqrt{1-\tau^{-1}}-1}{\sqrt{1-\tau^{-1}}+1}.
\label{eq: theta}
\end{align}
For the evaluation of the polylog function $\mathrm{Li}_{n}(x)$, we use \texttt{CHAPLIN}~\cite{Buehler:2011ev}.
The analytic expression for $I_{F}^{A}C_{2}^{A}(\tau)$ in Eq.~\eqref{eq: Gam_Agamgam} is given by~\cite{Harlander:2005rq}
\begin{align}
I_{F}^{A}(\tau)C_{2}^{A}(\tau) &= \frac{2}{\tau}\qty[f(\tau)-\tau f'(\tau)],
\end{align}
where $f(\tau)$ is given in Eq.~\eqref{eq: f_tau}.

\subsection{$A\to gg$}
The NLO QCD correction $E_{A}^{(1)}$ in \eqref{eq: Gam_Agg} can be decomposed as~\cite{Spira:1995rr}
\begin{align}
E_{A}^{(1)} = \eval{E_{A}^{\mathrm{virt}}}_{m_{t}\to \infty}
+\eval{E_{A}^{\mathrm{real}}}_{m_{t}\to \infty}
+\Delta E_{A}.
\end{align}
The first and second terms on the right-hand side denote the contributions from virtual gluon loops and those from real gluon emissions in the large top-mass limit, respectively.
At $\mu=m_{A}$, these are given by~\cite{Spira:1995rr}
\begin{align}
\eval{E_{A}^{\mathrm{virt}}}_{m_{t}\to \infty} = 6\qc
\eval{E_{A}^{\mathrm{real}}}_{m_{t}\to \infty} = \frac{73}{4}-\frac{7}{6}N_{f}.
\end{align}
The third term, $\Delta E_{A}$, vanishes in the large top-mass limit.
It can be decomposed as
\begin{align}
\Delta E_{A} = \Delta E_{A}^{\mathrm{virt}}+\Delta E_{A}^{ggg}+N_{f}\Delta E_{A}^{gq\bar{q}}.
\end{align}
The analytic expression for $\Delta E_{A}^{\mathrm{virt}}$ is given by~\cite{Harlander:2005rq}
\begin{align}
\Delta E_{A}^{\mathrm{virt}} = \Re{\frac{\sum_{q}2I_{f}\zeta_{f}I_{F}^{A}(\tau_{q})\qty(B_{1}^{A}(\tau_{q})+B_{2}^{A}(\tau_{q})\ln{\frac{m_{A}^{2}}{m_{q}^{2}}})}{\sum_{q}2I_{f}\zeta_{f}I_{F}^{A}(\tau_{q})}}-6,
\end{align}
with
\begin{align}
I_{F}^{A}(\tau)B_{1}^{A}(\tau) &=
\frac{\theta}{(1-\theta)^{2}}\bigg[48H(1,0,-1,-;\theta)+4\ln{(1-\theta)}\ln^{3}{\theta}-24\zeta(2)\mathrm{Li}_{2}(\theta)-24\zeta(2)\ln{(1-\theta)}\ln{\theta} \notag \\
&\quad
-72\zeta(3)\ln{(1-\theta)}-\frac{220}{3}\mathrm{Li}_{3}(\theta)-\frac{128}{3}\mathrm{Li}_{3}(-\theta)+68\mathrm{Li}_{2}(\theta)\ln{\theta}+\frac{64}{3}\mathrm{Li}_{2}(-\theta)\ln{\theta} \notag \\
&\quad
+\frac{94}{3}\ln{(1-\theta)}\ln^{2}{\theta^{2}}-\frac{16}{3}\zeta(2)\ln{\theta}+\frac{124}{3}\zeta(3)+3\ln^{2}{\theta}\bigg] \notag \\
&\quad
-\frac{24\theta(5+7\theta^{2})}{(1-\theta)^{3}(1+\theta)}\mathrm{Li}_{4}(\theta)
-\frac{24\theta(5+11\theta^{2})}{(1-\theta)^{3}(1+\theta)}\mathrm{Li}_{4}(-\theta) \notag \\
&\quad
+\frac{8\theta(23+41\theta^{2})}{3(1-\theta)^{3}(1+\theta)}\Big[\mathrm{Li}_{3}(\theta)+\mathrm{Li}_{3}(-\theta)\Big]\ln{\theta} \notag \\
&\quad
-\frac{4\theta(5+23\theta^{2})}{3(1-\theta)^{3}(1+\theta)}\mathrm{Li}_{2}(\theta)\ln^{2}{\theta}
-\frac{32\theta(1+\theta^{2})}{3(1-\theta)^{3}(1+\theta)}\mathrm{Li}_{2}(-\theta)\ln^{2}{\theta}  \notag \\
&\quad
+\frac{\theta(5-13\theta^{2})}{36(1-\theta)^{3}(1+\theta)}\ln^{4}{\theta}
+\frac{2\theta(1-17\theta^{2})}{3(1-\theta)^{3}(1+\theta)}\zeta(2)\ln^{2}{\theta}
+\frac{4\theta(11-43\theta^{2})}{3(1-\theta)^{3}(1+\theta)}\zeta(3)\ln{\theta} \notag \\
&\quad
+\frac{24\theta(1-3\theta^{2})}{(1-\theta)^{3}(1+\theta)}\zeta(4)
+\frac{2\theta(2+11\theta)}{3(1-\theta)^{3}}\ln^{3}{\theta}, \\
I_{F}^{A}(\tau)B_{2}^{A}(\tau) &= \frac{4}{\tau}\qty[f(\tau)-\tau f'(\tau)],
\end{align}
where $\theta$ is given in Eq.~\eqref{eq: theta}, and $f(\tau)$ is given in Eq.~\eqref{eq: f_tau}.
For the evaluation of the Harmonic Polylogarithm function $H(1,0,-1,-;\theta)$, we use \texttt{CHAPLIN}~\cite{Buehler:2011ev}.
According to Ref.~\cite{Spira:1995rr}, $\Delta E_{A}$ is dominantly determined by $\Delta E_{A}^{\mathrm{virt}}$.
Therefore, we neglect the contributions from the real emissions $\Delta E_{A}^{ggg}$ and $\Delta E_{A}^{gq\bar{q}}$ in our calculation.

The NNLO QCD correction $E_{A}^{(2)}$ in \eqref{eq: Gam_Agg} is evaluated in the large top-mass limit.
It is given by~\cite{Chetyrkin:1998mw}
\begin{align}
E_{A}^{(2)} &= \frac{51959}{96}-\frac{363}{8}\zeta(2)-\frac{495}{8}\zeta(3)
+N_{f}\qty[-\frac{473}{8}+\frac{11}{2}\zeta(2)+\frac{5}{4}\zeta(3)-\ln{\frac{\overline{m}_{t}^{2}(m_{A})}{m_{A}^{2}}}] \notag \\
&\quad
+N_{f}^{2}\qty(\frac{251}{216}-\frac{1}{6}\zeta(2)).
\end{align}

%% file: A05vertex_function.tex
\section{1PI diagram contributions for $Af\bar{f}$ and $AV\phi$ vertices} \label{app: vertex}
We give the analytic expressions for the 1PI diagram contributions in terms of the Passarino-Veltman functions~\cite{Passarino:1978jh}.
We calculate 1PI diagrams in the 't Hooft-Feynman gauge with $\xi=1$.
The coefficients of vector and axial-vector couplings of $Z$ boson are given by
\begin{align}
v_{f} = \frac{1}{2}I_{f}-Q_{f}s_{W}^{2}\qc
a_{f} = \frac{1}{2}I_{f}.
\end{align}

\subsection{$Af\bar{f}$ vertex}
The 1PI contributions to the form factor of $Af\bar{f}$ vertex are given by
\allowdisplaybreaks{\begin{align}
&(16\pi^{2})\Gamma_{Af\bar{f}}^{S, \mathrm{1PI}} \notag \\
&\quad
=
2I_{f}\lambda_{H^{+}G^{-}A}\frac{m_{f}m_{f'}^{2}}{v^{2}}(\zeta_{f}+\zeta_{f'})\qty[C_{0}(H^{\pm}, f', G^{\pm})-C_{0}(G^{\pm}, f', H^{\pm})] \notag \\
&\qquad
+i\frac{g_{Z}^{2}}{2}c_{\beta-\alpha}v_{f}\frac{m_{f}\zeta_{hff}}{v}\qty[C_{Af\bar{f}}^{SFV}(h, f, Z)-C_{Af\bar{f}}^{VFS}(Z, f, h)] \notag \\
&\qquad
-i\frac{g_{Z}^{2}}{2}s_{\beta-\alpha}v_{f}\frac{m_{f}\zeta_{Hff}}{v}\qty[C_{Af\bar{f}}^{SFV}(H, f, Z)-C_{Af\bar{f}}^{VFS}(Z, f, H)] \notag \\
&\qquad
+i\frac{g^{2}}{2}\frac{I_{f}m_{f}\zeta_{f}}{v}\qty[C_{Af\bar{f}}^{SFV}(H^{\pm}, f', W^{\pm})-C_{Af\bar{f}}^{VFS}(W^{\pm}, f', H^{\pm})], \\
\notag \\
%
%
%
&(16\pi^{2})\Gamma_{Af\bar{f}}^{P, \mathrm{1PI}} \notag \\
&\quad
=
i\frac{2I_{f}m_{f}\zeta_{f}}{v}\qty[g_{Z}^{2}(v_{f}^{2}-a_{f}^{2})C_{Af\bar{f}}^{FVF}(f, Z, f)
+e^{2}Q_{f}^{2}C_{Af\bar{f}}^{FVF}(f, \gamma, f)] \notag \\
&\qquad
+i\frac{2I_{f}m_{f}^{3}\zeta_{f}}{v^{2}}\bigg[\zeta_{hff}^{2}C_{Af\bar{f}}^{FSF}(f, h, f)
+\zeta_{Hff}^{2}C_{Af\bar{f}}^{FSF}(f, H, f)
-C_{Af\bar{f}}^{FSF}(f, G^{0}, f)
-\zeta_{f}^{2}C_{Af\bar{f}}^{FSF}(f, A, f)\bigg] \notag \\
&\qquad
-i\frac{4I_{f'}m_{f}m_{f'}^{2}\zeta_{f'}}{v^{3}}\qty[C_{Af\bar{f}}^{FSF}(f', G^{\pm}, f')
+\zeta_{f}\zeta_{f'}C_{Af\bar{f}}^{FSF}(f', H^{\pm}, f')] \notag \\
&\qquad
+i\lambda_{AG^{0}h}\frac{m_{f}\zeta_{hff}}{v}\frac{2I_{f}m_{f}^{2}}{v}\qty[C_{0}(h, f, G^{0})+C_{0}(G^{0}, f, h)] \notag \\
&\qquad
+i\lambda_{AG^{0}H}\frac{m_{f}\zeta_{Hff}}{v}\frac{2I_{f}m_{f}^{2}}{v}\qty[C_{0}(H, f, G^{0})+C_{0}(G^{0}, f, H)] \notag \\
&\qquad
+i2\lambda_{AAh}\frac{m_{f}\zeta_{hff}}{v}\frac{2I_{f}m_{f}^{2}\zeta_{f}}{v}\qty[C_{0}(h, f, A)+C_{0}(A, f, h)] \notag \\
&\qquad
+i2\lambda_{AAH}\frac{m_{f}\zeta_{Hff}}{v}\frac{2I_{f}m_{f}^{2}\zeta_{f}}{v}\qty[C_{0}(H, f, A)+C_{0}(A, f, H)] \notag \\
&\qquad
+2I_{f}\lambda_{H^{+}G^{-}A}\frac{m_{f}m_{f'}^{2}}{v^{2}}(\zeta_{f}-\zeta_{f'})\qty[C_{0}(H^{\pm}, f', G^{\pm})+C_{0}(G^{\pm}, f', H^{\pm})] \notag \\
&\qquad
+i\frac{g_{Z}^{2}}{2}c_{\beta-\alpha}a_{f}\frac{m_{f}\zeta_{hff}}{v}\qty[C_{Af\bar{f}}^{SFV}(h, f, Z)+C_{Af\bar{f}}^{VFS}(Z, f, h)] \notag \\
&\qquad
-i\frac{g_{Z}^{2}}{2}s_{\beta-\alpha}a_{f}\frac{m_{f}\zeta_{Hff}}{v}\qty[C_{Af\bar{f}}^{SFV}(H, f, Z)+C_{Af\bar{f}}^{VFS}(Z, f, H)] \notag \\
&\qquad
+i\frac{g^{2}}{2}\frac{I_{f}m_{f}\zeta_{f}}{v}\qty[C_{Af\bar{f}}^{SFV}(H^{\pm}, f', W^{\pm})+ C_{Af\bar{f}}^{VFS}(W^{\pm}, f', H^{\pm})], \\
\notag \\
%
%
%
&(16\pi^{2})\Gamma_{Af\bar{f}}^{V_{1}, \mathrm{1PI}} \notag \\
&\quad
=
i4g_{Z}^{2}v_{f}a_{f}\frac{2I_{f}m_{f}^{2}\zeta_{f}}{v}C_{0}(f, Z, f)
+i\frac{g^{2}}{2}\frac{2I_{f'}m_{f'}^{2}\zeta_{f'}}{v}C_{0}(f', W, f') \notag \\
&\qquad
+i\frac{2I_{f'}m_{f'}^{2}\zeta_{f'}}{v}\qty[\frac{m_{f}^{2}-m_{f'}^{2}}{v^{2}}C_{0}(f', G^{\pm}, f')+\frac{m_{f}^{2}\zeta_{f}^{2}-m_{f'}^{2}\zeta_{f'}^{2}}{v^{2}}C_{0}(f', H^{\pm}, f')] \notag \\
&\qquad
-2I_{f}\lambda_{H^{+}G^{-}A}\frac{m_{f}^{2}\zeta_{f}+m_{f'}^{2}\zeta_{f'}}{v^{2}}\qty[(C_{0}+C_{11})(H^{\pm}, f', G^{\pm})-(C_{0}+C_{11})(G^{\pm}, f', H^{\pm})] \notag \\
&\qquad
+i\frac{g_{Z}^{2}}{2}c_{\beta-\alpha}v_{f}\frac{m_{f}^{2}\zeta_{hff}}{v}\qty[(C_{11}-C_{0})(h, f, Z)-(C_{11}+2C_{0})(Z, f, h)] \notag \\
&\qquad
-i\frac{g_{Z}^{2}}{2}s_{\beta-\alpha}v_{f}\frac{m_{f}^{2}\zeta_{Hff}}{v}\qty[(C_{11}-C_{0})(H, f, Z)-(C_{11}+2C_{0})(Z, f, H)] \notag \\
&\qquad
+i\frac{g^{2}}{2}\frac{I_{f'}m_{f'}^{2}\zeta_{f'}}{v}\qty[(C_{11}-C_{0})(H^{\pm}, f', W^{\pm})-(C_{11}+2C_{0})(W^{\pm}, f', H^{\pm})], \\
\notag \\
%
%
%
&(16\pi^{2})\Gamma_{Af\bar{f}}^{V_{2}, \mathrm{1PI}} \notag \\
&\quad
=
i4g_{Z}^{2}v_{f}a_{f}\frac{2I_{f}m_{f}^{2}\zeta_{f}}{v}C_{0}(f, Z, f)
+i\frac{g^{2}}{2}\frac{2I_{f'}m_{f'}^{2}\zeta_{f'}}{v}C_{0}(f', W, f') \notag \\
&\qquad
+i\frac{2I_{f'}m_{f'}^{2}\zeta_{f'}}{v}\qty[\frac{m_{f}^{2}-m_{f'}^{2}}{v^{2}}C_{0}(f', G^{\pm}, f')
+\frac{m_{f}^{2}\zeta_{f}^{2}-m_{f'}^{2}\zeta_{f'}^{2}}{v^{2}}C_{0}(f', H^{\pm}, f')] \notag \\
&\qquad
-2I_{f}\lambda_{H^{+}G^{-}A}\frac{m_{f}^{2}\zeta_{f}+m_{f'}^{2}\zeta_{f'}}{v^{2}}\qty[C_{12}(H^{\pm}, f', G^{\pm})-C_{12}(G^{\pm}, f', H^{\pm})] \notag \\
&\qquad
+i\frac{g_{Z}^{2}}{2}c_{\beta-\alpha}v_{f}\frac{m_{f}^{2}\zeta_{hff}}{v}\qty[(C_{12}-C_{0})(h, f, Z)-(C_{12}+2C_{0})(Z, f, h)] \notag \\
&\qquad
-i\frac{g_{Z}^{2}}{2}s_{\beta-\alpha}v_{f}\frac{m_{f}^{2}\zeta_{Hff}}{v}\qty[(C_{12}-C_{0})(H, f, Z)-(C_{12}+2C_{0})(Z, f, H)] \notag \\
&\qquad
-i\frac{g^{2}}{2}\frac{I_{f'}m_{f'}^{2}\zeta_{f'}}{v}\qty[(C_{12}-C_{0})(H^{\pm}, f', W^{\pm})-(C_{12}+2C_{0})(W^{\pm}, f', H^{\pm})], \\
\notag \\
%
%
%
&(16\pi^{2})\Gamma_{Af\bar{f}}^{A_{1}, \mathrm{1PI}} \notag \\
&\quad
=
-i\frac{2I_{f}m_{f}^{2}\zeta_{f}}{v}\qty[2g_{Z}^{2}(v_{f}^{2}+a_{f}^{2})C_{0}(f, Z, f)
+2e^{2}Q_{f}^{2}C_{0}(f, \gamma, f)]
-i\frac{g^{2}}{2}\frac{2I_{f'}m_{f'}^{2}\zeta_{f'}}{v}C_{0}(f', W, f') \notag \\
&\qquad
+i\frac{2I_{f}m_{f}^{4}\zeta_{f}}{v^{3}}\bigg[\zeta_{hff}^{2}C_{0}(f, h, f)
+\zeta_{Hff}^{2}C_{0}(f, H, f)
+C_{0}(f, G^{0}, f)
+\zeta_{f}^{2}C_{0}(f, A, f)\bigg] \notag \\
&\qquad
+i\frac{2I_{f'}m_{f'}^{2}\zeta_{f'}}{v}\qty[\frac{m_{f}^{2}+m_{f'}^{2}}{v^{2}}C_{0}(f', G^{\pm}, f')
+\frac{m_{f}^{2}\zeta_{f}^{2}+m_{f'}^{2}\zeta_{f'}^{2}}{v^{2}}C_{0}(f', H^{\pm}, f')] \notag \\
&\qquad
-i\lambda_{AG^{0}h}\frac{m_{f}\zeta_{hff}}{v}\frac{2I_{f}m_{f}}{v}\qty[(C_{0}+C_{11})(h, f, G^{0})-(C_{0}+C_{11})(G^{0}, f, h)] \notag \\
&\qquad
-i\lambda_{AG^{0}H}\frac{m_{f}\zeta_{Hff}}{v}\frac{2I_{f}m_{f}}{v}\qty[(C_{0}+C_{11})(H, f, G^{0})-(C_{0}+C_{11})(G^{0}, f, H)] \notag \\
&\qquad
-i2\lambda_{AAh}\frac{m_{f}\zeta_{hff}}{v}\frac{2I_{f}m_{f}\zeta_{f}}{v}\qty[(C_{0}+C_{11})(h, f, A)-(C_{0}+C_{11})(A, f, h)] \notag \\
&\qquad
-i2\lambda_{AAH}\frac{m_{f}\zeta_{Hff}}{v}\frac{2I_{f}m_{f}\zeta_{f}}{v}\qty[(C_{0}+C_{11})(H, f, A)-(C_{0}+C_{11})(A, f, H)] \notag \\
&\qquad
-2I_{f}\lambda_{H^{+}G^{-}A}\frac{m_{f}^{2}\zeta_{f}-m_{f'}^{2}\zeta_{f'}}{v^{2}}\qty[(C_{0}+C_{11})(H^{\pm}, f', G^{\pm})-(C_{0}+C_{11})(G^{\pm}, f', H^{\pm})] \notag \\
&\qquad
+i\frac{g_{Z}^{2}}{2}c_{\beta-\alpha}a_{f}\frac{m_{f}^{2}\zeta_{hff}}{v}\qty[(-C_{11}+C_{0})(h, f, Z)+(C_{11}+2C_{0})(Z, f, h)] \notag \\
&\qquad
-i\frac{g_{Z}^{2}}{2}s_{\beta-\alpha}a_{f}\frac{m_{f}^{2}\zeta_{Hff}}{v}\qty[(-C_{11}+C_{0})(H, f, Z)+(C_{11}+2C_{0})(Z, f, H)] \notag \\
&\qquad
+i\frac{g^{2}}{2}\frac{I_{f'}m_{f'}^{2}\zeta_{f'}}{v}\qty[(-C_{11}+C_{0})(H^{\pm}, f', W^{\pm})+(C_{11}+2C_{0})(W^{\pm}, f', H^{\pm})], \\
\notag \\
%
%
%
&(16\pi^{2})\Gamma_{Af\bar{f}}^{A_{2}, \mathrm{1PI}} \notag \\
&\quad
=
-i\frac{2I_{f}m_{f}^{2}\zeta_{f}}{v}\qty[2g_{Z}^{2}(v_{f}^{2}+a_{f}^{2})C_{0}(f, Z, f)
+2e^{2}Q_{f}^{2}C_{0}(f, \gamma, f)]
-i\frac{g^{2}}{2}\frac{2I_{f'}m_{f'}^{2}\zeta_{f'}}{v}C_{0}(f', W, f') \notag \\
&\qquad
+i\frac{2I_{f}m_{f}^{4}\zeta_{f}}{v^{3}}\bigg[\zeta_{hff}^{2}C_{0}(f, h, f)
+\zeta_{Hff}^{2}C_{0}(f, H, f)
+C_{0}(f, G^{0}, f)
+\zeta_{f}^{2}C_{0}(f, A, f)\bigg] \notag \\
&\qquad
+i\frac{2I_{f'}m_{f'}^{2}\zeta_{f'}}{v}\qty[\frac{m_{f}^{2}+m_{f'}^{2}}{v^{2}}C_{0}(f', G^{\pm}, f')
+\frac{m_{f}^{2}\zeta_{f}^{2}+m_{f'}^{2}\zeta_{f'}^{2}}{v^{2}}C_{0}(f', H^{\pm}, f')] \notag \\
&\qquad
-i\lambda_{AG^{0}h}\frac{m_{f}\zeta_{hff}}{v}\frac{2I_{f}m_{f}}{v}\qty[C_{12}(h, f, G^{0})-C_{12}(G^{0}, f, h)] \notag \\
&\qquad
-i\lambda_{AG^{0}H}\frac{m_{f}\zeta_{Hff}}{v}\frac{2I_{f}m_{f}}{v}\qty[C_{12}(H, f, G^{0})-C_{12}(G^{0}, f, H)] \notag \\
&\qquad
-i2\lambda_{AAh}\frac{m_{f}\zeta_{hff}}{v}\frac{2I_{f}m_{f}\zeta_{f}}{v}\qty[C_{12}(h, f, A)-C_{12}(A, f, h)] \notag \\
&\qquad
-i2\lambda_{AAH}\frac{m_{f}\zeta_{Hff}}{v}\frac{2I_{f}m_{f}\zeta_{f}}{v}\qty[C_{12}(H, f, A)-C_{12}(A, f, H)] \notag \\
&\qquad
-2I_{f}\lambda_{H^{+}G^{-}A}\frac{m_{f}^{2}\zeta_{f}-m_{f'}^{2}\zeta_{f'}}{v^{2}}\qty[C_{12}(H^{\pm}, f', G^{\pm})-C_{12}(G^{\pm}, f', H^{\pm})] \notag \\
&\qquad
+i\frac{g_{Z}^{2}}{2}c_{\beta-\alpha}a_{f}\frac{m_{f}^{2}\zeta_{hff}}{v}\qty[(-C_{12}+C_{0})(h, f, Z)+(C_{12}+2C_{0})(Z, f, h)] \notag \\
&\qquad
-i\frac{g_{Z}^{2}}{2}s_{\beta-\alpha}a_{f}\frac{m_{f}^{2}\zeta_{Hff}}{v}\qty[(-C_{12}+C_{0})(H, f, Z)+(C_{12}+2C_{0})(Z, f, H)] \notag \\
&\qquad
-i\frac{g^{2}}{2}\frac{I_{f'}m_{f'}^{2}\zeta_{f'}}{v}\qty[(-C_{12}+C_{0})(H^{\pm}, f', W^{\pm})+(C_{12}+2C_{0})(W^{\pm}, f', H^{\pm})], \\
\notag \\
%
%
%
&(16\pi^{2})\Gamma_{Af\bar{f}}^{T, \mathrm{1PI}} \notag \\
&\quad
=
i\frac{g_{Z}^{2}}{2}c_{\beta-\alpha}v_{f}\frac{m_{f}\zeta_{hff}}{v}\qty[(C_{0}+C_{11}-2C_{12})(h, f, Z)-(2C_{0}+2C_{11}-C_{12})(Z, f, h)] \notag \\
&\qquad
-i\frac{g_{Z}^{2}}{2}s_{\beta-\alpha}v_{f}\frac{m_{f}\zeta_{Hff}}{v}\qty[(C_{0}+C_{11}-2C_{12})(H, f, Z)-(2C_{0}+2C_{11}-C_{12})(Z, f, H)] \notag \\
&\qquad
+i\frac{g^{2}}{2}\frac{I_{f}m_{f}\zeta_{f}}{v}\qty[(C_{0}+C_{11}-2C_{12})(H^{\pm}, f', W^{\pm})
-(2C_{0}+2C_{11}-C_{12})(W^{\pm}, f', H^{\pm})], \\
\notag \\
%
%
%
&(16\pi^{2})\Gamma_{Af\bar{f}}^{PT, \mathrm{1PI}} \notag \\
&\quad
=
i\frac{2I_{f}m_{f}^{3}\zeta_{f}}{v^{3}}\bigg[\zeta_{hff}^{2}(-C_{11}+C_{12})(f, h, f)
+\zeta_{Hff}^{2}(-C_{11}+C_{12})(f, H, f)
\notag \\
&\qquad
-(-C_{11}+C_{12})(f, G^{0}, f)
-\zeta_{f}^{2}(-C_{11}+C_{12})(f, A, f)\bigg] \notag \\
&\qquad
-i\frac{4I_{f'}m_{f}m_{f'}^{2}\zeta_{f'}}{v^{3}}\bigg[(-C_{11}+C_{12})(f', G^{\pm}, f')
+\zeta_{f}\zeta_{f'}(-C_{11}+C_{12})(f', H^{\pm}, f')\bigg] \notag \\
&\qquad
+i\frac{g_{Z}^{2}}{2}c_{\beta-\alpha}a_{f}\frac{m_{f}\zeta_{hff}}{v}\qty[(C_{0}+C_{11}-2C_{12})(h, f, Z)+(2C_{0}+2C_{11}-C_{12})(Z, f, h)] \notag \\
&\qquad
-i\frac{g_{Z}^{2}}{2}s_{\beta-\alpha}a_{f}\frac{m_{f}\zeta_{Hff}}{v}\qty[(C_{0}+C_{11}-2C_{12})(H, f, Z)+(2C_{0}+2C_{11}-C_{12})(Z, f, H)] \notag \\
&\qquad
+i\frac{g^{2}}{2}\frac{I_{f}m_{f}\zeta_{f}}{v}[(C_{0}+C_{11}-2C_{12})(H^{\pm}, f', W^{\pm})
+(2C_{0}+2C_{11}-C_{12})(W^{\pm}, f', H^{\pm})].
\end{align}}
%
%
%
where
\begin{align}
&C_{Af\bar{f}}^{SFV}(X, Y, Z)
= B_{0}(p_{2}^{2}; Y, Z) \notag \\
&\quad
+\bigg[
(m_{X}^{2}-q^{2}+p_{2}^{2})C_{0}
-(q^{2}-p_{1}^{2}-p_{2}^{2})C_{11} 
+(q^{2}-p_{1}^{2}-2p_{2}^{2})C_{12}
\bigg](p_{1}^{2}, p_{2}^{2}, q^{2}; X, Y, Z), \\
&C_{Af\bar{f}}^{VFS}(X, Y, Z)
= B_{0}(p_{2}^{2}; Y, Z) \notag \\
&\quad
+\bigg[
(m_{X}^{2}+2p_{1}^{2})C_{0}
+3p_{1}^{2}C_{11}
+2(q^{2}-p_{1}^{2})C_{12}
\bigg](p_{1}^{2}, p_{2}^{2}, q^{2}; X, Y, Z), \\
&C_{Af\bar{f}}^{FVF}(X, Y, Z) 
= 4B_{0}(p_{2}^{2}; Y, Z)-2 \notag \\
&\quad
+\bigg[
4m_{X}(m_{X}-m_{Z})C_{0}
+2(q^{2}+p_{1}^{2}-p_{2}^{2})C_{11}
+2(q^{2}-p_{1}^{2}+p_{2}^{2})C_{12}
\bigg](p_{1}^{2}, p_{2}^{2}, q^{2}; X, Y, Z), \\
&C_{Af\bar{f}}^{FSF}(X, Y, Z)
= B_{0}(p_{2}^{2}; Y, Z) \notag \\
&\quad
+\bigg[
m_{X}(m_{X}-m_{Z})C_{0}
+(q^{2}-p_{2}^{2})C_{11}
+p_{2}^{2}C_{12}
\bigg](p_{1}^{2}, p_{2}^{2}, q^{2}; X, Y, Z).
\end{align}
The loop functions satisfy following relations
\begin{align}
C_{Af\bar{f}}^{SFV}(X, Y, Z) &= C_{hf\bar{f}}^{SFV}(X, Y, Z), \\
C_{Af\bar{f}}^{VFS}(X, Y, Z) &= C_{hf\bar{f}}^{VFS}(X, Y, Z), \\
C_{Af\bar{f}}^{FVF}(X, Y, Z) &= C_{-}^{FVF}(X, Y, Z), \\
C_{Af\bar{f}}^{FSF}(X, Y, Z) &= C_{P}^{FSF}(X, Y, Z),
\end{align}
where $C_{hf\bar{f}}^{SFV}$ and $C_{hf\bar{f}}^{VFS}$ are given in Eq.~(C.44) in Ref.~\cite{Kanemura:2015mxa}, while $C_{-}^{FVF}$ and $C_{P}^{FSF}$ are given in Eqs.~(B.23) and (B.22) in Ref.~\cite{Aiko:2021can}.

\subsection{$AV\phi$ vertex}
The 1PI contributions to the form factor of $AZh$ vertex are given by
\allowdisplaybreaks{\begin{align}
&(16\pi^{2})\Gamma_{AZh}^{\mathrm{1PI}}(p_{1}^{2}, p_{2}^{2}, q^{2})_{F} \notag \\
&\quad=
i\sum_{f} 8I_{f}N_{c}^{f}g_{Z}a_{f}\frac{m_{f}^{2}\zeta_{f}\zeta_{hff}}{v^{2}}\bigg[
B_{0}(p_{2}^{2}; f, f)+2m_{f}^{2}C_{0}
+p_{1}^{2}C_{11}+(q^{2}-p_{1}^{2})C_{12}
\bigg](f, f, f), \\
\notag \\
%
%
%
&(16\pi^{2})\Gamma_{AZh}^{\mathrm{1PI}}(p_{1}^{2}, p_{2}^{2}, q^{2})_{B} \notag \\
&\quad=
i\frac{g^{3}}{2}c_{\beta-\alpha}c_{W}C_{AZ\phi}^{SVV}(H^{\pm}, W^{\pm}, W^{\pm}) \notag \\
&\qquad
+i\frac{g_{Z}^{3}}{8}c_{\beta-\alpha}\Big[
c_{\beta-\alpha}^{2}C_{AZ\phi}^{VSS}(Z, A, h)
+s_{\beta-\alpha}^{2}C_{AZ\phi}^{VSS}(Z, A, H) \notag \\
&\qquad
+s_{\beta-\alpha}^{2}C_{AZ\phi}^{VSS}(Z, G^{0}, h)
-s_{\beta-\alpha}^{2}C_{AZ\phi}^{VSS}(Z, G^{0}, H) 
-2c_{W}^{2}(c_{W}^{2}-s_{W}^{2})C_{AZ\phi}^{VSS}(W^{\pm}, H^{\pm}, H^{\pm})\Big] \notag \\
&\qquad
-i\frac{g_{Z}^{3}}{8}c_{\beta-\alpha}\Big[
(B_{0}-B_{1})(q^{2}; h, Z)
+2s_{W}^{2}c_{W}^{2}(B_{0}-B_{1})(q^{2}; H^{\pm}, W^{\pm}) \notag \\
&\qquad
+(B_{0}-B_{1})(p_{1}^{2}; A, Z)
+2s_{W}^{2}c_{W}^{2}(B_{0}-B_{1})(p_{1}^{2}; H^{\pm}, W^{\pm})\Big] \notag \\
&\qquad
-i\frac{g_{Z}^{3}m_{Z}^{2}}{4}s_{\beta-\alpha}^{2}c_{\beta-\alpha}\Big[
(2C_{0}+C_{11})(Z, Z, h)
-(2C_{0}+C_{11})(Z, Z, H)\Big] \notag \\
&\qquad
+i\frac{g_{Z}^{2}m_{Z}}{4}\Big[
6\lambda_{hhh}s_{\beta-\alpha}c_{\beta-\alpha}(C_{0}-C_{11})(h, h, Z)
+2\lambda_{hhH}c_{\beta-\alpha}^{2}(C_{0}-C_{11})(h, H, Z) \notag \\
&\qquad
-2\lambda_{hHh}s_{\beta-\alpha}^{2}(C_{0}-C_{11})(H, h, Z)
-2\lambda_{hHH}s_{\beta-\alpha}c_{\beta-\alpha}(C_{0}-C_{11})(H, H, Z) \notag \\
&\qquad
+2s_{W}^{2}c_{W}^{2}\lambda_{hH^{+}G^{-}}(C_{0}-C_{11})(H^{\pm}, G^{\pm}, W^{\pm})\Big] \notag \\
&\qquad
+i\frac{g_{Z}^{2}m_{Z}}{4}\Big[
2\lambda_{AAh}s_{\beta-\alpha}c_{\beta-\alpha}(C_{0}-C_{11})(A, Z, h)
+2\lambda_{AAH}c_{\beta-\alpha}^{2}(C_{0}-C_{11})(A, Z, H) \notag \\
&\qquad
+\lambda_{AG^{0}h}s_{\beta-\alpha}^{2}(C_{0}-C_{11})(G^{0}, Z, h)
+\lambda_{AG^{0}H}s_{\beta-\alpha}c_{\beta-\alpha}(C_{0}-C_{11})(G^{0}, Z, H) \notag \\
&\qquad
-is_{W}^{2}c_{W}^{2}\lambda_{AG^{+}H^{-}}c_{\beta-\alpha}(C_{0}-C_{11})(H^{\pm}, W^{\pm}, G^{\pm}) \notag \\
&\qquad
+is_{W}^{2}c_{W}^{2}\lambda_{AG^{-}H^{+}}c_{W}^{2}c_{W}^{2}(C_{0}-C_{11})(H^{\pm}, W^{\pm}, G^{\pm})\Big] \notag \\
&\qquad
+i\frac{g_{Z}}{2}\Big[
-12\lambda_{AhA}\lambda_{hhh}c_{\beta-\alpha}(C_{0}+C_{11})(h, h, A)
+4\lambda_{AhA}\lambda_{hhH}s_{\beta-\alpha}(C_{0}+C_{11})(h, H, A) \notag \\
&\qquad
-4\lambda_{AHA}\lambda_{hHh}c_{\beta-\alpha}(C_{0}+C_{11})(H, h, A)
-4\lambda_{AHA}\lambda_{hHH}s_{\beta-\alpha}(C_{0}+C_{11})(H, H, A) \notag \\
&\qquad
+4\lambda_{AAh}^{2}c_{\beta-\alpha}(C_{0}+C_{11})(A, A, h)
-4\lambda_{AAH}\lambda_{hAA}s_{\beta-\alpha}C_{0}+C_{11})(A, A, H)  \notag \\
&\qquad
-6\lambda_{AhG^{0}}\lambda_{hhh}s_{\beta-\alpha}(C_{0}+C_{11})(h, h, G^{0})
-2\lambda_{AhG^{0}}\lambda_{hhH}c_{\beta-\alpha}(C_{0}+C_{11})(h, H, G^{0})  \notag \\
&\qquad
-2\lambda_{AHG^{0}}\lambda_{hHh}s_{\beta-\alpha}(C_{0}+C_{11})(H, h, G^{0})
-2\lambda_{AHG^{0}}\lambda_{hHH}c_{\beta-\alpha}(C_{0}+C_{11})(H, H, G^{0})  \notag \\
&\qquad
+2\lambda_{AAh}\lambda_{hAG^{0}}s_{\beta-\alpha}(C_{0}+C_{11})(A, G^{0}, h)
+\lambda_{AG^{0}h}^{2}c_{\beta-\alpha}(C_{0}+C_{11})(G^{0}, A, h)  \notag \\
&\qquad
+2\lambda_{AAH}\lambda_{hAG^{0}}c_{\beta-\alpha}(C_{0}+C_{11})(A, G^{0}, H)
-\lambda_{AG^{0}H}\lambda_{hG^{0}A}s_{\beta-\alpha}(C_{0}+C_{11})(G^{0}, A, H)  \notag \\
&\qquad
+2\lambda_{AG^{0}h}\lambda_{hG^{0}G^{0}}s_{\beta-\alpha}(C_{0}+C_{11})(G^{0}, G^{0}, h)
+2\lambda_{AG^{0}H}\lambda_{hG^{0}G^{0}}c_{\beta-\alpha}(C_{0}+C_{11})(G^{0}, G^{0}, H)  \notag \\
&\qquad
+i(c_{W}^{2}-s_{W}^{2})\lambda_{AG^{-}H^{+}}\lambda_{hG^{+}H^{-}}(C_{0}+C_{11})(G^{\pm}, H^{\pm}, H^{\pm}) \notag \\
&\qquad
-i(c_{W}^{2}-s_{W}^{2})\lambda_{AG^{+}H^{-}}\lambda_{hG^{-}H^{+}}(C_{0}+C_{11})(G^{\pm}, H^{\pm}, H^{\pm}) \notag \\
&\qquad
+i(c_{W}^{2}-s_{W}^{2})\lambda_{AH^{-}G^{+}}\lambda_{hH^{+}G^{-}}(C_{0}+C_{11})(H^{\pm}, G^{\pm}, G^{\pm}) \notag \\
&\qquad
-i(c_{W}^{2}-s_{W}^{2})\lambda_{AH^{+}G^{-}}\lambda_{hH^{-}G^{+}}(C_{0}+C_{11})(H^{\pm}, G^{\pm}, G^{\pm})\Big].
\end{align}}
where
\begin{align}
&C_{AZ\phi}^{SVV}(X, Y, Z) = 2B_{0}(p_{2}^{2}; Y, Z)
+\bigg[-\frac{1}{2}(3p_{1}^{2}-p_{2}^{2}+q^{2})C_{21}+(p_{1}^{2}-q^{2})C_{23}-2C_{24} \notag \\
&\qquad
+\frac{1}{2}(-p_{1}^{2}-2p_{2}^{2}+q^{2})C_{11}
+(p_{1}^{2}-q^{2})C_{12}
+\frac{1}{2}(p_{2}^{2}+4m_{X}^{2})C_{0}\bigg](p_{1}^{2}, p_{2}^{2}, q^{2}; X, Y, Z), \\
&C_{AZ\phi}^{VSS}(X, Y, Z) = 
\bigg[(q^{2}+3p_{1}^{2}-p_{2}^{2})C_{21}+2(q^{2}-p_{1}^{2})C_{23}+4C_{24}
+(3q^{2}+5p_{1}^{2}-3p_{2}^{2}+m_{X}^{2})C_{11} \notag \\
&\qquad
+2(q^{2}-p_{1}^{2})C_{12}
+(2q^{2}+2p_{1}^{2}-2p_{2}^{2}+m_{X}^{2})C_{0}\bigg](p_{1}^{2}, p_{2}^{2}, q^{2}; X, Y, Z).
\end{align}

The 1PI contributions to the form factor of $AZH$ vertex are given by
\begin{align}
&(16\pi^{2})\Gamma_{AZH}^{\mathrm{1PI}}(p_{1}^{2}, p_{2}^{2}, q^{2})_{F} \notag \\
&\quad=
i\sum_{f} 8I_{f}N_{c}^{f}g_{Z}a_{f}\frac{m_{f}^{2}\zeta_{f}\zeta_{Hff}}{v^{2}}\bigg[
B_{0}(p_{2}^{2}; f, f)+2m_{f}^{2}C_{0}
+p_{1}^{2}C_{11}+(q^{2}-p_{1}^{2})C_{12}
\bigg](f, f, f), \\
\notag \\
%
%
%
&(16\pi^{2})\Gamma_{AZH}^{\mathrm{1PI}}(p_{1}^{2}, p_{2}^{2}, q^{2})_{B} \notag \\
&\quad=
-i\frac{g^{3}}{2}c_{W}s_{\beta-\alpha}
C_{AZ\phi}^{SVV}(H^{\pm}, W^{\pm}, W^{\pm}) \notag \\
&\qquad
+i\frac{g_{Z}^{3}}{8}s_{\beta-\alpha}\Big[
-c_{\beta-\alpha}^{2}C_{AZ\phi}^{VSS}(Z, A, h)
-s_{\beta-\alpha}^{2}C_{AZ\phi}^{VSS}(Z, A, H) \notag \\
&\qquad
+c_{\beta-\alpha}^{2}C_{AZ\phi}^{VSS}(Z, G^{0}, h)
-c_{\beta-\alpha}^{2}C_{AZ\phi}^{VSS}(Z, G^{0}, H)
+2c_{W}^{2}(c_{W}^{2}-s_{W}^{2})C_{AZ\phi}^{VSS}(W^{\pm}, H^{\pm}, H^{\pm})\Big] \notag \\
&\qquad
+i\frac{g_{Z}^{3}}{8}s_{\beta-\alpha}\Big[
(B_{0}-B_{1})(q^{2}; H, Z)
+2s_{W}^{2}c_{W}^{2}(B_{0}-B_{1})(q^{2}; H^{\pm}, W^{\pm}) \notag \\
&\qquad
+(B_{0}-B_{1})(p_{1}^{2}; A, Z)
+2s_{W}^{2}c_{W}^{2}(B_{0}-B_{1})(p_{1}^{2}; H^{\pm}, W^{\pm})\Big] \notag \\
&\qquad
-i\frac{g_{Z}^{3}m_{Z}^{2}}{4}s_{\beta-\alpha}c^{2}_{\beta-\alpha}\Big[
(2C_{0}+C_{11})(Z, Z, h)
-(2C_{0}+C_{11})(Z, Z, H)\Big] \notag \\
&\qquad
+i\frac{g_{Z}^{2}m_{Z}}{4}\Big[
2\lambda_{Hhh}s_{\beta-\alpha}c_{\beta-\alpha}(C_{0}-C_{11})(h, h, Z)
+2\lambda_{HhH}c_{\beta-\alpha}^{2}(C_{0}-C_{11})(h, H, Z) \notag \\
&\qquad
-2\lambda_{HHh}s_{\beta-\alpha}^{2}(C_{0}-C_{11})(H, h, Z)
-6\lambda_{HHH}s_{\beta-\alpha}c_{\beta-\alpha}(C_{0}-C_{11})(H, H, Z) \notag \\
&\qquad
+2s_{W}^{2}c_{W}^{2}\lambda_{HH^{+}G^{-}}(C_{0}-C_{11})(H^{\pm}, G^{\pm}, W^{\pm})\Big] \notag \\
&\qquad
-i\frac{g_{Z}^{2}m_{Z}}{4}\Big[
2\lambda_{AAh}s_{\beta-\alpha}^{2}(C_{0}-C_{11})(A, Z, h)
+2\lambda_{AAH}c_{\beta-\alpha}s_{\beta-\alpha}(C_{0}-C_{11})(A, Z, H) \notag \\
&\qquad
-\lambda_{AG^{0}h}c_{\beta-\alpha}s_{\beta-\alpha}(C_{0}-C_{11})(G^{0}, Z, h)
-\lambda_{AG^{0}H}c_{\beta-\alpha}^{2}(C_{0}-C_{11})(G^{0}, Z, H) \notag \\
&\qquad
-is_{W}^{2}c_{W}^{2}\lambda_{AG^{+}H^{-}}s_{\beta-\alpha}(C_{0}-C_{11})(H^{\pm}, W^{\pm}, G^{\pm}) \notag \\
&\qquad
+is_{W}^{2}c_{W}^{2}\lambda_{AG^{-}H^{+}}s_{\beta-\alpha}(C_{0}-C_{11})(H^{\pm}, W^{\pm}, G^{\pm})\Big] \notag \\
&\qquad
+i\frac{g_{Z}}{2}\Big[
-4\lambda_{AhA}\lambda_{Hhh}c_{\beta-\alpha}(C_{0}+C_{11})(h, h, A)
+4\lambda_{AhA}\lambda_{HhH}s_{\beta-\alpha}(C_{0}+C_{11})(h, H, A) \notag \\
&\qquad
-4\lambda_{AHA}\lambda_{HHh}c_{\beta-\alpha}(C_{0}+C_{11})(H, h, A)
12\lambda_{AHA}\lambda_{HHH}s_{\beta-\alpha}(C_{0}+C_{11})(H, H, A) \notag \\
&\qquad
+4\lambda_{AAh}\lambda_{HAA}c_{\beta-\alpha}(C_{0}+C_{11})(A, A, h)
-4\lambda_{AAH}^{2}s_{\beta-\alpha}(C_{0}+C_{11})(A, A, H)  \notag \\
&\qquad
-2\lambda_{AhG^{0}}\lambda_{Hhh}s_{\beta-\alpha}(C_{0}+C_{11})(h, h, G^{0})
-2\lambda_{AhG^{0}}\lambda_{HhH}c_{\beta-\alpha}(C_{0}+C_{11})(h, H, G^{0})  \notag \\
&\qquad
-2\lambda_{AHG^{0}}\lambda_{HHh}s_{\beta-\alpha}(C_{0}+C_{11})(H, h, G^{0})
-6\lambda_{AHG^{0}}\lambda_{HHH}c_{\beta-\alpha}(C_{0}+C_{11})(H, H, G^{0})  \notag \\
&\qquad
+2\lambda_{AAh}\lambda_{HAG^{0}}s_{\beta-\alpha}(C_{0}+C_{11})(A, G^{0}, h)
+\lambda_{AG^{0}h}\lambda_{HG^{0}A}c_{\beta-\alpha}(C_{0}+C_{11})(G^{0}, A, h)  \notag \\
&\qquad
+2\lambda_{AAH}\lambda_{HAG^{0}}c_{\beta-\alpha}(C_{0}+C_{11})(A, G^{0}, H)
-\lambda_{AG^{0}H}^{2}s_{\beta-\alpha}(C_{0}+C_{11})(G^{0}, A, H)  \notag \\
&\qquad
+2\lambda_{AG^{0}h}\lambda_{HG^{0}G^{0}}s_{\beta-\alpha}(C_{0}+C_{11})(G^{0}, G^{0}, h)
+2\lambda_{AG^{0}H}\lambda_{HG^{0}G^{0}}c_{\beta-\alpha}(C_{0}+C_{11})(G^{0}, G^{0}, H)  \notag \\
&\qquad
+i(c_{W}^{2}-s_{W}^{2})\lambda_{AG^{-}H^{+}}\lambda_{HG^{+}H^{-}}(C_{0}+C_{11})(G^{\pm}, H^{\pm}, H^{\pm}) \notag \\
&\qquad
-i(c_{W}^{2}-s_{W}^{2})\lambda_{AG^{+}H^{-}}\lambda_{HG^{-}H^{+}}(C_{0}+C_{11})(G^{\pm}, H^{\pm}, H^{\pm}) \notag \\
&\qquad
+i(c_{W}^{2}-s_{W}^{2})\lambda_{AH^{-}G^{+}}\lambda_{HH^{+}G^{-}}(C_{0}+C_{11})(H^{\pm}, G^{\pm}, G^{\pm}) \notag \\
&\qquad
-i(c_{W}^{2}-s_{W}^{2})\lambda_{AH^{+}G^{-}}\lambda_{HH^{-}G^{+}}(C_{0}+C_{11})(H^{\pm}, G^{\pm}, G^{\pm})\Big].
\end{align}

\subsection{$AV_{1}V_{2}$ vertex}
The 1PI contributions to the form factor of $AW^{+}W^{-}$ vertex are given by
\allowdisplaybreaks{\begin{align}
(16\pi^{2})\Gamma_{AW^{+}W^{-}}^{3, \mathrm{1PI}}(p_{1}^{2}, p_{2}^{2}, q^{2})_{F} 
=
ig^{2}\sum_{f} N_{c}^{f}\frac{2I_{f}m_{f}^{2}\zeta_{f}}{v}\bigg[
C_{0}+C_{11}-C_{12}
\bigg](f, f', f)+(f\leftrightarrow f'),
\end{align}}
where $f'$ is the $SU(2)_{L}$ partner of $f$.

The 1PI contributions to the form factor of $AZZ$ vertex are given by
\allowdisplaybreaks{\begin{align}
&(16\pi^{2})\Gamma_{AZZ}^{3, \mathrm{1PI}}(p_{1}^{2}, p_{2}^{2}, q^{2})_{F}
=
ig_{Z}^{2}\sum_{f} N_{c}^{f}\frac{16I_{f}m_{f}^{2}\zeta_{f}}{v}\bigg[
v_{f}^{2}C_{0}
+a_{f}^{2}\qty(C_{0}+2C_{11}-2C_{12})
\bigg](f, f, f).
\end{align}}

The 1PI contributions to the form factor of $AZ\gamma$ vertex are given by
\allowdisplaybreaks{\begin{align}
(16\pi^{2})\Gamma_{AZ\gamma}^{3, \mathrm{1PI}}(p_{1}^{2}, p_{2}^{2}, q^{2})_{F}
=
ieg_{Z}\sum_{f} N_{c}^{f}Q_{f}\frac{16I_{f}m_{f}^{2}\zeta_{f}}{v}v_{f}C_{0}(f, f, f).
\end{align}}

The 1PI contributions to the form factor of $A\gamma\gamma$ vertex are given by
\allowdisplaybreaks{\begin{align}
(16\pi^{2})\Gamma_{A\gamma\gamma}^{3, \mathrm{1PI}}(p_{1}^{2}, p_{2}^{2}, q^{2})_{F}
=
ie^{2}\sum_{f} N_{c}^{f}Q_{f}^{2}\frac{16I_{f}m_{f}^{2}\zeta_{f}}{v}C_{0}(f, f, f).
\end{align}}
Our results are consistent with Ref.~\cite{Gunion:1991cw}.\footnote{Difference in the overall sign comes from the notation of $\epsilon^{\mu\nu\rho\sigma}$. We take $\epsilon^{0123}=+1$, while $\epsilon^{0123}=-1$ in Ref.~\cite{Gunion:1991cw}.}

%% file: A06real_photon_emission.tex
\section{Formulae for the real photon emissions} \label{sec:real_emission}
We here give the decay rate of a massive particle with $p_{0}^{2}=m_{0}^{2}$ into two massive particles with $p_{1}^{2}=m_{1}^{2}$ and $p_{2}^{2}=m_{2}^{2}$ and a photon with $q^{2}=\mu^{2}$, where we introduce the small photon mass as an IR regulator.
We need to evaluate the following phase space integrals,
\begin{align}
I_{i_{1},\cdots,i_{n}}^{j_{1},\cdots j_{m}}(m_{0}, m_{1}, m_{2})
=
32\pi^{3}\int d\Phi_{3}\frac{(\pm 2q\vdot p_{j_{1}})\cdots(\pm 2q\vdot p_{j_{m}})}{(\pm 2q\vdot p_{i_{1}})\cdots(\pm 2q\vdot p_{i_{n}})}, \label{eq: brems_int}
\end{align}
where $j_{k}, i_{\ell}=0, 1, 2$ and the plus signs belong to $p_{1}, p_{2}$, the minus signs to $p_{0}$.
The analytic formulae for $I_{i_{1},\cdots,i_{n}}^{j_{1},\cdots j_{m}}(m_{0}, m_{1}, m_{2})$ are listed in Ref.~\cite{Denner:1991kt}.
\subsection{Decay rates of $A\to f\bar{f}\gamma$}
The decay rate for $A\to f\bar{f}\gamma$ is given by~\cite{Goodsell:2017pdq}
\begin{align}
\Gamma(A\to f\bar{f}\gamma) = N_{c}^{f}\frac{2\alpha_{\mathrm{em}}Q_{f}^{2}}{16\pi^{2}m_{A}}\qty(\frac{2I_{f}m_{f}\zeta_{f}}{v})^{2}\qty[\Omega_{11}+\Omega_{22}+\Omega_{12}],
\end{align}
where the functions $\Omega_{ij}$ are given by
\begin{align}
\Omega_{11} &=
-2I+2m_{A}^{2}I_{1}-4m_{f}^{2}m_{A}^{2}I_{11}, \\
\Omega_{22} &=
-2I+2m_{A}^{2}I_{2}-4m_{f}^{2}m_{A}^{2}I_{22}, \\
\Omega_{12} &=
8I+2(I_{1}^{2}+I_{2}^{1})-6m_{A}^{2}(I_{1}+I_{2})+4m_{A}^{2}(m_{A}^{2}-2m_{f}^{2})I_{12}.
\end{align}
These expressions can be obtained from the decay rate of $H^{\pm}\to f\bar{f'}\gamma$ given in Eq. (C.2) in Ref.~\cite{Aiko:2021can} by replacing
\begin{align}
c_{L}\to -i\frac{2I_{f}m_{f}\zeta_{f}}{v}\qc
c_{R}\to i\frac{2I_{f}m_{f}\zeta_{f}}{v},
\end{align}
and $H^{\pm}\to A,\, f'\to f$.
The functions $\Omega_{ij}$ are defined by
\begin{align}
\Omega_{ij} = \Omega^{LL}_{ij}-\Omega^{LR}_{ij},
\end{align}
where the functions $\Omega^{LL}_{ij}$ and $\Omega^{LR}_{ij}$ are given in Eqs. (C.3) to (C.14) in Ref.~\cite{Aiko:2021can}.

\subsection{Decay rate of $A\to H^{\pm}W^{\mp}\gamma$}
The decay rate of $A\to H^{\pm}W^{\mp}\gamma$ is given by~\cite{Goodsell:2017pdq}
\begin{align}
\Gamma(A\to H^{\pm}W^{\mp}\gamma)
&=
-\frac{\alpha_{\mathrm{em}}\sqrt{2}G_{F}m_{W}^{2}m_{A}^{3}}{2\pi^{2}}\Bigg\{ \lambda\qty(\frac{m_{H^{\pm}}^{2}}{m_{A}^{2}}, \frac{m_{W}^{2}}{m_{A}^{2}})
\bigg[I_{22}+\frac{m_{H^{\pm}}^{2}}{m_{W}^{2}}I_{11} \notag \\
&\qquad
+\qty(1+\frac{m_{H^{\pm}}^{2}-m_{A}^{2}}{m_{W}^{2}})I_{12}
+\frac{1}{m_{W}^{2}}(I_{1}+I_{2})\bigg]
-\frac{2}{m_{A}^{4}}\qty[I+2I_{2}^{1}+I_{22}^{11}]\Bigg\}.
\end{align}

%% file: A07numerical_result.tex
\section{Size of next-to-leading order electroweak corrections for fermionic decays}
\label{app: numerical_result}
In this section, we discuss the size of NLO EW corrections to the decay branching ratios $\Delta_{\mathrm{EW}}^{\mathrm{BR}}(A\to f\bar{f})$ in Scenario~A and Scenario~B.

\subsection{Scenario~A}
\begin{figure}[t]
	\centering
	\includegraphics[scale=0.475]{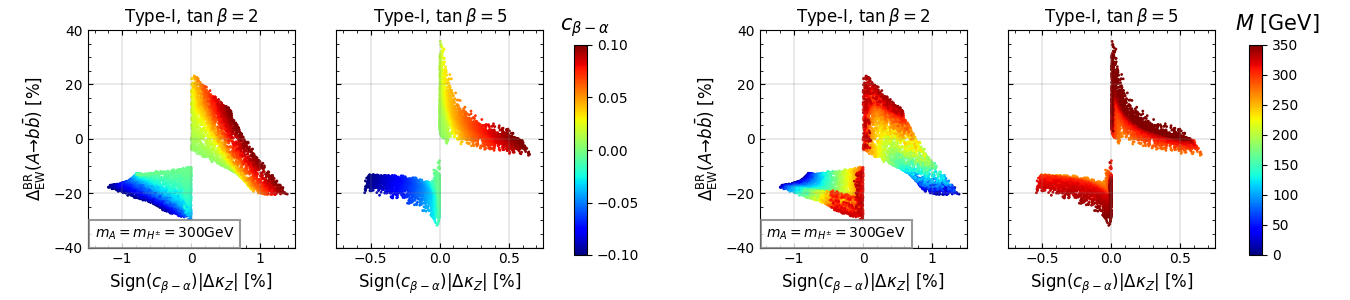}
	\caption{NLO corrections to the decay branching ratios of the CP-odd Higgs boson in Type-I 2HDM in Scenario~A.
	The color differences correspond to the values of $c_{\beta-\alpha}$ and $\sqrt{M^{2}}$, in the left and right panels, respectively.}
	\label{dBrA300_NLO_scan_ff}
\end{figure}

In Fig.~\ref{dBrA300_NLO_scan_ff}, we show the size of NLO EW corrections to the decay branching ratios $\Delta_{\mathrm{EW}}^{\mathrm{BR}}(A\to b\bar{b})$ in Scenario~A.
The color differences correspond to the values of $c_{\beta-\alpha}$ and $\sqrt{M^{2}}$, in the left and right panels, respectively.
$\Delta_{\mathrm{EW}}^{\mathrm{BR}}(A\to b\bar{b})$ can reach $+20\%\, (-30\%)$ at most when $c_{\beta-\alpha}>0\, (c_{\beta-\alpha}<0)$ with $\tan{\beta}=2$, while it can reach $+40\%\, (-30\%)$ at most when $c_{\beta-\alpha}>0\, (c_{\beta-\alpha}<0)$ with $\tan{\beta}=5$.
We note that $\Delta_{\mathrm{EW}}^{\mathrm{BR}}(A\to b\bar{b})$ takes non-zero value even if $c_{\beta-\alpha}\simeq 0$ because $\mathrm{BR}_{\mathrm{LO}}(A\to b\bar{b})$ does not become 100\%, and $\overline{\Delta}_{\mathrm{EW}}(A\to b\bar{b})$ and $\overline{\Delta}^{\mathrm{tot}}_{\mathrm{EW}}$ are not cancelled.

\subsection{Scenario~B}
\begin{figure}[t]
	\centering
	\includegraphics[scale=0.65]{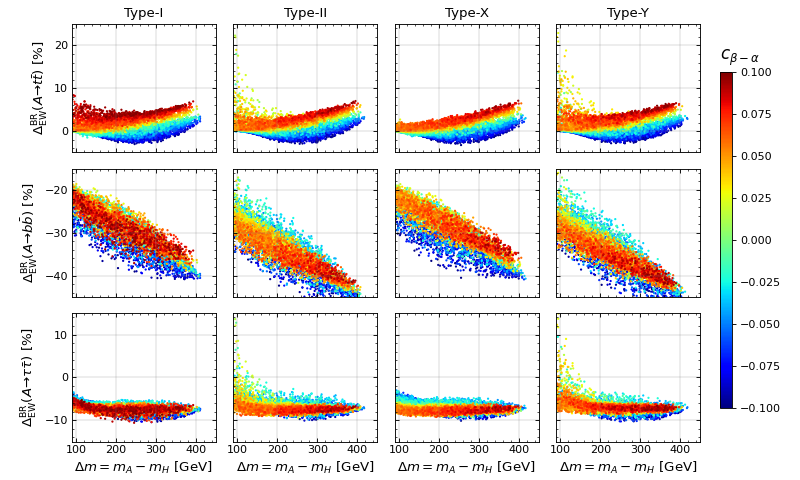}
	\caption{NLO corrections to the decay branching ratios for the CP-odd Higgs boson as a function of $m_{A}-m_{H}$ in Scenario~B.
	Predictions on Type-I, Type-II, Type-X and Type-Y are shown from the left to the right panels in order.
	The color differences correspond to the values of $c_{\beta-\alpha}$.}
	\label{dBrA800_NLO_B_ff}
\end{figure}
In Fig.~\ref{dBrA800_NLO_B_ff}, we show the size of $\Delta_{\mathrm{EW}}^{\mathrm{BR}}(A\to f\bar{f})$ as a function of $\Delta m$.
The color differences correspond to the values of $c_{\beta-\alpha}$.
$\Delta_{\mathrm{EW}}^{\mathrm{BR}}(A\to t\bar{t})$ tends to be positive, while $\overline{\Delta}_{\mathrm{EW}}(A\to t\bar{t})$ is negative in most case.
This is because $\overline{\Delta}^{\mathrm{tot}}_{\mathrm{EW}}$ is negative, and it gives positive contribution for $\Delta_{\mathrm{EW}}^{\mathrm{BR}}(A\to XY)$, as shown in Eq.~\eqref{eq: dBRA decomposed}.
On the other hand, both $\Delta_{\mathrm{EW}}^{\mathrm{BR}}(A\to b\bar{b})$ and $\overline{\Delta}_{\mathrm{EW}}(A\to b\bar{b})$ are negative because the magnitude of $\overline{\Delta}_{\mathrm{EW}}(A\to b\bar{b})$ is larger than that of $\overline{\Delta}^{\mathrm{tot}}_{\mathrm{EW}}$.
As similar to $\Delta_{\mathrm{EW}}^{\mathrm{BR}}(A\to b\bar{b})$, $\Delta_{\mathrm{EW}}^{\mathrm{BR}}(A\to \tau\bar{\tau})$ is negative except for the parameter regions with $\Delta m \lesssim 150$ GeV and $c_{\beta-\alpha}\simeq 0$ in Type-II and Type-Y, where $\tan{\beta}$ enhancement in $\overline{\Delta}^{\mathrm{tot}}_{\mathrm{EW}}$ gives large positive contribution.